\documentclass[aps,prd,superscriptaddress,preprintnumbers,notitlepage]{revtex4-1}

    \usepackage{graphicx}
    
    \usepackage{float}
    \usepackage{xcolor} 
    \usepackage{enumerate} 
    \usepackage{geometry} 
    \usepackage{amsmath} 
    \usepackage{amssymb} 
    \usepackage{textcomp} 
    \AtBeginDocument{%
        \def\PYZsq{\textquotesingle}
    }
    \usepackage{upquote} 
    \usepackage{eurosym} 
    \usepackage[mathletters]{ucs} 
    \usepackage{fancyvrb} 
    \usepackage[breakable]{tcolorbox}

    \definecolor{cellborder}{HTML}{CFCFCF}
    \definecolor{cellbackground}{HTML}{F7F7F7}
    \definecolor{cverbbg}{gray}{0.93}
        
    \usepackage{grffile} 
    \makeatletter 
    \@ifpackagelater{grffile}{2019/11/01}
    {
    }
    {
      \def\Gread@@xetex#1{%
        \IfFileExists{"\Gin@base".bb}%
        {\Gread@eps{\Gin@base.bb}}%
        {\Gread@@xetex@aux#1}%
      }
    }
    \makeatother
    \usepackage[Export]{adjustbox} 
    \adjustboxset{max size={0.9\linewidth}{0.9\paperheight}}

    \usepackage{hyperref}
    \usepackage{longtable} 
    \usepackage{booktabs}  
    \usepackage[inline]{enumitem} 
    \usepackage[normalem]{ulem} 
    \usepackage{mathrsfs}

    \definecolor{urlcolor}{rgb}{0,.145,.698}
    \definecolor{linkcolor}{rgb}{.71,0.21,0.01}
    \definecolor{citecolor}{rgb}{.12,.54,.11}

    \definecolor{ansi-black}{HTML}{3E424D}
    \definecolor{ansi-black-intense}{HTML}{282C36}
    \definecolor{ansi-red}{HTML}{E75C58}
    \definecolor{ansi-red-intense}{HTML}{B22B31}
    \definecolor{ansi-green}{HTML}{00A250}
    \definecolor{ansi-green-intense}{HTML}{007427}
    \definecolor{ansi-yellow}{HTML}{DDB62B}
    \definecolor{ansi-yellow-intense}{HTML}{B27D12}
    \definecolor{ansi-blue}{HTML}{208FFB}
    \definecolor{ansi-blue-intense}{HTML}{0065CA}
    \definecolor{ansi-magenta}{HTML}{D160C4}
    \definecolor{ansi-magenta-intense}{HTML}{A03196}
    \definecolor{ansi-cyan}{HTML}{60C6C8}
    \definecolor{ansi-cyan-intense}{HTML}{258F8F}
    \definecolor{ansi-white}{HTML}{C5C1B4}
    \definecolor{ansi-white-intense}{HTML}{A1A6B2}
    \definecolor{ansi-default-inverse-fg}{HTML}{FFFFFF}
    \definecolor{ansi-default-inverse-bg}{HTML}{000000}

    \definecolor{outerrorbackground}{HTML}{FFDFDF}

    \providecommand{\tightlist}{%
      \setlength{\itemsep}{0pt}\setlength{\parskip}{0pt}}
    \DefineVerbatimEnvironment{Highlighting}{Verbatim}{commandchars=\\\{\}}

    \let\Oldtex\TeX
    \let\Oldlatex\LaTeX
    \renewcommand{\TeX}{\textrm{\Oldtex}}
    \renewcommand{\LaTeX}{\textrm{\Oldlatex}}

\makeatletter
\def\PY@reset{\let\PY@it=\relax \let\PY@bf=\relax%
    \let\PY@ul=\relax \let\PY@tc=\relax%
    \let\PY@bc=\relax \let\PY@ff=\relax}
\def\PY@tok#1{\csname PY@tok@#1\endcsname}
\def\PY@toks#1+{\ifx\relax#1\empty\else%
    \PY@tok{#1}\expandafter\PY@toks\fi}
\def\PY@do#1{\PY@bc{\PY@tc{\PY@ul{%
    \PY@it{\PY@bf{\PY@ff{#1}}}}}}}
\def\PY#1#2{\PY@reset\PY@toks#1+\relax+\PY@do{#2}}

\@namedef{PY@tok@w}{\def\PY@tc##1{\textcolor[rgb]{0.73,0.73,0.73}{##1}}}
\@namedef{PY@tok@c}{\let\PY@it=\textit\def\PY@tc##1{\textcolor[rgb]{0.25,0.50,0.50}{##1}}}
\@namedef{PY@tok@cp}{\def\PY@tc##1{\textcolor[rgb]{0.74,0.48,0.00}{##1}}}
\@namedef{PY@tok@k}{\let\PY@bf=\textbf\def\PY@tc##1{\textcolor[rgb]{0.00,0.50,0.00}{##1}}}
\@namedef{PY@tok@kp}{\def\PY@tc##1{\textcolor[rgb]{0.00,0.50,0.00}{##1}}}
\@namedef{PY@tok@kt}{\def\PY@tc##1{\textcolor[rgb]{0.69,0.00,0.25}{##1}}}
\@namedef{PY@tok@o}{\def\PY@tc##1{\textcolor[rgb]{0.40,0.40,0.40}{##1}}}
\@namedef{PY@tok@ow}{\let\PY@bf=\textbf\def\PY@tc##1{\textcolor[rgb]{0.67,0.13,1.00}{##1}}}
\@namedef{PY@tok@nb}{\def\PY@tc##1{\textcolor[rgb]{0.00,0.50,0.00}{##1}}}
\@namedef{PY@tok@nf}{\def\PY@tc##1{\textcolor[rgb]{0.00,0.00,1.00}{##1}}}
\@namedef{PY@tok@nc}{\let\PY@bf=\textbf\def\PY@tc##1{\textcolor[rgb]{0.00,0.00,1.00}{##1}}}
\@namedef{PY@tok@nn}{\let\PY@bf=\textbf\def\PY@tc##1{\textcolor[rgb]{0.00,0.00,1.00}{##1}}}
\@namedef{PY@tok@ne}{\let\PY@bf=\textbf\def\PY@tc##1{\textcolor[rgb]{0.82,0.25,0.23}{##1}}}
\@namedef{PY@tok@nv}{\def\PY@tc##1{\textcolor[rgb]{0.10,0.09,0.49}{##1}}}
\@namedef{PY@tok@no}{\def\PY@tc##1{\textcolor[rgb]{0.53,0.00,0.00}{##1}}}
\@namedef{PY@tok@nl}{\def\PY@tc##1{\textcolor[rgb]{0.63,0.63,0.00}{##1}}}
\@namedef{PY@tok@ni}{\let\PY@bf=\textbf\def\PY@tc##1{\textcolor[rgb]{0.60,0.60,0.60}{##1}}}
\@namedef{PY@tok@na}{\def\PY@tc##1{\textcolor[rgb]{0.49,0.56,0.16}{##1}}}
\@namedef{PY@tok@nt}{\let\PY@bf=\textbf\def\PY@tc##1{\textcolor[rgb]{0.00,0.50,0.00}{##1}}}
\@namedef{PY@tok@nd}{\def\PY@tc##1{\textcolor[rgb]{0.67,0.13,1.00}{##1}}}
\@namedef{PY@tok@s}{\def\PY@tc##1{\textcolor[rgb]{0.73,0.13,0.13}{##1}}}
\@namedef{PY@tok@sd}{\let\PY@it=\textit\def\PY@tc##1{\textcolor[rgb]{0.73,0.13,0.13}{##1}}}
\@namedef{PY@tok@si}{\let\PY@bf=\textbf\def\PY@tc##1{\textcolor[rgb]{0.73,0.40,0.53}{##1}}}
\@namedef{PY@tok@se}{\let\PY@bf=\textbf\def\PY@tc##1{\textcolor[rgb]{0.73,0.40,0.13}{##1}}}
\@namedef{PY@tok@sr}{\def\PY@tc##1{\textcolor[rgb]{0.73,0.40,0.53}{##1}}}
\@namedef{PY@tok@ss}{\def\PY@tc##1{\textcolor[rgb]{0.10,0.09,0.49}{##1}}}
\@namedef{PY@tok@sx}{\def\PY@tc##1{\textcolor[rgb]{0.00,0.50,0.00}{##1}}}
\@namedef{PY@tok@m}{\def\PY@tc##1{\textcolor[rgb]{0.40,0.40,0.40}{##1}}}
\@namedef{PY@tok@gh}{\let\PY@bf=\textbf\def\PY@tc##1{\textcolor[rgb]{0.00,0.00,0.50}{##1}}}
\@namedef{PY@tok@gu}{\let\PY@bf=\textbf\def\PY@tc##1{\textcolor[rgb]{0.50,0.00,0.50}{##1}}}
\@namedef{PY@tok@gd}{\def\PY@tc##1{\textcolor[rgb]{0.63,0.00,0.00}{##1}}}
\@namedef{PY@tok@gi}{\def\PY@tc##1{\textcolor[rgb]{0.00,0.63,0.00}{##1}}}
\@namedef{PY@tok@gr}{\def\PY@tc##1{\textcolor[rgb]{1.00,0.00,0.00}{##1}}}
\@namedef{PY@tok@ge}{\let\PY@it=\textit}
\@namedef{PY@tok@gs}{\let\PY@bf=\textbf}
\@namedef{PY@tok@gp}{\let\PY@bf=\textbf\def\PY@tc##1{\textcolor[rgb]{0.00,0.00,0.50}{##1}}}
\@namedef{PY@tok@go}{\def\PY@tc##1{\textcolor[rgb]{0.53,0.53,0.53}{##1}}}
\@namedef{PY@tok@gt}{\def\PY@tc##1{\textcolor[rgb]{0.00,0.27,0.87}{##1}}}
\@namedef{PY@tok@err}{\def\PY@bc##1{{\setlength{\fboxsep}{\string -\fboxrule}\fcolorbox[rgb]{1.00,0.00,0.00}{1,1,1}{\strut ##1}}}}
\@namedef{PY@tok@kc}{\let\PY@bf=\textbf\def\PY@tc##1{\textcolor[rgb]{0.00,0.50,0.00}{##1}}}
\@namedef{PY@tok@kd}{\let\PY@bf=\textbf\def\PY@tc##1{\textcolor[rgb]{0.00,0.50,0.00}{##1}}}
\@namedef{PY@tok@kn}{\let\PY@bf=\textbf\def\PY@tc##1{\textcolor[rgb]{0.00,0.50,0.00}{##1}}}
\@namedef{PY@tok@kr}{\let\PY@bf=\textbf\def\PY@tc##1{\textcolor[rgb]{0.00,0.50,0.00}{##1}}}
\@namedef{PY@tok@bp}{\def\PY@tc##1{\textcolor[rgb]{0.00,0.50,0.00}{##1}}}
\@namedef{PY@tok@fm}{\def\PY@tc##1{\textcolor[rgb]{0.00,0.00,1.00}{##1}}}
\@namedef{PY@tok@vc}{\def\PY@tc##1{\textcolor[rgb]{0.10,0.09,0.49}{##1}}}
\@namedef{PY@tok@vg}{\def\PY@tc##1{\textcolor[rgb]{0.10,0.09,0.49}{##1}}}
\@namedef{PY@tok@vi}{\def\PY@tc##1{\textcolor[rgb]{0.10,0.09,0.49}{##1}}}
\@namedef{PY@tok@vm}{\def\PY@tc##1{\textcolor[rgb]{0.10,0.09,0.49}{##1}}}
\@namedef{PY@tok@sa}{\def\PY@tc##1{\textcolor[rgb]{0.73,0.13,0.13}{##1}}}
\@namedef{PY@tok@sb}{\def\PY@tc##1{\textcolor[rgb]{0.73,0.13,0.13}{##1}}}
\@namedef{PY@tok@sc}{\def\PY@tc##1{\textcolor[rgb]{0.73,0.13,0.13}{##1}}}
\@namedef{PY@tok@dl}{\def\PY@tc##1{\textcolor[rgb]{0.73,0.13,0.13}{##1}}}
\@namedef{PY@tok@s2}{\def\PY@tc##1{\textcolor[rgb]{0.73,0.13,0.13}{##1}}}
\@namedef{PY@tok@sh}{\def\PY@tc##1{\textcolor[rgb]{0.73,0.13,0.13}{##1}}}
\@namedef{PY@tok@s1}{\def\PY@tc##1{\textcolor[rgb]{0.73,0.13,0.13}{##1}}}
\@namedef{PY@tok@mb}{\def\PY@tc##1{\textcolor[rgb]{0.40,0.40,0.40}{##1}}}
\@namedef{PY@tok@mf}{\def\PY@tc##1{\textcolor[rgb]{0.40,0.40,0.40}{##1}}}
\@namedef{PY@tok@mh}{\def\PY@tc##1{\textcolor[rgb]{0.40,0.40,0.40}{##1}}}
\@namedef{PY@tok@mi}{\def\PY@tc##1{\textcolor[rgb]{0.40,0.40,0.40}{##1}}}
\@namedef{PY@tok@il}{\def\PY@tc##1{\textcolor[rgb]{0.40,0.40,0.40}{##1}}}
\@namedef{PY@tok@mo}{\def\PY@tc##1{\textcolor[rgb]{0.40,0.40,0.40}{##1}}}
\@namedef{PY@tok@ch}{\let\PY@it=\textit\def\PY@tc##1{\textcolor[rgb]{0.25,0.50,0.50}{##1}}}
\@namedef{PY@tok@cm}{\let\PY@it=\textit\def\PY@tc##1{\textcolor[rgb]{0.25,0.50,0.50}{##1}}}
\@namedef{PY@tok@cpf}{\let\PY@it=\textit\def\PY@tc##1{\textcolor[rgb]{0.25,0.50,0.50}{##1}}}
\@namedef{PY@tok@c1}{\let\PY@it=\textit\def\PY@tc##1{\textcolor[rgb]{0.25,0.50,0.50}{##1}}}
\@namedef{PY@tok@cs}{\let\PY@it=\textit\def\PY@tc##1{\textcolor[rgb]{0.25,0.50,0.50}{##1}}}

\def\PYZbs{\char`\\}
\def\PYZus{\char`\_}
\def\PYZob{\char`\{}
\def\PYZcb{\char`\}}
\def\PYZca{\char`\^}

\def\PYZlt{\char`\<}
\def\PYZgt{\char`\>}
\def\PYZsh{\char`\#}
\def\PYZpc{\char`\%}
\def\PYZdl{\char`\$}
\def\PYZhy{\char`\-}
\def\PYZsq{\char`\'}
\def\PYZdq{\char`\"}
\def\PYZti{\char`\~}

\makeatother

    \hypersetup{
      breaklinks=true,  
      colorlinks=true,
      urlcolor=urlcolor,
      linkcolor=linkcolor,
      citecolor=citecolor,
      }
    
\begin{document}
    
  \renewcommand{\tightlist}{}
  \title{Introduction to Normalizing Flows for Lattice Field Theory}
  
    \date{\today}

        \author{Michael~S.~Albergo}\email{albergo@nyu.edu}\affiliation{Center for Cosmology and Particle Physics, New York University, New York, NY 10003, USA}
        \author{Denis Boyda}\email{boyda@mit.edu}\affiliation{Argonne Leadership Computing Facility, Argonne National Laboratory, Lemont IL 60439, USA}\affiliation{Center for Theoretical Physics, Massachusetts Institute of Technology, Cambridge, MA 02139, USA}\affiliation{The NSF AI Institute for Artificial Intelligence and Fundamental Interactions}
        \author{Daniel C.~Hackett}\email{dhackett@mit.edu}\affiliation{Center for Theoretical Physics, Massachusetts Institute of Technology, Cambridge, MA 02139, USA}\affiliation{The NSF AI Institute for Artificial Intelligence and Fundamental Interactions}
        \author{Gurtej~Kanwar}\email{gurtej@mit.edu}\affiliation{Center for Theoretical Physics, Massachusetts Institute of Technology, Cambridge, MA 02139, USA}\affiliation{The NSF AI Institute for Artificial Intelligence and Fundamental Interactions}
        \author{Kyle~Cranmer}\affiliation{Center for Cosmology and Particle Physics, New York University, New York, NY 10003, USA}
        \author{S\'{e}bastien~Racani\`{e}re}\affiliation{DeepMind, London, UK}
        \author{Danilo~Jimenez~Rezende}\affiliation{DeepMind, London, UK}
        \author{Phiala~E.~Shanahan}\affiliation{Center for Theoretical Physics, Massachusetts Institute of Technology, Cambridge, MA 02139, USA}\affiliation{The NSF AI Institute for Artificial Intelligence and Fundamental Interactions}

    \preprint{MIT-CTP/5272}

  \begin{abstract}
      This notebook tutorial demonstrates a method for sampling Boltzmann distributions of lattice field theories using a class of machine learning models known as normalizing flows. The ideas and approaches proposed in arXiv:1904.12072, arXiv:2002.02428, and arXiv:2003.06413 are reviewed and a concrete implementation of the framework is presented. We apply this framework to a lattice scalar field theory and to U(1) gauge theory, explicitly encoding gauge symmetries in the flow-based approach to the latter. This presentation is intended to be interactive and working with the attached Jupyter notebook is recommended.
      \end{abstract}
  \maketitle

  \hypertarget{introduction-to-normalizing-flows-for-lattice-field-theory}{%
\section{Introduction to Normalizing Flows for Lattice Field
Theory}\label{introduction-to-normalizing-flows-for-lattice-field-theory}}

  A central challenge in lattice field theory is devising algorithms to efficiently generate field configurations. In recent works \cite{Albergo:2019eim,Rezende:2020hrd,Kanwar:2020xzo} we have demonstrated a promising new method based on normalizing flows, a class of probabilistic machine-learning models for which both direct sampling and exact likelihood evaluation are computationally tractable. The aims of this tutorial are to introduce the reader to the normalizing flow method and its application to scalar and gauge field theory.

We first work through some toy examples which illustrate the underlying concept of normalizing flows as a change of variables. From there, we straightforwardly generalize to more expressive forms that can parametrize samplers for close approximations of our distributions of interest. We detail how such approximations can be corrected using MCMC methods, yielding provably correct statistics. As an important part of our toolkit, we show how we can dramatically reduce the complexity of these models by constraining them to be equivariant with respect to physical symmetries: (a subgroup of) lattice translational symmetries and, for U(1) gauge theory, local gauge invariance. Readers unfamiliar with the notebook format should read this document as a single annotated program, wherein all code is executed sequentially from start to finish without clearing the scope.

  We begin by defining a few utilities and importing common packages.
Readers may safely execute and skip over the remainder of this section.

  \begin{tcolorbox}[breakable, size=fbox, boxrule=1pt, pad at break*=1mm,colback=cellbackground, colframe=cellborder]
    \begin{Verbatim}[commandchars=\\\{\},fontsize=\small]
\PY{k+kn}{import} \PY{n+nn}{base64}
\PY{k+kn}{import} \PY{n+nn}{io}
\PY{k+kn}{import} \PY{n+nn}{pickle}
\PY{k+kn}{import} \PY{n+nn}{numpy} \PY{k}{as} \PY{n+nn}{np}
\PY{k+kn}{import} \PY{n+nn}{torch}
\PY{n+nb}{print}\PY{p}{(}\PY{l+s+sa}{f}\PY{l+s+s1}{\PYZsq{}}\PY{l+s+s1}{TORCH VERSION: }\PY{l+s+si}{\PYZob{}}\PY{n}{torch}\PY{o}{.}\PY{n}{\PYZus{}\PYZus{}version\PYZus{}\PYZus{}}\PY{l+s+si}{\PYZcb{}}\PY{l+s+s1}{\PYZsq{}}\PY{p}{)}
\PY{k+kn}{import} \PY{n+nn}{packaging}\PY{n+nn}{.}\PY{n+nn}{version}
\PY{k}{if} \PY{n}{packaging}\PY{o}{.}\PY{n}{version}\PY{o}{.}\PY{n}{parse}\PY{p}{(}\PY{n}{torch}\PY{o}{.}\PY{n}{\PYZus{}\PYZus{}version\PYZus{}\PYZus{}}\PY{p}{)} \PY{o}{\PYZlt{}} \PY{n}{packaging}\PY{o}{.}\PY{n}{version}\PY{o}{.}\PY{n}{parse}\PY{p}{(}\PY{l+s+s1}{\PYZsq{}}\PY{l+s+s1}{1.5.0}\PY{l+s+s1}{\PYZsq{}}\PY{p}{)}\PY{p}{:}
    \PY{k}{raise} \PY{n+ne}{RuntimeError}\PY{p}{(}\PY{l+s+s1}{\PYZsq{}}\PY{l+s+s1}{Torch versions lower than 1.5.0 not supported}\PY{l+s+s1}{\PYZsq{}}\PY{p}{)}

\PY{o}{\PYZpc{}}\PY{k}{matplotlib} inline
\PY{k+kn}{import} \PY{n+nn}{matplotlib}\PY{n+nn}{.}\PY{n+nn}{pyplot} \PY{k}{as} \PY{n+nn}{plt}
\PY{k+kn}{import} \PY{n+nn}{seaborn} \PY{k}{as} \PY{n+nn}{sns}
\PY{n}{sns}\PY{o}{.}\PY{n}{set\PYZus{}style}\PY{p}{(}\PY{l+s+s1}{\PYZsq{}}\PY{l+s+s1}{whitegrid}\PY{l+s+s1}{\PYZsq{}}\PY{p}{)}
    \end{Verbatim}
  \end{tcolorbox}

{\color{gray}
    \begin{Verbatim}[commandchars=\\\{\},fontsize=\small]
>>> TORCH VERSION: 1.6.0
    \end{Verbatim}
}

  \begin{tcolorbox}[breakable, size=fbox, boxrule=1pt, pad at break*=1mm,colback=cellbackground, colframe=cellborder]
    \begin{Verbatim}[commandchars=\\\{\},fontsize=\small]
\PY{k}{if} \PY{n}{torch}\PY{o}{.}\PY{n}{cuda}\PY{o}{.}\PY{n}{is\PYZus{}available}\PY{p}{(}\PY{p}{)}\PY{p}{:}
    \PY{n}{torch\PYZus{}device} \PY{o}{=} \PY{l+s+s1}{\PYZsq{}}\PY{l+s+s1}{cuda}\PY{l+s+s1}{\PYZsq{}}
    \PY{n}{float\PYZus{}dtype} \PY{o}{=} \PY{n}{np}\PY{o}{.}\PY{n}{float32} \PY{c+c1}{\PYZsh{} single}
    \PY{n}{torch}\PY{o}{.}\PY{n}{set\PYZus{}default\PYZus{}tensor\PYZus{}type}\PY{p}{(}\PY{n}{torch}\PY{o}{.}\PY{n}{cuda}\PY{o}{.}\PY{n}{FloatTensor}\PY{p}{)}
\PY{k}{else}\PY{p}{:}
    \PY{n}{torch\PYZus{}device} \PY{o}{=} \PY{l+s+s1}{\PYZsq{}}\PY{l+s+s1}{cpu}\PY{l+s+s1}{\PYZsq{}}
    \PY{n}{float\PYZus{}dtype} \PY{o}{=} \PY{n}{np}\PY{o}{.}\PY{n}{float64} \PY{c+c1}{\PYZsh{} double}
    \PY{n}{torch}\PY{o}{.}\PY{n}{set\PYZus{}default\PYZus{}tensor\PYZus{}type}\PY{p}{(}\PY{n}{torch}\PY{o}{.}\PY{n}{DoubleTensor}\PY{p}{)}
\PY{n+nb}{print}\PY{p}{(}\PY{l+s+sa}{f}\PY{l+s+s2}{\PYZdq{}}\PY{l+s+s2}{TORCH DEVICE: }\PY{l+s+si}{\PYZob{}}\PY{n}{torch\PYZus{}device}\PY{l+s+si}{\PYZcb{}}\PY{l+s+s2}{\PYZdq{}}\PY{p}{)}
    \end{Verbatim}
  \end{tcolorbox}

{\color{gray}
    \begin{Verbatim}[commandchars=\\\{\},fontsize=\small]
>>> TORCH DEVICE: cpu
    \end{Verbatim}
}

  \begin{tcolorbox}[breakable, size=fbox, boxrule=1pt, pad at break*=1mm,colback=cellbackground, colframe=cellborder]
    \begin{Verbatim}[commandchars=\\\{\},fontsize=\small]
\PY{k}{def} \PY{n+nf}{torch\PYZus{}mod}\PY{p}{(}\PY{n}{x}\PY{p}{)}\PY{p}{:}
    \PY{k}{return} \PY{n}{torch}\PY{o}{.}\PY{n}{remainder}\PY{p}{(}\PY{n}{x}\PY{p}{,} \PY{l+m+mi}{2}\PY{o}{*}\PY{n}{np}\PY{o}{.}\PY{n}{pi}\PY{p}{)}
\PY{k}{def} \PY{n+nf}{torch\PYZus{}wrap}\PY{p}{(}\PY{n}{x}\PY{p}{)}\PY{p}{:}
    \PY{k}{return} \PY{n}{torch\PYZus{}mod}\PY{p}{(}\PY{n}{x}\PY{o}{+}\PY{n}{np}\PY{o}{.}\PY{n}{pi}\PY{p}{)} \PY{o}{\PYZhy{}} \PY{n}{np}\PY{o}{.}\PY{n}{pi}
    \end{Verbatim}
  \end{tcolorbox}

  Often we want to detach tensors from the computational graph and pull
them to the CPU as a numpy array.

  \begin{tcolorbox}[breakable, size=fbox, boxrule=1pt, pad at break*=1mm,colback=cellbackground, colframe=cellborder]
    \begin{Verbatim}[commandchars=\\\{\},fontsize=\small]
\PY{k}{def} \PY{n+nf}{grab}\PY{p}{(}\PY{n}{var}\PY{p}{)}\PY{p}{:}
    \PY{k}{return} \PY{n}{var}\PY{o}{.}\PY{n}{detach}\PY{p}{(}\PY{p}{)}\PY{o}{.}\PY{n}{cpu}\PY{p}{(}\PY{p}{)}\PY{o}{.}\PY{n}{numpy}\PY{p}{(}\PY{p}{)}
    \end{Verbatim}
  \end{tcolorbox}

  The code below makes a live-updating plot during training.

  \begin{tcolorbox}[breakable, size=fbox, boxrule=1pt, pad at break*=1mm,colback=cellbackground, colframe=cellborder]
    \begin{Verbatim}[commandchars=\\\{\},fontsize=\small]
\PY{k+kn}{from} \PY{n+nn}{IPython}\PY{n+nn}{.}\PY{n+nn}{display} \PY{k+kn}{import} \PY{n}{display}

\PY{k}{def} \PY{n+nf}{init\PYZus{}live\PYZus{}plot}\PY{p}{(}\PY{n}{dpi}\PY{o}{=}\PY{l+m+mi}{125}\PY{p}{,} \PY{n}{figsize}\PY{o}{=}\PY{p}{(}\PY{l+m+mi}{8}\PY{p}{,}\PY{l+m+mi}{4}\PY{p}{)}\PY{p}{)}\PY{p}{:}
    \PY{n}{fig}\PY{p}{,} \PY{n}{ax\PYZus{}ess} \PY{o}{=} \PY{n}{plt}\PY{o}{.}\PY{n}{subplots}\PY{p}{(}\PY{l+m+mi}{1}\PY{p}{,}\PY{l+m+mi}{1}\PY{p}{,} \PY{n}{dpi}\PY{o}{=}\PY{n}{dpi}\PY{p}{,} \PY{n}{figsize}\PY{o}{=}\PY{n}{figsize}\PY{p}{)}
    \PY{n}{plt}\PY{o}{.}\PY{n}{xlim}\PY{p}{(}\PY{l+m+mi}{0}\PY{p}{,} \PY{n}{N\PYZus{}era}\PY{o}{*}\PY{n}{N\PYZus{}epoch}\PY{p}{)}
    \PY{n}{plt}\PY{o}{.}\PY{n}{ylim}\PY{p}{(}\PY{l+m+mi}{0}\PY{p}{,} \PY{l+m+mi}{1}\PY{p}{)}
    
    \PY{n}{ess\PYZus{}line} \PY{o}{=} \PY{n}{plt}\PY{o}{.}\PY{n}{plot}\PY{p}{(}\PY{p}{[}\PY{l+m+mi}{0}\PY{p}{]}\PY{p}{,}\PY{p}{[}\PY{l+m+mi}{0}\PY{p}{]}\PY{p}{,} \PY{n}{alpha}\PY{o}{=}\PY{l+m+mf}{0.5}\PY{p}{)} \PY{c+c1}{\PYZsh{} dummy}
    \PY{n}{plt}\PY{o}{.}\PY{n}{grid}\PY{p}{(}\PY{k+kc}{False}\PY{p}{)}
    \PY{n}{plt}\PY{o}{.}\PY{n}{ylabel}\PY{p}{(}\PY{l+s+s1}{\PYZsq{}}\PY{l+s+s1}{ESS}\PY{l+s+s1}{\PYZsq{}}\PY{p}{)}
    
    \PY{n}{ax\PYZus{}loss} \PY{o}{=} \PY{n}{ax\PYZus{}ess}\PY{o}{.}\PY{n}{twinx}\PY{p}{(}\PY{p}{)}
    \PY{n}{loss\PYZus{}line} \PY{o}{=} \PY{n}{plt}\PY{o}{.}\PY{n}{plot}\PY{p}{(}\PY{p}{[}\PY{l+m+mi}{0}\PY{p}{]}\PY{p}{,}\PY{p}{[}\PY{l+m+mi}{0}\PY{p}{]}\PY{p}{,} \PY{n}{alpha}\PY{o}{=}\PY{l+m+mf}{0.5}\PY{p}{,} \PY{n}{c}\PY{o}{=}\PY{l+s+s1}{\PYZsq{}}\PY{l+s+s1}{orange}\PY{l+s+s1}{\PYZsq{}}\PY{p}{)} \PY{c+c1}{\PYZsh{} dummy}
    \PY{n}{plt}\PY{o}{.}\PY{n}{grid}\PY{p}{(}\PY{k+kc}{False}\PY{p}{)}
    \PY{n}{plt}\PY{o}{.}\PY{n}{ylabel}\PY{p}{(}\PY{l+s+s1}{\PYZsq{}}\PY{l+s+s1}{Loss}\PY{l+s+s1}{\PYZsq{}}\PY{p}{)}
    
    \PY{n}{plt}\PY{o}{.}\PY{n}{xlabel}\PY{p}{(}\PY{l+s+s1}{\PYZsq{}}\PY{l+s+s1}{Epoch}\PY{l+s+s1}{\PYZsq{}}\PY{p}{)}

    \PY{n}{display\PYZus{}id} \PY{o}{=} \PY{n}{display}\PY{p}{(}\PY{n}{fig}\PY{p}{,} \PY{n}{display\PYZus{}id}\PY{o}{=}\PY{k+kc}{True}\PY{p}{)}

    \PY{k}{return} \PY{n+nb}{dict}\PY{p}{(}
        \PY{n}{fig}\PY{o}{=}\PY{n}{fig}\PY{p}{,} \PY{n}{ax\PYZus{}ess}\PY{o}{=}\PY{n}{ax\PYZus{}ess}\PY{p}{,} \PY{n}{ax\PYZus{}loss}\PY{o}{=}\PY{n}{ax\PYZus{}loss}\PY{p}{,}
        \PY{n}{ess\PYZus{}line}\PY{o}{=}\PY{n}{ess\PYZus{}line}\PY{p}{,} \PY{n}{loss\PYZus{}line}\PY{o}{=}\PY{n}{loss\PYZus{}line}\PY{p}{,}
        \PY{n}{display\PYZus{}id}\PY{o}{=}\PY{n}{display\PYZus{}id}
    \PY{p}{)}

\PY{k}{def} \PY{n+nf}{moving\PYZus{}average}\PY{p}{(}\PY{n}{x}\PY{p}{,} \PY{n}{window}\PY{o}{=}\PY{l+m+mi}{10}\PY{p}{)}\PY{p}{:}
    \PY{k}{if} \PY{n+nb}{len}\PY{p}{(}\PY{n}{x}\PY{p}{)} \PY{o}{\PYZlt{}} \PY{n}{window}\PY{p}{:}
        \PY{k}{return} \PY{n}{np}\PY{o}{.}\PY{n}{mean}\PY{p}{(}\PY{n}{x}\PY{p}{,} \PY{n}{keepdims}\PY{o}{=}\PY{k+kc}{True}\PY{p}{)}
    \PY{k}{else}\PY{p}{:}
        \PY{k}{return} \PY{n}{np}\PY{o}{.}\PY{n}{convolve}\PY{p}{(}\PY{n}{x}\PY{p}{,} \PY{n}{np}\PY{o}{.}\PY{n}{ones}\PY{p}{(}\PY{n}{window}\PY{p}{)}\PY{p}{,} \PY{l+s+s1}{\PYZsq{}}\PY{l+s+s1}{valid}\PY{l+s+s1}{\PYZsq{}}\PY{p}{)} \PY{o}{/} \PY{n}{window}

\PY{k}{def} \PY{n+nf}{update\PYZus{}plots}\PY{p}{(}\PY{n}{history}\PY{p}{,} \PY{n}{fig}\PY{p}{,} \PY{n}{ax\PYZus{}ess}\PY{p}{,} \PY{n}{ax\PYZus{}loss}\PY{p}{,} \PY{n}{ess\PYZus{}line}\PY{p}{,} \PY{n}{loss\PYZus{}line}\PY{p}{,} \PY{n}{display\PYZus{}id}\PY{p}{)}\PY{p}{:}
    \PY{n}{Y} \PY{o}{=} \PY{n}{np}\PY{o}{.}\PY{n}{array}\PY{p}{(}\PY{n}{history}\PY{p}{[}\PY{l+s+s1}{\PYZsq{}}\PY{l+s+s1}{ess}\PY{l+s+s1}{\PYZsq{}}\PY{p}{]}\PY{p}{)}
    \PY{n}{Y} \PY{o}{=} \PY{n}{moving\PYZus{}average}\PY{p}{(}\PY{n}{Y}\PY{p}{,} \PY{n}{window}\PY{o}{=}\PY{l+m+mi}{15}\PY{p}{)}
    \PY{n}{ess\PYZus{}line}\PY{p}{[}\PY{l+m+mi}{0}\PY{p}{]}\PY{o}{.}\PY{n}{set\PYZus{}ydata}\PY{p}{(}\PY{n}{Y}\PY{p}{)}
    \PY{n}{ess\PYZus{}line}\PY{p}{[}\PY{l+m+mi}{0}\PY{p}{]}\PY{o}{.}\PY{n}{set\PYZus{}xdata}\PY{p}{(}\PY{n}{np}\PY{o}{.}\PY{n}{arange}\PY{p}{(}\PY{n+nb}{len}\PY{p}{(}\PY{n}{Y}\PY{p}{)}\PY{p}{)}\PY{p}{)}
    \PY{n}{Y} \PY{o}{=} \PY{n}{history}\PY{p}{[}\PY{l+s+s1}{\PYZsq{}}\PY{l+s+s1}{loss}\PY{l+s+s1}{\PYZsq{}}\PY{p}{]}
    \PY{n}{Y} \PY{o}{=} \PY{n}{moving\PYZus{}average}\PY{p}{(}\PY{n}{Y}\PY{p}{,} \PY{n}{window}\PY{o}{=}\PY{l+m+mi}{15}\PY{p}{)}
    \PY{n}{loss\PYZus{}line}\PY{p}{[}\PY{l+m+mi}{0}\PY{p}{]}\PY{o}{.}\PY{n}{set\PYZus{}ydata}\PY{p}{(}\PY{n}{np}\PY{o}{.}\PY{n}{array}\PY{p}{(}\PY{n}{Y}\PY{p}{)}\PY{p}{)}
    \PY{n}{loss\PYZus{}line}\PY{p}{[}\PY{l+m+mi}{0}\PY{p}{]}\PY{o}{.}\PY{n}{set\PYZus{}xdata}\PY{p}{(}\PY{n}{np}\PY{o}{.}\PY{n}{arange}\PY{p}{(}\PY{n+nb}{len}\PY{p}{(}\PY{n}{Y}\PY{p}{)}\PY{p}{)}\PY{p}{)}
    \PY{n}{ax\PYZus{}loss}\PY{o}{.}\PY{n}{relim}\PY{p}{(}\PY{p}{)}
    \PY{n}{ax\PYZus{}loss}\PY{o}{.}\PY{n}{autoscale\PYZus{}view}\PY{p}{(}\PY{p}{)}
    \PY{n}{fig}\PY{o}{.}\PY{n}{canvas}\PY{o}{.}\PY{n}{draw}\PY{p}{(}\PY{p}{)}
    \PY{n}{display\PYZus{}id}\PY{o}{.}\PY{n}{update}\PY{p}{(}\PY{n}{fig}\PY{p}{)} \PY{c+c1}{\PYZsh{} need to force colab to update plot}
    \end{Verbatim}
  \end{tcolorbox}

  \hypertarget{notation}{%
\section{Notation}\label{notation}}

This section is intended as a reference. The phrases and notation listed
here will be defined in detail in the remainder of the notebook.

\begin{enumerate}
\def\labelenumi{\arabic{enumi}.}
\tightlist
\item
  \textbf{Notation for generic normalizing flows}
\end{enumerate}

\begin{itemize}
\tightlist
\item
  Coordinates \(z, x \in\) some manifold \(\mathcal{X}\) (a space with
  local \(\mathbb{R}^n\) structure)\\
  The manifolds used here are \(\mathcal{X} = \mathbb{R}^n\) (for scalar
  field theory) and \(\mathcal{X} = \mathbb{T}^n\) (for
  \(\mathrm{U}(1)\) gauge theory) where \(\mathbb{T}^n\) refers to the
  n-dimensional torus.
\item
  Probability densities over those manifolds,

  \begin{itemize}
  \tightlist
  \item
    Prior density \(r(z)\)
  \item
    Model density \(q(x)\)
  \item
    Target density \(p(x)\)
  \end{itemize}
\item
  Normalizing flow \(f: \mathcal{X} \rightarrow \mathcal{X}\),
  invertible and differentiable
\item
  Jacobian factor \(J(z) = |\det_{ij} \partial f_i(z) / \partial z_j|\)
\item
  Coupling layer \(g: \mathcal{X} \rightarrow \mathcal{X}\), invertible
  and differentiable
\item
  Subsets of the components of the coordinate \(x = (x_1, x_2)\), where
  the choice of subsets will be clear from context
\end{itemize}

\begin{enumerate}
\def\labelenumi{\arabic{enumi}.}
\setcounter{enumi}{1}
\tightlist
\item
  \textbf{Notation for lattice field theories}
\end{enumerate}

\begin{itemize}
\tightlist
\item
  Lattice spacing \(a\)\\
  We work in ``lattice units'' where \(a=1\).
\item
  Spacetime dimension \(N_d\)\\
  We work in this notebook with \(N_d=2\).
\item
  Lattice extent \(L\), with volume \(V = L^{N_d} = L^2\), in lattice
  units where \(a=1\).
\item
  Lattice position \(\vec{x} = a\vec{n} \equiv (an_x, an_y)\), with
  \(\vec{x}=\vec{n}\) in lattice units where \(a=1\). We use
  \(n_x, n_y \in [0, L-1]\).
\end{itemize}

\begin{enumerate}
\def\labelenumi{\arabic{enumi}.}
\setcounter{enumi}{2}
\tightlist
\item
  \textbf{Notation for normalizing flows targeting scalar lattice field
  theory}
\end{enumerate}

\begin{itemize}
\tightlist
\item
  Field configurations \(z \in \mathbb{R}^V\) or
  \(\phi \in \mathbb{R}^V\), corresponding to \(z\) or \(x\) in the
  generic notation
\item
  \(\phi(\vec{n})\) denotes the field configuration which lives on the
  sites of the lattice, while \(\phi_{\vec{n}}\) denotes the unraveled
  1D vector of lattice DOF
\item
  Action \(S[\phi] \in \mathbb{R}\)
\item
  Discretized path integral measure \(\prod_{\vec{n}} d\phi_{\vec{n}}\)
\end{itemize}

\begin{enumerate}
\def\labelenumi{\arabic{enumi}.}
\setcounter{enumi}{3}
\tightlist
\item
  \textbf{Notation for normalizing flows targeting U(1) lattice gauge
  theory}
\end{enumerate}

\begin{itemize}
\tightlist
\item
  Field configurations \(U \in \mathbb{T}^{N_d V}\) or
  \(U' \in \mathbb{T}^{N_d V}\), corresponding to \(z\) or \(x\) in the
  generic notation
\item
  \(U_\mu(\vec{n})\) denotes the component of field configuration \(U\)
  which lives on the link \((n, n+\hat{\mu})\) of the lattice, where
  \(\mu \in [0, N_d-1]\) indicates the Cartesian direction.
  \(U_{\mu,\vec{n}}\) denotes the unraveled 1D vector of lattice DOF
\item
  Action \(S[U] \in \mathbb{R}\)
\item
  Angular parameterization of each component
  \(U_{\mu, \vec{n}} \equiv e^{i\theta_{\mu, \vec{n}}}\)
\item
  Discretized path integral measure
  \(\prod_{\mu,\vec{n}} dU_{\mu,\vec{n}}\), where
  \(dU_{\mu, \vec{n}} = d\theta_{\mu, \vec{n}}\) is the Haar measure for
  \(\mathrm{U}(1)\)
\end{itemize}

  \hypertarget{normalizing-flows-for-lattice-qfts}{%
\section{Normalizing flows (for lattice
QFTs)}\label{normalizing-flows-for-lattice-qfts}}

A powerful method to generate samples from complicated distributions is
to combine (1) sampling from a simpler / tractable distribution with (2)
applying a deterministic change-of-variables (a \emph{normalizing flow})
to the output samples. The transformed samples are distributed according
to a new distribution which is determined by the initial distribution
and change-of-variables. These two components together define a
\emph{normalizing flow model}. See \cite{papamakarios2019normalizing}
for a review.

  \hypertarget{a-simple-example}{%
\subsection{\texorpdfstring{\textbf{A simple
example}}{A simple example}}\label{a-simple-example}}

The Box-Muller transform is an example of this trick in practice: to
produce Gaussian random variables, draw two variables \(U_1\) and
\(U_2\) from \(\text{unif}(0,1)\) then change variables to

\begin{equation}
    Z_1 = \sqrt{-2 \ln{U_1}} \cos(2\pi U_2)
    \quad \text{and} \quad
    Z_2 = \sqrt{-2 \ln{U_1}} \sin(2\pi U_2).
\end{equation}

The resulting variables \(Z_1, Z_2\) are then distributed according to
an uncorrelated, unit-variance Gaussian distribution.

  \begin{tcolorbox}[breakable, size=fbox, boxrule=1pt, pad at break*=1mm,colback=cellbackground, colframe=cellborder]
    \begin{Verbatim}[commandchars=\\\{\},fontsize=\small]
\PY{n}{batch\PYZus{}size} \PY{o}{=} \PY{l+m+mi}{2}\PY{o}{*}\PY{o}{*}\PY{l+m+mi}{14}
\PY{n}{u} \PY{o}{=} \PY{n}{np}\PY{o}{.}\PY{n}{random}\PY{o}{.}\PY{n}{random}\PY{p}{(}\PY{n}{size}\PY{o}{=}\PY{p}{(}\PY{n}{batch\PYZus{}size}\PY{p}{,} \PY{l+m+mi}{2}\PY{p}{)}\PY{p}{)}
\PY{n}{z} \PY{o}{=} \PY{n}{np}\PY{o}{.}\PY{n}{sqrt}\PY{p}{(}\PY{o}{\PYZhy{}}\PY{l+m+mi}{2}\PY{o}{*}\PY{n}{np}\PY{o}{.}\PY{n}{log}\PY{p}{(}\PY{n}{u}\PY{p}{[}\PY{p}{:}\PY{p}{,}\PY{l+m+mi}{0}\PY{p}{]}\PY{p}{)}\PY{p}{)}\PY{p}{[}\PY{p}{:}\PY{p}{,}\PY{n}{np}\PY{o}{.}\PY{n}{newaxis}\PY{p}{]} \PY{o}{*} \PY{n}{np}\PY{o}{.}\PY{n}{stack}\PY{p}{(}
    \PY{p}{(}\PY{n}{np}\PY{o}{.}\PY{n}{cos}\PY{p}{(}\PY{l+m+mi}{2}\PY{o}{*}\PY{n}{np}\PY{o}{.}\PY{n}{pi}\PY{o}{*}\PY{n}{u}\PY{p}{[}\PY{p}{:}\PY{p}{,}\PY{l+m+mi}{1}\PY{p}{]}\PY{p}{)}\PY{p}{,} \PY{n}{np}\PY{o}{.}\PY{n}{sin}\PY{p}{(}\PY{l+m+mi}{2}\PY{o}{*}\PY{n}{np}\PY{o}{.}\PY{n}{pi}\PY{o}{*}\PY{n}{u}\PY{p}{[}\PY{p}{:}\PY{p}{,}\PY{l+m+mi}{1}\PY{p}{]}\PY{p}{)}\PY{p}{)}\PY{p}{,} \PY{n}{axis}\PY{o}{=}\PY{o}{\PYZhy{}}\PY{l+m+mi}{1}\PY{p}{)}

\PY{n}{fig}\PY{p}{,} \PY{n}{ax} \PY{o}{=} \PY{n}{plt}\PY{o}{.}\PY{n}{subplots}\PY{p}{(}\PY{l+m+mi}{1}\PY{p}{,}\PY{l+m+mi}{2}\PY{p}{,} \PY{n}{dpi}\PY{o}{=}\PY{l+m+mi}{125}\PY{p}{,} \PY{n}{figsize}\PY{o}{=}\PY{p}{(}\PY{l+m+mi}{4}\PY{p}{,}\PY{l+m+mi}{2}\PY{p}{)}\PY{p}{)}
\PY{k}{for} \PY{n}{a} \PY{o+ow}{in} \PY{n}{ax}\PY{p}{:}
    \PY{n}{a}\PY{o}{.}\PY{n}{set\PYZus{}xticks}\PY{p}{(}\PY{p}{[}\PY{o}{\PYZhy{}}\PY{l+m+mi}{2}\PY{p}{,} \PY{l+m+mi}{0}\PY{p}{,} \PY{l+m+mi}{2}\PY{p}{]}\PY{p}{)}
    \PY{n}{a}\PY{o}{.}\PY{n}{set\PYZus{}yticks}\PY{p}{(}\PY{p}{[}\PY{o}{\PYZhy{}}\PY{l+m+mi}{2}\PY{p}{,} \PY{l+m+mi}{0}\PY{p}{,} \PY{l+m+mi}{2}\PY{p}{]}\PY{p}{)}
    \PY{n}{a}\PY{o}{.}\PY{n}{set\PYZus{}aspect}\PY{p}{(}\PY{l+s+s1}{\PYZsq{}}\PY{l+s+s1}{equal}\PY{l+s+s1}{\PYZsq{}}\PY{p}{)}
\PY{n}{ax}\PY{p}{[}\PY{l+m+mi}{0}\PY{p}{]}\PY{o}{.}\PY{n}{hist2d}\PY{p}{(}\PY{n}{u}\PY{p}{[}\PY{p}{:}\PY{p}{,}\PY{l+m+mi}{0}\PY{p}{]}\PY{p}{,} \PY{n}{u}\PY{p}{[}\PY{p}{:}\PY{p}{,}\PY{l+m+mi}{1}\PY{p}{]}\PY{p}{,} \PY{n}{bins}\PY{o}{=}\PY{l+m+mi}{30}\PY{p}{,} \PY{n+nb}{range}\PY{o}{=}\PY{p}{[}\PY{p}{[}\PY{o}{\PYZhy{}}\PY{l+m+mf}{3.0}\PY{p}{,}\PY{l+m+mf}{3.0}\PY{p}{]}\PY{p}{,} \PY{p}{[}\PY{o}{\PYZhy{}}\PY{l+m+mf}{3.0}\PY{p}{,}\PY{l+m+mf}{3.0}\PY{p}{]}\PY{p}{]}\PY{p}{)}
\PY{n}{ax}\PY{p}{[}\PY{l+m+mi}{0}\PY{p}{]}\PY{o}{.}\PY{n}{set\PYZus{}xlabel}\PY{p}{(}\PY{l+s+sa}{r}\PY{l+s+s2}{\PYZdq{}}\PY{l+s+s2}{\PYZdl{}U\PYZus{}1\PYZdl{}}\PY{l+s+s2}{\PYZdq{}}\PY{p}{)}
\PY{n}{ax}\PY{p}{[}\PY{l+m+mi}{0}\PY{p}{]}\PY{o}{.}\PY{n}{set\PYZus{}ylabel}\PY{p}{(}\PY{l+s+sa}{r}\PY{l+s+s2}{\PYZdq{}}\PY{l+s+s2}{\PYZdl{}U\PYZus{}2\PYZdl{}}\PY{l+s+s2}{\PYZdq{}}\PY{p}{,} \PY{n}{rotation}\PY{o}{=}\PY{l+m+mi}{0}\PY{p}{,} \PY{n}{y}\PY{o}{=}\PY{l+m+mf}{.46}\PY{p}{)}
\PY{n}{ax}\PY{p}{[}\PY{l+m+mi}{1}\PY{p}{]}\PY{o}{.}\PY{n}{hist2d}\PY{p}{(}\PY{n}{z}\PY{p}{[}\PY{p}{:}\PY{p}{,}\PY{l+m+mi}{0}\PY{p}{]}\PY{p}{,} \PY{n}{z}\PY{p}{[}\PY{p}{:}\PY{p}{,}\PY{l+m+mi}{1}\PY{p}{]}\PY{p}{,} \PY{n}{bins}\PY{o}{=}\PY{l+m+mi}{30}\PY{p}{,} \PY{n+nb}{range}\PY{o}{=}\PY{p}{[}\PY{p}{[}\PY{o}{\PYZhy{}}\PY{l+m+mf}{3.0}\PY{p}{,}\PY{l+m+mf}{3.0}\PY{p}{]}\PY{p}{,} \PY{p}{[}\PY{o}{\PYZhy{}}\PY{l+m+mf}{3.0}\PY{p}{,}\PY{l+m+mf}{3.0}\PY{p}{]}\PY{p}{]}\PY{p}{)}
\PY{n}{ax}\PY{p}{[}\PY{l+m+mi}{1}\PY{p}{]}\PY{o}{.}\PY{n}{set\PYZus{}yticklabels}\PY{p}{(}\PY{p}{[}\PY{p}{]}\PY{p}{)}
\PY{n}{ax}\PY{p}{[}\PY{l+m+mi}{1}\PY{p}{]}\PY{o}{.}\PY{n}{set\PYZus{}xlabel}\PY{p}{(}\PY{l+s+sa}{r}\PY{l+s+s2}{\PYZdq{}}\PY{l+s+s2}{\PYZdl{}Z\PYZus{}1\PYZdl{}}\PY{l+s+s2}{\PYZdq{}}\PY{p}{)}
\PY{n}{ax}\PY{p}{[}\PY{l+m+mi}{1}\PY{p}{]}\PY{o}{.}\PY{n}{set\PYZus{}ylabel}\PY{p}{(}\PY{l+s+sa}{r}\PY{l+s+s2}{\PYZdq{}}\PY{l+s+s2}{\PYZdl{}Z\PYZus{}2\PYZdl{}}\PY{l+s+s2}{\PYZdq{}}\PY{p}{,} \PY{n}{rotation}\PY{o}{=}\PY{l+m+mi}{0}\PY{p}{,} \PY{n}{y}\PY{o}{=}\PY{l+m+mf}{.53}\PY{p}{)}
\PY{n}{ax}\PY{p}{[}\PY{l+m+mi}{1}\PY{p}{]}\PY{o}{.}\PY{n}{yaxis}\PY{o}{.}\PY{n}{set\PYZus{}label\PYZus{}position}\PY{p}{(}\PY{l+s+s2}{\PYZdq{}}\PY{l+s+s2}{right}\PY{l+s+s2}{\PYZdq{}}\PY{p}{)}
\PY{n}{ax}\PY{p}{[}\PY{l+m+mi}{1}\PY{p}{]}\PY{o}{.}\PY{n}{yaxis}\PY{o}{.}\PY{n}{tick\PYZus{}right}\PY{p}{(}\PY{p}{)}
\PY{n}{plt}\PY{o}{.}\PY{n}{show}\PY{p}{(}\PY{p}{)}
    \end{Verbatim}
  \end{tcolorbox}

    \begin{center}
    \adjustimage{max size={0.9\linewidth}{0.9\paperheight}}{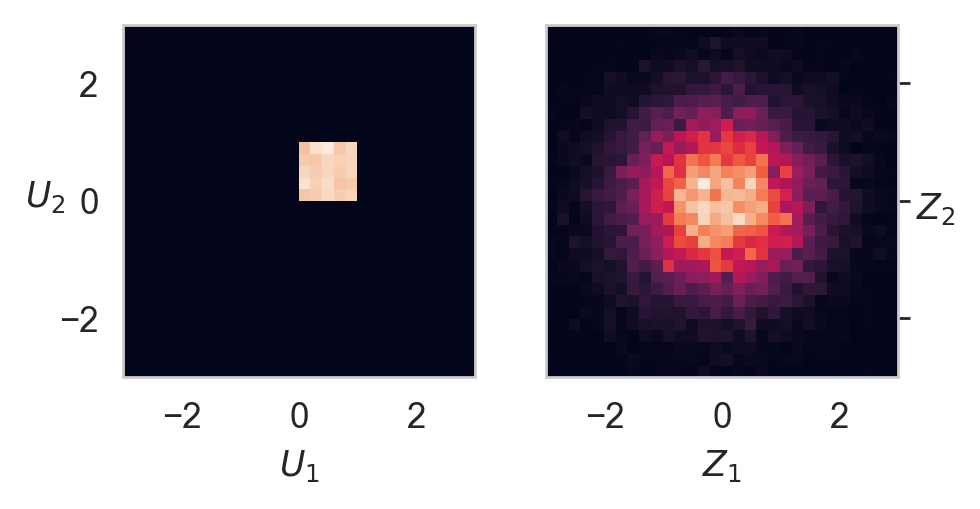}
    \end{center}
    { \hspace*{\fill} \\}
    
  We can analytically compute the density associated with output samples
by the \textbf{change-of-variables formula} relating the \emph{prior
density} \(r(U_1, U_2) = 1\) to the \emph{output density}
\(q(Z_1, Z_2)\):

\begin{equation}
\begin{split}
    q(Z_1, Z_2) &= r(U_1, U_2) \left| \det_{kl} \frac{\partial Z_k(U_1, U_2)}{\partial U_l} \right|^{-1} \\
    &= 1 \times \left| \det \left( \begin{matrix}
        \frac{-1}{U_1 \sqrt{-2 \ln{U_1}}} \cos(2\pi U_2) &
        - 2\pi \sqrt{-2 \ln{U_1}} \sin(2\pi U_2) \\
        \frac{-1}{U_1 \sqrt{-2 \ln{U_1}}} \sin(2\pi U_2) &
        2\pi \sqrt{-2 \ln{U_1}} \cos(2\pi U_2)
        \end{matrix} \right) \right|^{-1} \\
    &= \left| \frac{2 \pi}{U_1} \right|^{-1}.
\end{split}
\end{equation}

Here, the term
\(J(U_1, U_2) \equiv \left| \det_{kl} \frac{\partial Z_k(U_1, U_2)}{\partial U_l} \right|\)
is the determinant of the Jacobian of the transformation from
\((U_1,U_2)\) to \((Z_1,Z_2)\). Intuitively, the Jacobian factor can be
thought of as a change in volume element, therefore the
change-of-variables formula must contain the inverse of this factor
(spreading out volume decreases density). To complete the example, we
can rearrange the change of variables to find
\(U_1 = \exp(-(Z_1^2 + Z_2^2) / 2)\) and therefore \begin{equation}
    q(Z_1, Z_2) = \frac{1}{2\pi} e^{-(Z_1^2 + Z_2^2)/2}.
\end{equation}

\textbf{NOTE}: In this example, the model has no free parameters because
we didn't need any to create a transform that exactly reproduced our
target distribution (independent, unit-variance Gaussian). In general,
we may not know a normalizing flow that exactly produces our desired
distribution, and so instead construct parametrized models that we can
variationally optimize to \emph{approximate} that target distribution,
and because we can compute the density these can be corrected to
nevertheless guarantee exactness.

  \hypertarget{the-general-approach}{%
\subsection{\texorpdfstring{\textbf{The general
approach}}{The general approach}}\label{the-general-approach}}

Generalizing this example, it is clear that any invertible and
differentiable function \(f(z)\) will transform a prior density \(r(z)\)
on the (possibly multi-dimensional) random variable \(z\) to an output
density \(q(x)\) on \(x \equiv f(z)\). If the Jacobian factor
\(J(z) \equiv |\det_{kl} \partial f_k(z) / \partial z_l |\) is
efficiently calculable, we can compute the output density
\textbf{alongside} any samples drawn using the change-of-variables
formula, \begin{equation}
    q(x) = r(z) [J(z)]^{-1} = r(z) \left|\det_{kl} \frac{\partial f_k(z)}{ \partial z_l} \right|^{-1}.
\end{equation}

In some cases, it is easy to compute the Jacobian factor even when the
whole Jacobian matrix is intractable; for example, only the diagonal
elements are needed if the Jacobian matrix is known to be triangular.
Below we will see how to construct \(f\) with a triangular Jacobian
using \emph{coupling layers}.

In lattice field theory simulations, our goal is to draw samples from a
distribution over lattice field configurations defined by the
imaginary-time path integral. By optimizing the function \(f\) we hope
to find an output distribution that closely models this desired physical
distribution. If the family of functions is \textbf{expressive}
(i.e.~includes a wide variety of possible functions) we expect the
optimal choice to be a good approximation to the true distribution.
Moreover, we can make the task of searching for the optimal choice more
efficient by restricting to functions that guarantee certain
\textbf{symmetries} in the output distribution. Once we have a good
approximation to the output distribution, we can draw samples from it
and use MCMC methods or reweighting to correct their statistics to the
exact distribution of interest.

  \hypertarget{prior-distributions}{%
\subsection{\texorpdfstring{\textbf{Prior
distributions}}{Prior distributions}}\label{prior-distributions}}

Any probability distribution that is easy to sample from and has
calculable density \(r(z)\) can be used as the prior distribution.

In code, our interface mimics a subset of the pytorch
\texttt{Distribution} interface. For example, below we define a prior
distribution corresponding to uncorrelated Gaussians (one per component
of the field). Any other distribution you may want to define should
provide analogous methods \texttt{log\_prob} and \texttt{sample\_n}.

  \begin{tcolorbox}[breakable, size=fbox, boxrule=1pt, pad at break*=1mm,colback=cellbackground, colframe=cellborder]
    \begin{Verbatim}[commandchars=\\\{\},fontsize=\small]
\PY{k}{class} \PY{n+nc}{SimpleNormal}\PY{p}{:}
    \PY{k}{def} \PY{n+nf+fm}{\PYZus{}\PYZus{}init\PYZus{}\PYZus{}}\PY{p}{(}\PY{n+nb+bp}{self}\PY{p}{,} \PY{n}{loc}\PY{p}{,} \PY{n}{var}\PY{p}{)}\PY{p}{:}
        \PY{n+nb+bp}{self}\PY{o}{.}\PY{n}{dist} \PY{o}{=} \PY{n}{torch}\PY{o}{.}\PY{n}{distributions}\PY{o}{.}\PY{n}{normal}\PY{o}{.}\PY{n}{Normal}\PY{p}{(}
            \PY{n}{torch}\PY{o}{.}\PY{n}{flatten}\PY{p}{(}\PY{n}{loc}\PY{p}{)}\PY{p}{,} \PY{n}{torch}\PY{o}{.}\PY{n}{flatten}\PY{p}{(}\PY{n}{var}\PY{p}{)}\PY{p}{)}
        \PY{n+nb+bp}{self}\PY{o}{.}\PY{n}{shape} \PY{o}{=} \PY{n}{loc}\PY{o}{.}\PY{n}{shape}
    \PY{k}{def} \PY{n+nf}{log\PYZus{}prob}\PY{p}{(}\PY{n+nb+bp}{self}\PY{p}{,} \PY{n}{x}\PY{p}{)}\PY{p}{:}
        \PY{n}{logp} \PY{o}{=} \PY{n+nb+bp}{self}\PY{o}{.}\PY{n}{dist}\PY{o}{.}\PY{n}{log\PYZus{}prob}\PY{p}{(}\PY{n}{x}\PY{o}{.}\PY{n}{reshape}\PY{p}{(}\PY{n}{x}\PY{o}{.}\PY{n}{shape}\PY{p}{[}\PY{l+m+mi}{0}\PY{p}{]}\PY{p}{,} \PY{o}{\PYZhy{}}\PY{l+m+mi}{1}\PY{p}{)}\PY{p}{)}
        \PY{k}{return} \PY{n}{torch}\PY{o}{.}\PY{n}{sum}\PY{p}{(}\PY{n}{logp}\PY{p}{,} \PY{n}{dim}\PY{o}{=}\PY{l+m+mi}{1}\PY{p}{)}
    \PY{k}{def} \PY{n+nf}{sample\PYZus{}n}\PY{p}{(}\PY{n+nb+bp}{self}\PY{p}{,} \PY{n}{batch\PYZus{}size}\PY{p}{)}\PY{p}{:}
        \PY{n}{x} \PY{o}{=} \PY{n+nb+bp}{self}\PY{o}{.}\PY{n}{dist}\PY{o}{.}\PY{n}{sample}\PY{p}{(}\PY{p}{(}\PY{n}{batch\PYZus{}size}\PY{p}{,}\PY{p}{)}\PY{p}{)}
        \PY{k}{return} \PY{n}{x}\PY{o}{.}\PY{n}{reshape}\PY{p}{(}\PY{n}{batch\PYZus{}size}\PY{p}{,} \PY{o}{*}\PY{n+nb+bp}{self}\PY{o}{.}\PY{n}{shape}\PY{p}{)}
    \end{Verbatim}
  \end{tcolorbox}

  The shape of \texttt{loc} and \texttt{var} determine the shape of
samples drawn.

  \begin{tcolorbox}[breakable, size=fbox, boxrule=1pt, pad at break*=1mm,colback=cellbackground, colframe=cellborder]
    \begin{Verbatim}[commandchars=\\\{\},fontsize=\small]
\PY{n}{normal\PYZus{}prior} \PY{o}{=} \PY{n}{SimpleNormal}\PY{p}{(}\PY{n}{torch}\PY{o}{.}\PY{n}{zeros}\PY{p}{(}\PY{p}{(}\PY{l+m+mi}{3}\PY{p}{,}\PY{l+m+mi}{4}\PY{p}{,}\PY{l+m+mi}{5}\PY{p}{)}\PY{p}{)}\PY{p}{,} \PY{n}{torch}\PY{o}{.}\PY{n}{ones}\PY{p}{(}\PY{p}{(}\PY{l+m+mi}{3}\PY{p}{,}\PY{l+m+mi}{4}\PY{p}{,}\PY{l+m+mi}{5}\PY{p}{)}\PY{p}{)}\PY{p}{)}
\PY{n}{z} \PY{o}{=} \PY{n}{normal\PYZus{}prior}\PY{o}{.}\PY{n}{sample\PYZus{}n}\PY{p}{(}\PY{l+m+mi}{17}\PY{p}{)}
\PY{n+nb}{print}\PY{p}{(}\PY{l+s+sa}{f}\PY{l+s+s1}{\PYZsq{}}\PY{l+s+s1}{z.shape = }\PY{l+s+si}{\PYZob{}}\PY{n}{z}\PY{o}{.}\PY{n}{shape}\PY{l+s+si}{\PYZcb{}}\PY{l+s+s1}{\PYZsq{}}\PY{p}{)}
\PY{n+nb}{print}\PY{p}{(}\PY{l+s+sa}{f}\PY{l+s+s1}{\PYZsq{}}\PY{l+s+s1}{log r(z) = }\PY{l+s+si}{\PYZob{}}\PY{n}{grab}\PY{p}{(}\PY{n}{normal\PYZus{}prior}\PY{o}{.}\PY{n}{log\PYZus{}prob}\PY{p}{(}\PY{n}{z}\PY{p}{)}\PY{p}{)}\PY{l+s+si}{\PYZcb{}}\PY{l+s+s1}{\PYZsq{}}\PY{p}{)}
    \end{Verbatim}
  \end{tcolorbox}

{\color{gray}
    \begin{Verbatim}[commandchars=\\\{\},fontsize=\small]
>>> z.shape = torch.Size([17, 3, 4, 5])
... log r(z) = [-82.15322048 -92.57309315 -81.64606318 -81.02122974 -82.402781
...  -87.32209986 -89.71201831 -80.20303503 -84.56155853 -89.50678784
...  -84.31995159 -90.90156267 -82.82711487 -80.48685168 -88.90803586
...  -80.86843481 -87.73830259]
    \end{Verbatim}
}

  We use \texttt{SimpleNormal} as the prior distribution for scalar field
theory, and later define a uniform distribution as the prior
distribution for \(\mathrm{U}(1)\) gauge theory.

  \hypertarget{designing-the-flow-f}{%
\subsection{\texorpdfstring{\textbf{Designing the flow
\(f\)}}{Designing the flow f}}\label{designing-the-flow-f}}

As a reminder, a normalizing flow \(f\) must be \textbf{invertible} and
\textbf{differentiable}. To be useful, it should also be efficient to
compute the Jacobian factor and be expressive.

Expressive functions can be built through composition of simpler ones.
When each simpler function is invertible and differentiable, the
composed function is as well. Schematically, this subdivides the task of
learning a complicated map as below:

  \begin{figure}[H]
      \centering
      \includegraphics{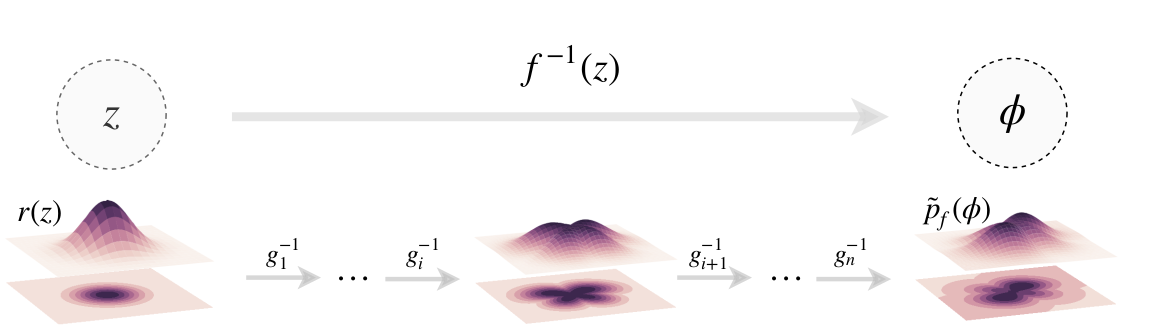}
      \caption{Fig.~1 of \cite{Albergo:2019eim}. The notation superficially differs from what we present here.}
    \end{figure}

  \textbf{Coupling layers} are one approach to defining the \(g_i\) in the
composed function. These functions are defined to update only a subset
of the input variables, conditioned on the complimentary (``frozen'')
subset. For example, if the input to a coupling layer was a lattice with
one real number per site, the layer could be defined to update only the
odd sites in a checkerboard pattern. To ensure all variables are
updated, we could then compose coupling layers that alternatingly update
odd sites and even sites.

In a coupling layer, the transform applied to the updated subset of
variables is manifestly a simply inverted operation such as a scaling
(\(x \rightarrow e^s x\)) or affine transformation
(\(x \rightarrow e^s x + t\)). For example, a coupling layer
\(g(x_1, x_2) = (x_1', x_2')\) based on an scaling transformation looks
like

\begin{equation}
\begin{split}
  x_1 '  &= e^{s(x_2)}x_1 \\
  x_2 '  &= x_2
\end{split}
\end{equation}

where \(x_1, x_2\) are subsets of the components of \(x\). We say that
\(x_1\) is \emph{updated} based on the \emph{frozen} subset \(x_2\),
which is not changed by the coupling layer. \(e^{s(x_2)}\) is a vector
of the same shape as \(x_1\), and \(e^{x(s_2)} x_1\) denotes an
elementwise product. The \emph{parameters defining the transform},
\(s(x_2)\), can be complicated, non-invertible functions of the frozen
subset of variables. However, the inverse of this transformation
\(g^{-1} (x_1', x_2') = (x_1, x_2)\) is simply computed using the same
parameters,

\begin{equation}
\begin{split}
x_1  &= e^{-s(x_2')} x_1' \\
x_2  &= x_2'.
\end{split}
\end{equation}

Here the key to guaranteeing invertibility is that \(x_2 = x_2'\). This
``trick'' is exactly what guarantees invertibility for leapfrog
integrators, which alternately update position and momentum variables.

This also ensures a triangular Jacobian, \begin{equation}
\frac{\partial g(x_1, x_2)}{\partial x} =
\left( \begin{matrix}
    \frac{\partial x_1'}{\partial x_1} & \frac{\partial x_1'}{\partial x_2} \\
    0 & 1
\end{matrix} \right),
\end{equation} which in the scaling example takes the form
\begin{equation}
\frac{\partial g(x_1, x_2)}{\partial x} =
\left( \begin{array}{ccc|ccc}
    e^{[s(x_2)]_1} & & & \cdot & \cdot & \cdot \\
    & e^{[s(x_2)]_2} & & \cdot & \cdot & \cdot \\
    & & \ddots & \cdot & \cdot & \cdot \\
    \hline
    & & & 1 & & \\
    & 0 & & & 1 & \\
    & & & & & \ddots
\end{array} \right)
\end{equation} where we have expanded the blocks over
\((x_1', x_2') \times (x_1, x_2)\) from the first expression. Therefore
\(J(x)\) is efficiently computed as \begin{equation}
J(x) = \left|\det_{kl} \frac{\partial [g(x_1, x_2)]_k}{\partial x_l} \right| = \prod_{k} e^{[s(x_2)]_k}
\end{equation} where \(k\) runs over the components in \(s(x_2)\). The
Jacobian of the inverse transformation is simply
\(J_\text{reverse}(x') = 1/J(x') = \prod_k e^{[-s(x_2')]_k}\); note that
here we were able to compute the reverse Jacobian in terms of the
forward Jacobian applied to \(x'\) because of the simplicity of the
coupling layer.

The coupling layer architecture makes it easier to guarantee
invertibility while retaining expressivity: the functions which provide
the parameters of the transformation are flexible while the inverse and
Jacobian factor of such coupling transformations are easy to compute.
Many such coupling layers can be stacked to compose expressive functions
\(f\) efficiently.

  \hypertarget{simple-coupling-layer-demo}{%
\subsection{\texorpdfstring{\textbf{Simple coupling layer
demo}}{Simple coupling layer demo}}\label{simple-coupling-layer-demo}}

To demonstrate coupling layers in practice, we define a coupling layer
using scaling (see above) for two-dimensional inputs
\(x \equiv (x_1,x_2)\) {[}i.e.~in comparison to the previous section,
here \(x_1\) and \(x_2\) are just scalars{]}. Because we have the
freedom to make the function \(s(x_2)\) arbitrarily complex without
sacrificing invertibility, we parametrize \(s(x_2)\) as a neural net
made of alternating layers of linear transformations and ReLU
(``rectified linear unit'') activation functions, with a \(\tanh\)
activation function after the final linear transform.

  We implement coupling layers as an extension of \texttt{torch.nn.Module}
to include application of \(g\) (see \texttt{forward}) and inverse
\(g^{-1}\) (see \texttt{reverse}). These both map from the domain of
lattice degrees of freedom to itself,
\(\mathcal{X} \rightarrow \mathcal{X}\); in this case, this is just
\(\mathbb{R}^2 \rightarrow \mathbb{R}^2\). The superclass automatically
holds references to all tunable parameters (weights) that will later be
optimized.

  \begin{tcolorbox}[breakable, size=fbox, boxrule=1pt, pad at break*=1mm,colback=cellbackground, colframe=cellborder]
    \begin{Verbatim}[commandchars=\\\{\},fontsize=\small]
\PY{k}{class} \PY{n+nc}{SimpleCouplingLayer}\PY{p}{(}\PY{n}{torch}\PY{o}{.}\PY{n}{nn}\PY{o}{.}\PY{n}{Module}\PY{p}{)}\PY{p}{:}
    \PY{k}{def} \PY{n+nf+fm}{\PYZus{}\PYZus{}init\PYZus{}\PYZus{}}\PY{p}{(}\PY{n+nb+bp}{self}\PY{p}{)}\PY{p}{:}
        \PY{n+nb}{super}\PY{p}{(}\PY{p}{)}\PY{o}{.}\PY{n+nf+fm}{\PYZus{}\PYZus{}init\PYZus{}\PYZus{}}\PY{p}{(}\PY{p}{)}
        \PY{n+nb+bp}{self}\PY{o}{.}\PY{n}{s} \PY{o}{=} \PY{n}{torch}\PY{o}{.}\PY{n}{nn}\PY{o}{.}\PY{n}{Sequential}\PY{p}{(}
            \PY{n}{torch}\PY{o}{.}\PY{n}{nn}\PY{o}{.}\PY{n}{Linear}\PY{p}{(}\PY{l+m+mi}{1}\PY{p}{,} \PY{l+m+mi}{8}\PY{p}{)}\PY{p}{,}
            \PY{n}{torch}\PY{o}{.}\PY{n}{nn}\PY{o}{.}\PY{n}{ReLU}\PY{p}{(}\PY{p}{)}\PY{p}{,}
            \PY{n}{torch}\PY{o}{.}\PY{n}{nn}\PY{o}{.}\PY{n}{Linear}\PY{p}{(}\PY{l+m+mi}{8}\PY{p}{,} \PY{l+m+mi}{8}\PY{p}{)}\PY{p}{,}
            \PY{n}{torch}\PY{o}{.}\PY{n}{nn}\PY{o}{.}\PY{n}{ReLU}\PY{p}{(}\PY{p}{)}\PY{p}{,}
            \PY{n}{torch}\PY{o}{.}\PY{n}{nn}\PY{o}{.}\PY{n}{Linear}\PY{p}{(}\PY{l+m+mi}{8}\PY{p}{,} \PY{l+m+mi}{1}\PY{p}{)}\PY{p}{,}
            \PY{n}{torch}\PY{o}{.}\PY{n}{nn}\PY{o}{.}\PY{n}{Tanh}\PY{p}{(}\PY{p}{)}
        \PY{p}{)}
    \PY{k}{def} \PY{n+nf}{forward}\PY{p}{(}\PY{n+nb+bp}{self}\PY{p}{,} \PY{n}{x}\PY{p}{)}\PY{p}{:}
        \PY{n}{x1}\PY{p}{,} \PY{n}{x2} \PY{o}{=} \PY{n}{x}\PY{p}{[}\PY{p}{:}\PY{p}{,}\PY{l+m+mi}{0}\PY{p}{]}\PY{p}{,} \PY{n}{x}\PY{p}{[}\PY{p}{:}\PY{p}{,}\PY{l+m+mi}{1}\PY{p}{]}
        \PY{n}{s} \PY{o}{=} \PY{n+nb+bp}{self}\PY{o}{.}\PY{n}{s}\PY{p}{(}\PY{n}{x2}\PY{o}{.}\PY{n}{unsqueeze}\PY{p}{(}\PY{o}{\PYZhy{}}\PY{l+m+mi}{1}\PY{p}{)}\PY{p}{)}\PY{o}{.}\PY{n}{squeeze}\PY{p}{(}\PY{o}{\PYZhy{}}\PY{l+m+mi}{1}\PY{p}{)}
        \PY{n}{fx1} \PY{o}{=} \PY{n}{torch}\PY{o}{.}\PY{n}{exp}\PY{p}{(}\PY{n}{s}\PY{p}{)} \PY{o}{*} \PY{n}{x1}
        \PY{n}{fx2} \PY{o}{=} \PY{n}{x2}
        \PY{n}{logJ} \PY{o}{=} \PY{n}{s}
        \PY{k}{return} \PY{n}{torch}\PY{o}{.}\PY{n}{stack}\PY{p}{(}\PY{p}{(}\PY{n}{fx1}\PY{p}{,} \PY{n}{fx2}\PY{p}{)}\PY{p}{,} \PY{n}{dim}\PY{o}{=}\PY{o}{\PYZhy{}}\PY{l+m+mi}{1}\PY{p}{)}\PY{p}{,} \PY{n}{logJ}
    \PY{k}{def} \PY{n+nf}{reverse}\PY{p}{(}\PY{n+nb+bp}{self}\PY{p}{,} \PY{n}{fx}\PY{p}{)}\PY{p}{:}
        \PY{n}{fx1}\PY{p}{,} \PY{n}{fx2} \PY{o}{=} \PY{n}{fx}\PY{p}{[}\PY{p}{:}\PY{p}{,}\PY{l+m+mi}{0}\PY{p}{]}\PY{p}{,} \PY{n}{fx}\PY{p}{[}\PY{p}{:}\PY{p}{,}\PY{l+m+mi}{1}\PY{p}{]}
        \PY{n}{x2} \PY{o}{=} \PY{n}{fx2}
        \PY{n}{s} \PY{o}{=} \PY{n+nb+bp}{self}\PY{o}{.}\PY{n}{s}\PY{p}{(}\PY{n}{x2}\PY{o}{.}\PY{n}{unsqueeze}\PY{p}{(}\PY{o}{\PYZhy{}}\PY{l+m+mi}{1}\PY{p}{)}\PY{p}{)}\PY{o}{.}\PY{n}{squeeze}\PY{p}{(}\PY{o}{\PYZhy{}}\PY{l+m+mi}{1}\PY{p}{)}
        \PY{n}{logJ} \PY{o}{=} \PY{o}{\PYZhy{}}\PY{n}{s}
        \PY{n}{x1} \PY{o}{=} \PY{n}{torch}\PY{o}{.}\PY{n}{exp}\PY{p}{(}\PY{o}{\PYZhy{}}\PY{n}{s}\PY{p}{)} \PY{o}{*} \PY{n}{fx1}
        \PY{k}{return} \PY{n}{torch}\PY{o}{.}\PY{n}{stack}\PY{p}{(}\PY{p}{(}\PY{n}{x1}\PY{p}{,} \PY{n}{x2}\PY{p}{)}\PY{p}{,} \PY{n}{dim}\PY{o}{=}\PY{o}{\PYZhy{}}\PY{l+m+mi}{1}\PY{p}{)}\PY{p}{,} \PY{n}{logJ}

\PY{n}{coupling\PYZus{}layer} \PY{o}{=} \PY{n}{SimpleCouplingLayer}\PY{p}{(}\PY{p}{)}

\PY{c+c1}{\PYZsh{} init weights in a way that gives interesting behavior without training}
\PY{k}{def} \PY{n+nf}{set\PYZus{}weights}\PY{p}{(}\PY{n}{m}\PY{p}{)}\PY{p}{:}
    \PY{k}{if} \PY{n+nb}{hasattr}\PY{p}{(}\PY{n}{m}\PY{p}{,} \PY{l+s+s1}{\PYZsq{}}\PY{l+s+s1}{weight}\PY{l+s+s1}{\PYZsq{}}\PY{p}{)} \PY{o+ow}{and} \PY{n}{m}\PY{o}{.}\PY{n}{weight} \PY{o+ow}{is} \PY{o+ow}{not} \PY{k+kc}{None}\PY{p}{:}
        \PY{n}{torch}\PY{o}{.}\PY{n}{nn}\PY{o}{.}\PY{n}{init}\PY{o}{.}\PY{n}{normal\PYZus{}}\PY{p}{(}\PY{n}{m}\PY{o}{.}\PY{n}{weight}\PY{p}{,} \PY{n}{mean}\PY{o}{=}\PY{l+m+mi}{1}\PY{p}{,} \PY{n}{std}\PY{o}{=}\PY{l+m+mi}{2}\PY{p}{)}
    \PY{k}{if} \PY{n+nb}{hasattr}\PY{p}{(}\PY{n}{m}\PY{p}{,} \PY{l+s+s1}{\PYZsq{}}\PY{l+s+s1}{bias}\PY{l+s+s1}{\PYZsq{}}\PY{p}{)} \PY{o+ow}{and} \PY{n}{m}\PY{o}{.}\PY{n}{bias} \PY{o+ow}{is} \PY{o+ow}{not} \PY{k+kc}{None}\PY{p}{:}
        \PY{n}{m}\PY{o}{.}\PY{n}{bias}\PY{o}{.}\PY{n}{data}\PY{o}{.}\PY{n}{fill\PYZus{}}\PY{p}{(}\PY{o}{\PYZhy{}}\PY{l+m+mi}{1}\PY{p}{)}
\PY{n}{torch}\PY{o}{.}\PY{n}{manual\PYZus{}seed}\PY{p}{(}\PY{l+m+mi}{1234}\PY{p}{)}
\PY{n}{coupling\PYZus{}layer}\PY{o}{.}\PY{n}{s}\PY{o}{.}\PY{n}{apply}\PY{p}{(}\PY{n}{set\PYZus{}weights}\PY{p}{)}\PY{p}{;}
    \end{Verbatim}
  \end{tcolorbox}

  Let's see what our simple coupling layer \(g\) does. We draw a batch of
samples \(x\) from an arbitrary input distribution (uniform in
\([0,1]^2\)), feed it through the coupling layer forwards to get samples
\(g(x)\) from a new distribution, then feed \(g(x)\) backwards through
the coupling layer to double-check that we recover our original sample
\(x' = g^{-1}(g(x)) \overset{!}{=} x\).

  \begin{tcolorbox}[breakable, size=fbox, boxrule=1pt, pad at break*=1mm,colback=cellbackground, colframe=cellborder]
    \begin{Verbatim}[commandchars=\\\{\},fontsize=\small]
\PY{n}{batch\PYZus{}size} \PY{o}{=} \PY{l+m+mi}{1024}
\PY{n}{np\PYZus{}x} \PY{o}{=} \PY{p}{(}\PY{l+m+mi}{2}\PY{o}{*}\PY{n}{np}\PY{o}{.}\PY{n}{random}\PY{o}{.}\PY{n}{random}\PY{p}{(}\PY{n}{size}\PY{o}{=}\PY{p}{(}\PY{n}{batch\PYZus{}size}\PY{p}{,} \PY{l+m+mi}{2}\PY{p}{)}\PY{p}{)} \PY{o}{\PYZhy{}} \PY{l+m+mi}{1}\PY{p}{)}\PY{o}{.}\PY{n}{astype}\PY{p}{(}\PY{n}{float\PYZus{}dtype}\PY{p}{)}
\PY{n}{x} \PY{o}{=} \PY{n}{torch}\PY{o}{.}\PY{n}{from\PYZus{}numpy}\PY{p}{(}\PY{n}{np\PYZus{}x}\PY{p}{)}\PY{o}{.}\PY{n}{to}\PY{p}{(}\PY{n}{torch\PYZus{}device}\PY{p}{)}

\PY{n}{gx}\PY{p}{,} \PY{n}{fwd\PYZus{}logJ} \PY{o}{=} \PY{n}{coupling\PYZus{}layer}\PY{o}{.}\PY{n}{forward}\PY{p}{(}\PY{n}{x}\PY{p}{)}
\PY{n}{xp}\PY{p}{,} \PY{n}{bwd\PYZus{}logJ} \PY{o}{=} \PY{n}{coupling\PYZus{}layer}\PY{o}{.}\PY{n}{reverse}\PY{p}{(}\PY{n}{gx}\PY{p}{)}

\PY{n}{fig}\PY{p}{,} \PY{n}{ax} \PY{o}{=} \PY{n}{plt}\PY{o}{.}\PY{n}{subplots}\PY{p}{(}\PY{l+m+mi}{1}\PY{p}{,}\PY{l+m+mi}{3}\PY{p}{,} \PY{n}{dpi}\PY{o}{=}\PY{l+m+mi}{125}\PY{p}{,} \PY{n}{figsize}\PY{o}{=}\PY{p}{(}\PY{l+m+mi}{6}\PY{p}{,}\PY{l+m+mf}{2.3}\PY{p}{)}\PY{p}{,} \PY{n}{sharex}\PY{o}{=}\PY{k+kc}{True}\PY{p}{,} \PY{n}{sharey}\PY{o}{=}\PY{k+kc}{True}\PY{p}{)}
\PY{n}{np\PYZus{}gx}\PY{p}{,} \PY{n}{np\PYZus{}xp} \PY{o}{=} \PY{n}{grab}\PY{p}{(}\PY{n}{gx}\PY{p}{)}\PY{p}{,} \PY{n}{grab}\PY{p}{(}\PY{n}{xp}\PY{p}{)}
\PY{k}{for} \PY{n}{a} \PY{o+ow}{in} \PY{n}{ax}\PY{p}{:}
    \PY{n}{a}\PY{o}{.}\PY{n}{set\PYZus{}xlim}\PY{p}{(}\PY{o}{\PYZhy{}}\PY{l+m+mf}{1.1}\PY{p}{,}\PY{l+m+mf}{1.1}\PY{p}{)}
    \PY{n}{a}\PY{o}{.}\PY{n}{set\PYZus{}ylim}\PY{p}{(}\PY{o}{\PYZhy{}}\PY{l+m+mf}{1.1}\PY{p}{,}\PY{l+m+mf}{1.1}\PY{p}{)}
\PY{n}{ax}\PY{p}{[}\PY{l+m+mi}{0}\PY{p}{]}\PY{o}{.}\PY{n}{scatter}\PY{p}{(}\PY{n}{np\PYZus{}x}\PY{p}{[}\PY{p}{:}\PY{p}{,}\PY{l+m+mi}{0}\PY{p}{]}\PY{p}{,} \PY{n}{np\PYZus{}x}\PY{p}{[}\PY{p}{:}\PY{p}{,}\PY{l+m+mi}{1}\PY{p}{]}\PY{p}{,} \PY{n}{marker}\PY{o}{=}\PY{l+s+s1}{\PYZsq{}}\PY{l+s+s1}{.}\PY{l+s+s1}{\PYZsq{}}\PY{p}{)}
\PY{n}{ax}\PY{p}{[}\PY{l+m+mi}{0}\PY{p}{]}\PY{o}{.}\PY{n}{set\PYZus{}title}\PY{p}{(}\PY{l+s+sa}{r}\PY{l+s+s1}{\PYZsq{}}\PY{l+s+s1}{\PYZdl{}x\PYZdl{}}\PY{l+s+s1}{\PYZsq{}}\PY{p}{)}
\PY{n}{ax}\PY{p}{[}\PY{l+m+mi}{1}\PY{p}{]}\PY{o}{.}\PY{n}{scatter}\PY{p}{(}\PY{n}{np\PYZus{}gx}\PY{p}{[}\PY{p}{:}\PY{p}{,}\PY{l+m+mi}{0}\PY{p}{]}\PY{p}{,} \PY{n}{np\PYZus{}gx}\PY{p}{[}\PY{p}{:}\PY{p}{,}\PY{l+m+mi}{1}\PY{p}{]}\PY{p}{,} \PY{n}{marker}\PY{o}{=}\PY{l+s+s1}{\PYZsq{}}\PY{l+s+s1}{.}\PY{l+s+s1}{\PYZsq{}}\PY{p}{,} \PY{n}{color}\PY{o}{=}\PY{l+s+s1}{\PYZsq{}}\PY{l+s+s1}{tab:orange}\PY{l+s+s1}{\PYZsq{}}\PY{p}{)}
\PY{n}{ax}\PY{p}{[}\PY{l+m+mi}{1}\PY{p}{]}\PY{o}{.}\PY{n}{set\PYZus{}title}\PY{p}{(}\PY{l+s+sa}{r}\PY{l+s+s1}{\PYZsq{}}\PY{l+s+s1}{\PYZdl{}g(x)\PYZdl{}}\PY{l+s+s1}{\PYZsq{}}\PY{p}{)}
\PY{n}{ax}\PY{p}{[}\PY{l+m+mi}{2}\PY{p}{]}\PY{o}{.}\PY{n}{scatter}\PY{p}{(}\PY{n}{np\PYZus{}xp}\PY{p}{[}\PY{p}{:}\PY{p}{,}\PY{l+m+mi}{0}\PY{p}{]}\PY{p}{,} \PY{n}{np\PYZus{}xp}\PY{p}{[}\PY{p}{:}\PY{p}{,}\PY{l+m+mi}{1}\PY{p}{]}\PY{p}{,} \PY{n}{marker}\PY{o}{=}\PY{l+s+s1}{\PYZsq{}}\PY{l+s+s1}{.}\PY{l+s+s1}{\PYZsq{}}\PY{p}{)}
\PY{n}{ax}\PY{p}{[}\PY{l+m+mi}{2}\PY{p}{]}\PY{o}{.}\PY{n}{set\PYZus{}title}\PY{p}{(}\PY{l+s+sa}{r}\PY{l+s+s2}{\PYZdq{}}\PY{l+s+s2}{\PYZdl{}g\PYZca{}}\PY{l+s+s2}{\PYZob{}}\PY{l+s+s2}{\PYZhy{}1\PYZcb{}(g(x))\PYZdl{}}\PY{l+s+s2}{\PYZdq{}}\PY{p}{)}
\PY{n}{fig}\PY{o}{.}\PY{n}{set\PYZus{}tight\PYZus{}layout}\PY{p}{(}\PY{k+kc}{True}\PY{p}{)}
\PY{n}{plt}\PY{o}{.}\PY{n}{show}\PY{p}{(}\PY{p}{)}
    \end{Verbatim}
  \end{tcolorbox}

    \begin{center}
    \adjustimage{max size={0.9\linewidth}{0.9\paperheight}}{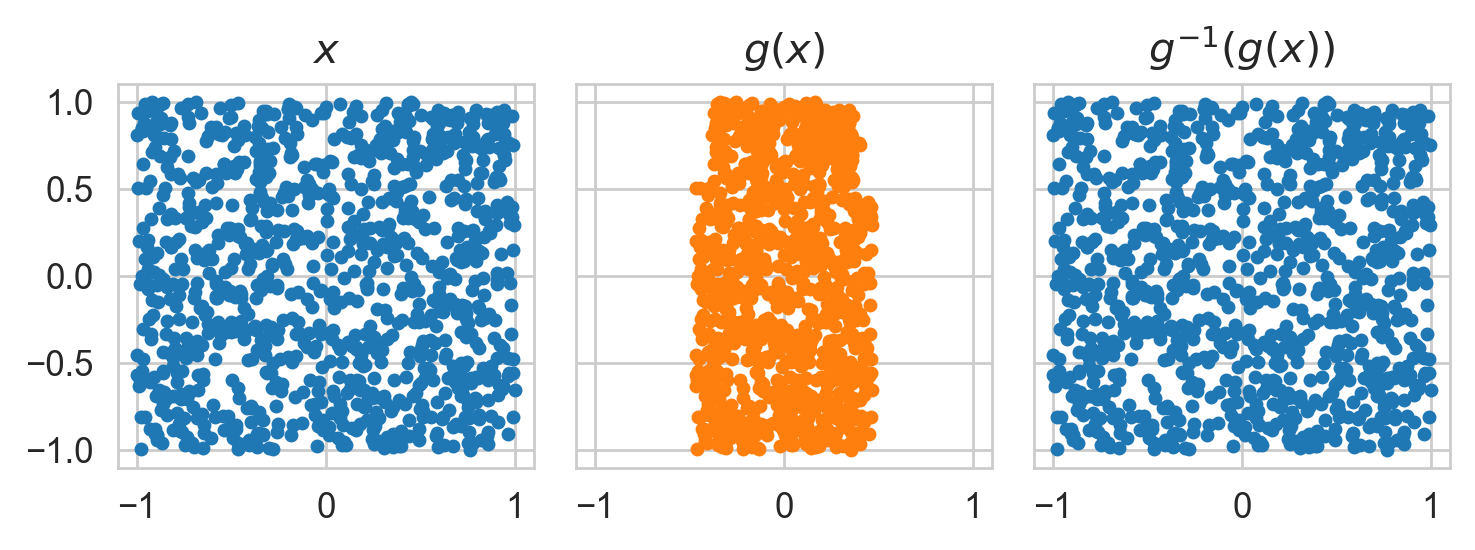}
    \end{center}
    { \hspace*{\fill} \\}
    
  \hypertarget{composition}{%
\subsection{\texorpdfstring{\textbf{Composition}}{Composition}}\label{composition}}

The Jacobian factors \(J_i\) from each coupling layer simply multiply
together to define the Jacobian factor of the composed function, so that
the final density is \begin{equation}
\begin{split}
    q(x) &= r(z) \left| \det \frac{\partial f(z)}{\partial z} \right|^{-1} = r(z) \prod_{i} J_i^{-1}.
\end{split}
\end{equation} In practice, we'll add together log Jacobians instead.
Altogether, sampling and computing the density is simple composition.

  \begin{tcolorbox}[breakable, size=fbox, boxrule=1pt, pad at break*=1mm,colback=cellbackground, colframe=cellborder]
    \begin{Verbatim}[commandchars=\\\{\},fontsize=\small]
\PY{k}{def} \PY{n+nf}{apply\PYZus{}flow\PYZus{}to\PYZus{}prior}\PY{p}{(}\PY{n}{prior}\PY{p}{,} \PY{n}{coupling\PYZus{}layers}\PY{p}{,} \PY{o}{*}\PY{p}{,} \PY{n}{batch\PYZus{}size}\PY{p}{)}\PY{p}{:}
    \PY{n}{x} \PY{o}{=} \PY{n}{prior}\PY{o}{.}\PY{n}{sample\PYZus{}n}\PY{p}{(}\PY{n}{batch\PYZus{}size}\PY{p}{)}
    \PY{n}{logq} \PY{o}{=} \PY{n}{prior}\PY{o}{.}\PY{n}{log\PYZus{}prob}\PY{p}{(}\PY{n}{x}\PY{p}{)}
    \PY{k}{for} \PY{n}{layer} \PY{o+ow}{in} \PY{n}{coupling\PYZus{}layers}\PY{p}{:}
        \PY{n}{x}\PY{p}{,} \PY{n}{logJ} \PY{o}{=} \PY{n}{layer}\PY{o}{.}\PY{n}{forward}\PY{p}{(}\PY{n}{x}\PY{p}{)}
        \PY{n}{logq} \PY{o}{=} \PY{n}{logq} \PY{o}{\PYZhy{}} \PY{n}{logJ}
    \PY{k}{return} \PY{n}{x}\PY{p}{,} \PY{n}{logq}
    \end{Verbatim}
  \end{tcolorbox}

  \hypertarget{application-1-phi4-lattice-scalar-field-theory-in-2d}{%
\section{\texorpdfstring{Application 1: \(\phi^4\) lattice scalar field
theory in
2d}{Application 1: \textbackslash phi\^{}4 lattice scalar field theory in 2d}}\label{application-1-phi4-lattice-scalar-field-theory-in-2d}}

As an example, we consider applying normalizing flows to sampling the
distributions associated with scalar field theory in two spacetime
dimensions with a \(\phi^4\) interaction. See \cite{Albergo:2019eim} for
details.

  \hypertarget{physical-theory}{%
\subsection{\texorpdfstring{\textbf{Physical
theory}}{Physical theory}}\label{physical-theory}}

The continuum theory consists of a single real scalar field
\(\phi(\vec{x})\) as a function of 2D coordinates \(\vec{x}\). To access
non-perturbative results, such as behavior in the strong-coupling
regime, we can regularize the theory on a 2D lattice, assigning one real
degree of freedom per site of the lattice. Let's initialize some
configurations of an example lattice of size \(8\times8\) and generate
two random configurations:

  \begin{tcolorbox}[breakable, size=fbox, boxrule=1pt, pad at break*=1mm,colback=cellbackground, colframe=cellborder]
    \begin{Verbatim}[commandchars=\\\{\},fontsize=\small]
\PY{n}{L} \PY{o}{=} \PY{l+m+mi}{8}
\PY{n}{lattice\PYZus{}shape} \PY{o}{=} \PY{p}{(}\PY{n}{L}\PY{p}{,}\PY{n}{L}\PY{p}{)}

\PY{n}{phi\PYZus{}ex1} \PY{o}{=} \PY{n}{np}\PY{o}{.}\PY{n}{random}\PY{o}{.}\PY{n}{normal}\PY{p}{(}\PY{n}{size}\PY{o}{=}\PY{n}{lattice\PYZus{}shape}\PY{p}{)}\PY{o}{.}\PY{n}{astype}\PY{p}{(}\PY{n}{float\PYZus{}dtype}\PY{p}{)}
\PY{n}{phi\PYZus{}ex2} \PY{o}{=} \PY{n}{np}\PY{o}{.}\PY{n}{random}\PY{o}{.}\PY{n}{normal}\PY{p}{(}\PY{n}{size}\PY{o}{=}\PY{n}{lattice\PYZus{}shape}\PY{p}{)}\PY{o}{.}\PY{n}{astype}\PY{p}{(}\PY{n}{float\PYZus{}dtype}\PY{p}{)}
\PY{n}{cfgs} \PY{o}{=} \PY{n}{torch}\PY{o}{.}\PY{n}{from\PYZus{}numpy}\PY{p}{(}\PY{n}{np}\PY{o}{.}\PY{n}{stack}\PY{p}{(}\PY{p}{(}\PY{n}{phi\PYZus{}ex1}\PY{p}{,} \PY{n}{phi\PYZus{}ex2}\PY{p}{)}\PY{p}{,} \PY{n}{axis}\PY{o}{=}\PY{l+m+mi}{0}\PY{p}{)}\PY{p}{)}\PY{o}{.}\PY{n}{to}\PY{p}{(}\PY{n}{torch\PYZus{}device}\PY{p}{)}
    \end{Verbatim}
  \end{tcolorbox}

  A simple discretization of the derivatives in the continuum Euclidean
action gives rise to a valid lattice Euclidean action, \begin{equation}
\begin{split}
S^E_{\text{cont}}[\phi] &= \int d^2\vec{x} ~ (\partial_\mu \phi(\vec{x}))^2 + m^2 \phi(\vec{x})^2 + \lambda \phi(\vec{x})^4 \\
\rightarrow S^E_{\text{latt}}(\phi) &= \sum_{\vec{n}} \phi(\vec{n}) \left[ \sum_{\mu \in \{1,2\}} 2\phi(\vec{n}) - \phi(\vec{n}+\hat{\mu}) - \phi(\vec{n}-\hat{\mu}) \right] + m^2 \phi(\vec{n})^2 + \lambda \phi(\vec{n})^4
\end{split}
\end{equation} where now \(\phi(\vec{n})\) is only defined on the sites
of the \(L_x \times L_y\) lattice, \(\vec{n} = (n_x, n_y)\), with
integer \(n_x, n_y\). We have implicitly moved to ``lattice units''
where \(a=1\) such that \(L_x, L_y, V\) are integers and all quantities
are unitless. The discretized field \(\phi\) can therefore be thought of
as an \((L_x \times L_y)\)-dimensional vector. We use periodic boundary
conditions in all directions, i.e.~\(\phi(L_x, y) \equiv \phi(0, y)\),
etc. For convenience, we typically abbreviate
\(S^E_{\text{latt}} \equiv S\).

More details on \(\phi^4\) lattice scalar field theory can be found in
\cite{vierhaus2010simulation}.

  The lattice action then defines a probability distribution over
configurations \(\phi\), \begin{equation}
p(\phi) = \frac{1}{Z} e^{-S(\phi)}, \quad
Z \equiv \int \prod_{\vec{n}} d\phi(\vec{n}) ~ e^{-S(\phi)},
\end{equation} where \(\prod_{\vec{n}}\) runs over all lattice sites
\(\vec{n}\). This is the distribution we are training the normalizing
flows to reproduce. While \(Z\) is difficult to calculate, in practice
we only need \(p(\phi)\) up to a constant. The action can be efficiently
calculated on arbitrary configurations using Pytorch. Note that while
the theory describes 2D spacetime, the dimensionality of distribution
\(p(\phi)\) is the number of lattice sites, scaling with the volume of
the lattice.

  \begin{tcolorbox}[breakable, size=fbox, boxrule=1pt, pad at break*=1mm,colback=cellbackground, colframe=cellborder]
    \begin{Verbatim}[commandchars=\\\{\},fontsize=\small]
\PY{k}{class} \PY{n+nc}{ScalarPhi4Action}\PY{p}{:}
    \PY{k}{def} \PY{n+nf+fm}{\PYZus{}\PYZus{}init\PYZus{}\PYZus{}}\PY{p}{(}\PY{n+nb+bp}{self}\PY{p}{,} \PY{n}{M2}\PY{p}{,} \PY{n}{lam}\PY{p}{)}\PY{p}{:}
        \PY{n+nb+bp}{self}\PY{o}{.}\PY{n}{M2} \PY{o}{=} \PY{n}{M2}
        \PY{n+nb+bp}{self}\PY{o}{.}\PY{n}{lam} \PY{o}{=} \PY{n}{lam}
    \PY{k}{def} \PY{n+nf+fm}{\PYZus{}\PYZus{}call\PYZus{}\PYZus{}}\PY{p}{(}\PY{n+nb+bp}{self}\PY{p}{,} \PY{n}{cfgs}\PY{p}{)}\PY{p}{:}
        \PY{c+c1}{\PYZsh{} potential term}
        \PY{n}{action\PYZus{}density} \PY{o}{=} \PY{n+nb+bp}{self}\PY{o}{.}\PY{n}{M2}\PY{o}{*}\PY{n}{cfgs}\PY{o}{*}\PY{o}{*}\PY{l+m+mi}{2} \PY{o}{+} \PY{n+nb+bp}{self}\PY{o}{.}\PY{n}{lam}\PY{o}{*}\PY{n}{cfgs}\PY{o}{*}\PY{o}{*}\PY{l+m+mi}{4}
        \PY{c+c1}{\PYZsh{} kinetic term (discrete Laplacian)}
        \PY{n}{Nd} \PY{o}{=} \PY{n+nb}{len}\PY{p}{(}\PY{n}{cfgs}\PY{o}{.}\PY{n}{shape}\PY{p}{)}\PY{o}{\PYZhy{}}\PY{l+m+mi}{1}
        \PY{n}{dims} \PY{o}{=} \PY{n+nb}{range}\PY{p}{(}\PY{l+m+mi}{1}\PY{p}{,}\PY{n}{Nd}\PY{o}{+}\PY{l+m+mi}{1}\PY{p}{)}
        \PY{k}{for} \PY{n}{mu} \PY{o+ow}{in} \PY{n}{dims}\PY{p}{:}
            \PY{n}{action\PYZus{}density} \PY{o}{+}\PY{o}{=} \PY{l+m+mi}{2}\PY{o}{*}\PY{n}{cfgs}\PY{o}{*}\PY{o}{*}\PY{l+m+mi}{2}
            \PY{n}{action\PYZus{}density} \PY{o}{\PYZhy{}}\PY{o}{=} \PY{n}{cfgs}\PY{o}{*}\PY{n}{torch}\PY{o}{.}\PY{n}{roll}\PY{p}{(}\PY{n}{cfgs}\PY{p}{,} \PY{o}{\PYZhy{}}\PY{l+m+mi}{1}\PY{p}{,} \PY{n}{mu}\PY{p}{)}
            \PY{n}{action\PYZus{}density} \PY{o}{\PYZhy{}}\PY{o}{=} \PY{n}{cfgs}\PY{o}{*}\PY{n}{torch}\PY{o}{.}\PY{n}{roll}\PY{p}{(}\PY{n}{cfgs}\PY{p}{,} \PY{l+m+mi}{1}\PY{p}{,} \PY{n}{mu}\PY{p}{)}
        \PY{k}{return} \PY{n}{torch}\PY{o}{.}\PY{n}{sum}\PY{p}{(}\PY{n}{action\PYZus{}density}\PY{p}{,} \PY{n}{dim}\PY{o}{=}\PY{n+nb}{tuple}\PY{p}{(}\PY{n}{dims}\PY{p}{)}\PY{p}{)}

\PY{n+nb}{print}\PY{p}{(}\PY{l+s+s2}{\PYZdq{}}\PY{l+s+s2}{Actions for example configs:}\PY{l+s+s2}{\PYZdq{}}\PY{p}{,} \PY{n}{ScalarPhi4Action}\PY{p}{(}\PY{n}{M2}\PY{o}{=}\PY{l+m+mf}{1.0}\PY{p}{,} \PY{n}{lam}\PY{o}{=}\PY{l+m+mf}{1.0}\PY{p}{)}\PY{p}{(}\PY{n}{cfgs}\PY{p}{)}\PY{p}{)}
    \end{Verbatim}
  \end{tcolorbox}

{\color{gray}
    \begin{Verbatim}[commandchars=\\\{\},fontsize=\small]
>>> Actions for example configs: tensor([682.8262, 295.3547])
    \end{Verbatim}
}

  The theory has a symmetric phase and a broken symmetry phase,
corresponding respectively to nearly one mode of the distribution or two
widely separated modes (with intermediate configurations suppressed
exponentially in volume). The broken symmetry phase can be accessed for
\(m^2 < 0\) and \(\lambda\) less than a critical \(\lambda_c\). For
simplicity, we restrict focus to the \textbf{symmetric phase}, but
remain close to this phase transition such that the system has a
non-trivial correlation length.

  \begin{tcolorbox}[breakable, size=fbox, boxrule=1pt, pad at break*=1mm,colback=cellbackground, colframe=cellborder]
    \begin{Verbatim}[commandchars=\\\{\},fontsize=\small]
\PY{n}{M2} \PY{o}{=} \PY{o}{\PYZhy{}}\PY{l+m+mf}{4.0}
\PY{n}{lam} \PY{o}{=} \PY{l+m+mf}{8.0}
\PY{n}{phi4\PYZus{}action} \PY{o}{=} \PY{n}{ScalarPhi4Action}\PY{p}{(}\PY{n}{M2}\PY{o}{=}\PY{n}{M2}\PY{p}{,} \PY{n}{lam}\PY{o}{=}\PY{n}{lam}\PY{p}{)}
    \end{Verbatim}
  \end{tcolorbox}

  \hypertarget{prior-distribution}{%
\subsection{\texorpdfstring{\textbf{Prior
distribution}}{Prior distribution}}\label{prior-distribution}}

We choose the prior distribution to be I.I.D. Gaussians at each lattice
site. This is easy to sample from, and intuitively gives the coupling
layers a ``blank slate'' from which to build in correlations.

  \begin{tcolorbox}[breakable, size=fbox, boxrule=1pt, pad at break*=1mm,colback=cellbackground, colframe=cellborder]
    \begin{Verbatim}[commandchars=\\\{\},fontsize=\small]
\PY{n}{prior} \PY{o}{=} \PY{n}{SimpleNormal}\PY{p}{(}\PY{n}{torch}\PY{o}{.}\PY{n}{zeros}\PY{p}{(}\PY{n}{lattice\PYZus{}shape}\PY{p}{)}\PY{p}{,} \PY{n}{torch}\PY{o}{.}\PY{n}{ones}\PY{p}{(}\PY{n}{lattice\PYZus{}shape}\PY{p}{)}\PY{p}{)}
    \end{Verbatim}
  \end{tcolorbox}

  We can use the \texttt{draw} function to acquire samples from the prior.
Some samples drawn from the prior are visualized below.

  \begin{tcolorbox}[breakable, size=fbox, boxrule=1pt, pad at break*=1mm,colback=cellbackground, colframe=cellborder]
    \begin{Verbatim}[commandchars=\\\{\},fontsize=\small]
\PY{n}{torch\PYZus{}z} \PY{o}{=} \PY{n}{prior}\PY{o}{.}\PY{n}{sample\PYZus{}n}\PY{p}{(}\PY{l+m+mi}{1024}\PY{p}{)}
\PY{n}{z} \PY{o}{=} \PY{n}{grab}\PY{p}{(}\PY{n}{torch\PYZus{}z}\PY{p}{)}
\PY{n+nb}{print}\PY{p}{(}\PY{l+s+sa}{f}\PY{l+s+s1}{\PYZsq{}}\PY{l+s+s1}{z.shape = }\PY{l+s+si}{\PYZob{}}\PY{n}{z}\PY{o}{.}\PY{n}{shape}\PY{l+s+si}{\PYZcb{}}\PY{l+s+s1}{\PYZsq{}}\PY{p}{)}

\PY{n}{fig}\PY{p}{,} \PY{n}{ax} \PY{o}{=} \PY{n}{plt}\PY{o}{.}\PY{n}{subplots}\PY{p}{(}\PY{l+m+mi}{4}\PY{p}{,}\PY{l+m+mi}{4}\PY{p}{,} \PY{n}{dpi}\PY{o}{=}\PY{l+m+mi}{125}\PY{p}{,} \PY{n}{figsize}\PY{o}{=}\PY{p}{(}\PY{l+m+mi}{4}\PY{p}{,}\PY{l+m+mi}{4}\PY{p}{)}\PY{p}{)}
\PY{k}{for} \PY{n}{i} \PY{o+ow}{in} \PY{n+nb}{range}\PY{p}{(}\PY{l+m+mi}{4}\PY{p}{)}\PY{p}{:}
    \PY{k}{for} \PY{n}{j} \PY{o+ow}{in} \PY{n+nb}{range}\PY{p}{(}\PY{l+m+mi}{4}\PY{p}{)}\PY{p}{:}
        \PY{n}{ind} \PY{o}{=} \PY{n}{i}\PY{o}{*}\PY{l+m+mi}{4} \PY{o}{+} \PY{n}{j}
        \PY{n}{ax}\PY{p}{[}\PY{n}{i}\PY{p}{,}\PY{n}{j}\PY{p}{]}\PY{o}{.}\PY{n}{imshow}\PY{p}{(}\PY{n}{np}\PY{o}{.}\PY{n}{tanh}\PY{p}{(}\PY{n}{z}\PY{p}{[}\PY{n}{ind}\PY{p}{]}\PY{p}{)}\PY{p}{,} \PY{n}{vmin}\PY{o}{=}\PY{o}{\PYZhy{}}\PY{l+m+mi}{1}\PY{p}{,} \PY{n}{vmax}\PY{o}{=}\PY{l+m+mi}{1}\PY{p}{,} \PY{n}{cmap}\PY{o}{=}\PY{l+s+s1}{\PYZsq{}}\PY{l+s+s1}{viridis}\PY{l+s+s1}{\PYZsq{}}\PY{p}{)}
        \PY{n}{ax}\PY{p}{[}\PY{n}{i}\PY{p}{,}\PY{n}{j}\PY{p}{]}\PY{o}{.}\PY{n}{axes}\PY{o}{.}\PY{n}{xaxis}\PY{o}{.}\PY{n}{set\PYZus{}visible}\PY{p}{(}\PY{k+kc}{False}\PY{p}{)}
        \PY{n}{ax}\PY{p}{[}\PY{n}{i}\PY{p}{,}\PY{n}{j}\PY{p}{]}\PY{o}{.}\PY{n}{axes}\PY{o}{.}\PY{n}{yaxis}\PY{o}{.}\PY{n}{set\PYZus{}visible}\PY{p}{(}\PY{k+kc}{False}\PY{p}{)}
\PY{n}{plt}\PY{o}{.}\PY{n}{show}\PY{p}{(}\PY{p}{)}
    \end{Verbatim}
  \end{tcolorbox}

{\color{gray}
    \begin{Verbatim}[commandchars=\\\{\},fontsize=\small]
>>> z.shape = (1024, 8, 8)
    \end{Verbatim}
}

    \begin{center}
    \adjustimage{max size={0.9\linewidth}{0.9\paperheight}}{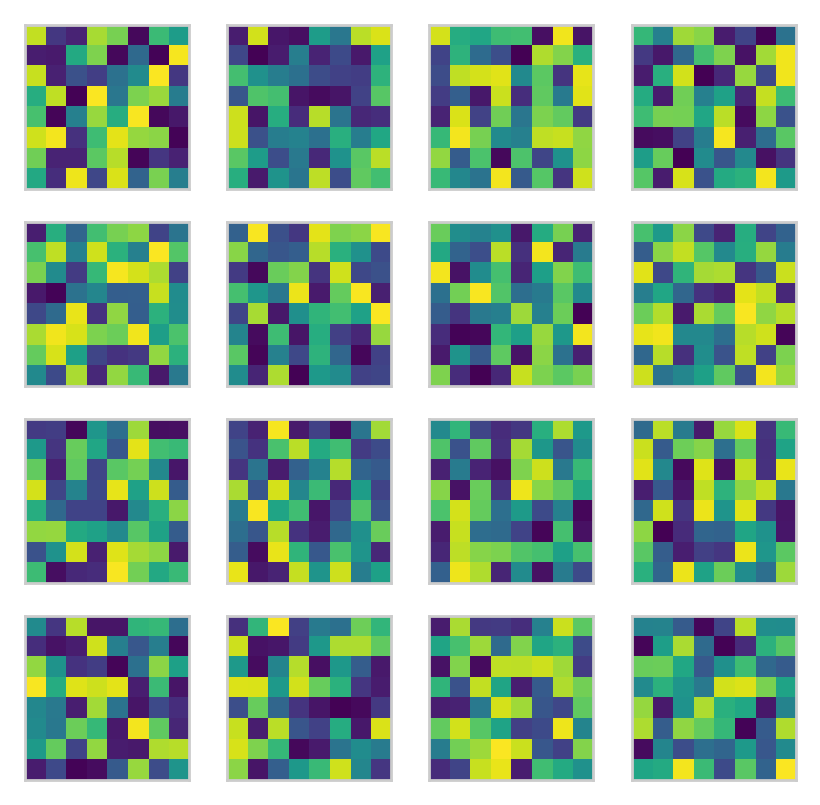}
    \end{center}
    { \hspace*{\fill} \\}
    
  These samples look nothing like typical lattice configurations for the
theory, which should exhibit patches of correlated positive and negative
fluctuations (see configurations drawn from the true distribution in
later sections). Samples from the prior do not have any correlations.

Below we show two-way histograms between various sites, demonstrating
the uncorrelated nature of the prior distribution.

  \begin{tcolorbox}[breakable, size=fbox, boxrule=1pt, pad at break*=1mm,colback=cellbackground, colframe=cellborder]
    \begin{Verbatim}[commandchars=\\\{\},fontsize=\small]
\PY{n}{fig}\PY{p}{,} \PY{n}{ax} \PY{o}{=} \PY{n}{plt}\PY{o}{.}\PY{n}{subplots}\PY{p}{(}\PY{l+m+mi}{4}\PY{p}{,}\PY{l+m+mi}{4}\PY{p}{,} \PY{n}{dpi}\PY{o}{=}\PY{l+m+mi}{125}\PY{p}{,} \PY{n}{figsize}\PY{o}{=}\PY{p}{(}\PY{l+m+mi}{4}\PY{p}{,}\PY{l+m+mi}{4}\PY{p}{)}\PY{p}{)}
\PY{k}{for} \PY{n}{x1} \PY{o+ow}{in} \PY{n+nb}{range}\PY{p}{(}\PY{l+m+mi}{2}\PY{p}{)}\PY{p}{:}
    \PY{k}{for} \PY{n}{y1} \PY{o+ow}{in} \PY{n+nb}{range}\PY{p}{(}\PY{l+m+mi}{2}\PY{p}{)}\PY{p}{:}
        \PY{n}{i1} \PY{o}{=} \PY{n}{x1}\PY{o}{*}\PY{l+m+mi}{2} \PY{o}{+} \PY{n}{y1}
        \PY{k}{for} \PY{n}{x2} \PY{o+ow}{in} \PY{n+nb}{range}\PY{p}{(}\PY{l+m+mi}{2}\PY{p}{)}\PY{p}{:}
            \PY{k}{for} \PY{n}{y2} \PY{o+ow}{in} \PY{n+nb}{range}\PY{p}{(}\PY{l+m+mi}{2}\PY{p}{)}\PY{p}{:}
                \PY{n}{i2} \PY{o}{=} \PY{n}{x2}\PY{o}{*}\PY{l+m+mi}{2} \PY{o}{+} \PY{n}{y2}
                \PY{n}{ax}\PY{p}{[}\PY{n}{i1}\PY{p}{,}\PY{n}{i2}\PY{p}{]}\PY{o}{.}\PY{n}{hist2d}\PY{p}{(}\PY{n}{z}\PY{p}{[}\PY{p}{:}\PY{p}{,}\PY{n}{x1}\PY{p}{,}\PY{n}{y1}\PY{p}{]}\PY{p}{,} \PY{n}{z}\PY{p}{[}\PY{p}{:}\PY{p}{,}\PY{n}{x2}\PY{p}{,}\PY{n}{y2}\PY{p}{]}\PY{p}{,} \PY{n+nb}{range}\PY{o}{=}\PY{p}{[}\PY{p}{[}\PY{o}{\PYZhy{}}\PY{l+m+mi}{3}\PY{p}{,}\PY{l+m+mi}{3}\PY{p}{]}\PY{p}{,}\PY{p}{[}\PY{o}{\PYZhy{}}\PY{l+m+mi}{3}\PY{p}{,}\PY{l+m+mi}{3}\PY{p}{]}\PY{p}{]}\PY{p}{,} \PY{n}{bins}\PY{o}{=}\PY{l+m+mi}{20}\PY{p}{)}
                \PY{n}{ax}\PY{p}{[}\PY{n}{i1}\PY{p}{,}\PY{n}{i2}\PY{p}{]}\PY{o}{.}\PY{n}{set\PYZus{}xticks}\PY{p}{(}\PY{p}{[}\PY{p}{]}\PY{p}{)}
                \PY{n}{ax}\PY{p}{[}\PY{n}{i1}\PY{p}{,}\PY{n}{i2}\PY{p}{]}\PY{o}{.}\PY{n}{set\PYZus{}yticks}\PY{p}{(}\PY{p}{[}\PY{p}{]}\PY{p}{)}
                \PY{k}{if} \PY{n}{i1} \PY{o}{==} \PY{l+m+mi}{3}\PY{p}{:}
                    \PY{n}{ax}\PY{p}{[}\PY{n}{i1}\PY{p}{,}\PY{n}{i2}\PY{p}{]}\PY{o}{.}\PY{n}{set\PYZus{}xlabel}\PY{p}{(}\PY{l+s+sa}{rf}\PY{l+s+s1}{\PYZsq{}}\PY{l+s+s1}{\PYZdl{}}\PY{l+s+s1}{\PYZbs{}}\PY{l+s+s1}{phi(}\PY{l+s+si}{\PYZob{}}\PY{n}{x2}\PY{l+s+si}{\PYZcb{}}\PY{l+s+s1}{,}\PY{l+s+si}{\PYZob{}}\PY{n}{y2}\PY{l+s+si}{\PYZcb{}}\PY{l+s+s1}{)\PYZdl{}}\PY{l+s+s1}{\PYZsq{}}\PY{p}{)}
                \PY{k}{if} \PY{n}{i2} \PY{o}{==} \PY{l+m+mi}{0}\PY{p}{:}
                    \PY{n}{ax}\PY{p}{[}\PY{n}{i1}\PY{p}{,}\PY{n}{i2}\PY{p}{]}\PY{o}{.}\PY{n}{set\PYZus{}ylabel}\PY{p}{(}\PY{l+s+sa}{rf}\PY{l+s+s1}{\PYZsq{}}\PY{l+s+s1}{\PYZdl{}}\PY{l+s+s1}{\PYZbs{}}\PY{l+s+s1}{phi(}\PY{l+s+si}{\PYZob{}}\PY{n}{x1}\PY{l+s+si}{\PYZcb{}}\PY{l+s+s1}{,}\PY{l+s+si}{\PYZob{}}\PY{n}{y1}\PY{l+s+si}{\PYZcb{}}\PY{l+s+s1}{)\PYZdl{}}\PY{l+s+s1}{\PYZsq{}}\PY{p}{)}
\PY{n}{fig}\PY{o}{.}\PY{n}{suptitle}\PY{p}{(}\PY{l+s+s2}{\PYZdq{}}\PY{l+s+s2}{Correlations in Various Lattice Sites}\PY{l+s+s2}{\PYZdq{}}\PY{p}{)}
\PY{n}{plt}\PY{o}{.}\PY{n}{show}\PY{p}{(}\PY{p}{)}
    \end{Verbatim}
  \end{tcolorbox}

    \begin{center}
    \adjustimage{max size={0.9\linewidth}{0.9\paperheight}}{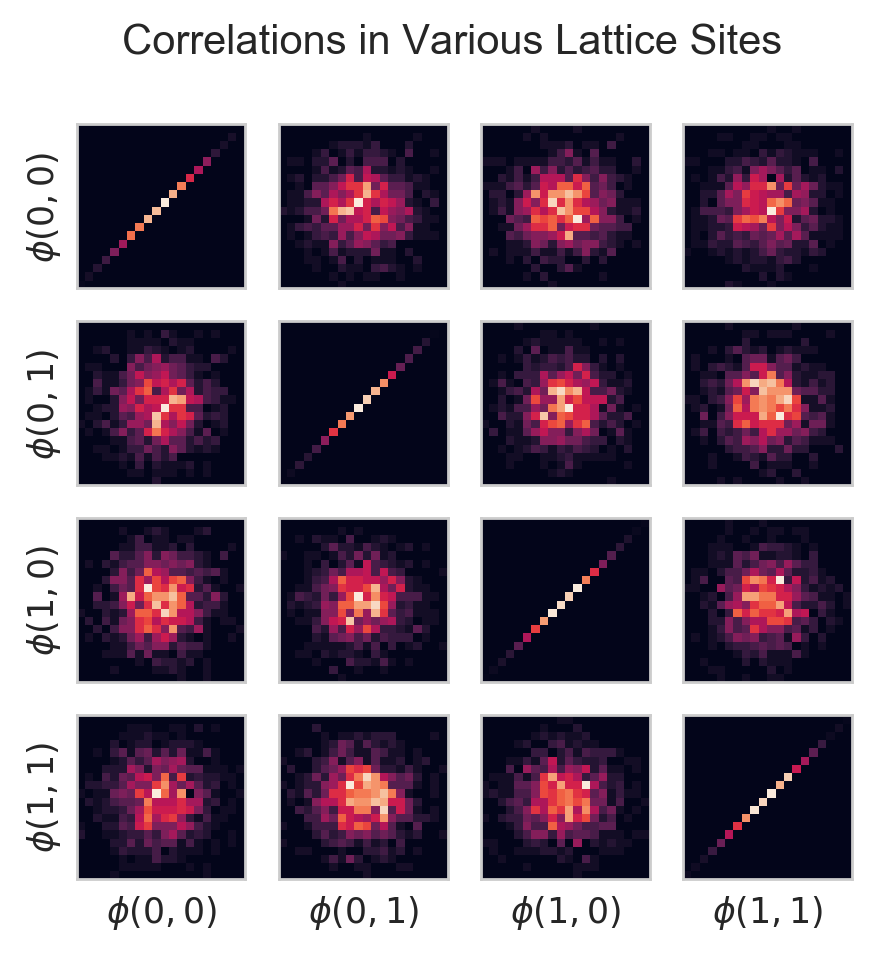}
    \end{center}
    { \hspace*{\fill} \\}
    
  We can also investigate the correlation between the ``effective action''
defining the model distribution (here, \(-\log{r}(z)\)) and the true
action (\(S(z)\)). If the prior distribution was already a good model
for the true distribution, all samples should have identical action
under the prior and true distributions, up to an overall shift. In other
words, these should have linear correlation with slope \(1\).

  \begin{tcolorbox}[breakable, size=fbox, boxrule=1pt, pad at break*=1mm,colback=cellbackground, colframe=cellborder]
    \begin{Verbatim}[commandchars=\\\{\},fontsize=\small]
\PY{n}{S\PYZus{}eff} \PY{o}{=} \PY{o}{\PYZhy{}}\PY{n}{grab}\PY{p}{(}\PY{n}{prior}\PY{o}{.}\PY{n}{log\PYZus{}prob}\PY{p}{(}\PY{n}{torch\PYZus{}z}\PY{p}{)}\PY{p}{)}
\PY{n}{S} \PY{o}{=} \PY{n}{grab}\PY{p}{(}\PY{n}{phi4\PYZus{}action}\PY{p}{(}\PY{n}{torch\PYZus{}z}\PY{p}{)}\PY{p}{)}
\PY{n}{fit\PYZus{}b} \PY{o}{=} \PY{n}{np}\PY{o}{.}\PY{n}{mean}\PY{p}{(}\PY{n}{S}\PY{p}{)} \PY{o}{\PYZhy{}} \PY{n}{np}\PY{o}{.}\PY{n}{mean}\PY{p}{(}\PY{n}{S\PYZus{}eff}\PY{p}{)}
\PY{n+nb}{print}\PY{p}{(}\PY{l+s+sa}{f}\PY{l+s+s1}{\PYZsq{}}\PY{l+s+s1}{slope 1 linear regression S = \PYZhy{}logr + }\PY{l+s+si}{\PYZob{}}\PY{n}{fit\PYZus{}b}\PY{l+s+si}{:}\PY{l+s+s1}{.4f}\PY{l+s+si}{\PYZcb{}}\PY{l+s+s1}{\PYZsq{}}\PY{p}{)}
\PY{n}{fig}\PY{p}{,} \PY{n}{ax} \PY{o}{=} \PY{n}{plt}\PY{o}{.}\PY{n}{subplots}\PY{p}{(}\PY{l+m+mi}{1}\PY{p}{,}\PY{l+m+mi}{1}\PY{p}{,} \PY{n}{dpi}\PY{o}{=}\PY{l+m+mi}{125}\PY{p}{,} \PY{n}{figsize}\PY{o}{=}\PY{p}{(}\PY{l+m+mi}{4}\PY{p}{,}\PY{l+m+mi}{4}\PY{p}{)}\PY{p}{)}
\PY{n}{ax}\PY{o}{.}\PY{n}{hist2d}\PY{p}{(}\PY{n}{S\PYZus{}eff}\PY{p}{,} \PY{n}{S}\PY{p}{,} \PY{n}{bins}\PY{o}{=}\PY{l+m+mi}{20}\PY{p}{,} \PY{n+nb}{range}\PY{o}{=}\PY{p}{[}\PY{p}{[}\PY{o}{\PYZhy{}}\PY{l+m+mi}{800}\PY{p}{,} \PY{l+m+mi}{800}\PY{p}{]}\PY{p}{,} \PY{p}{[}\PY{l+m+mi}{200}\PY{p}{,}\PY{l+m+mi}{1800}\PY{p}{]}\PY{p}{]}\PY{p}{)}
\PY{n}{xs} \PY{o}{=} \PY{n}{np}\PY{o}{.}\PY{n}{linspace}\PY{p}{(}\PY{o}{\PYZhy{}}\PY{l+m+mi}{800}\PY{p}{,} \PY{l+m+mi}{800}\PY{p}{,} \PY{n}{num}\PY{o}{=}\PY{l+m+mi}{4}\PY{p}{,} \PY{n}{endpoint}\PY{o}{=}\PY{k+kc}{True}\PY{p}{)}
\PY{n}{ax}\PY{o}{.}\PY{n}{plot}\PY{p}{(}\PY{n}{xs}\PY{p}{,} \PY{n}{xs} \PY{o}{+} \PY{n}{fit\PYZus{}b}\PY{p}{,} \PY{l+s+s1}{\PYZsq{}}\PY{l+s+s1}{:}\PY{l+s+s1}{\PYZsq{}}\PY{p}{,} \PY{n}{color}\PY{o}{=}\PY{l+s+s1}{\PYZsq{}}\PY{l+s+s1}{w}\PY{l+s+s1}{\PYZsq{}}\PY{p}{,} \PY{n}{label}\PY{o}{=}\PY{l+s+s1}{\PYZsq{}}\PY{l+s+s1}{slope 1 fit}\PY{l+s+s1}{\PYZsq{}}\PY{p}{)}
\PY{n}{ax}\PY{o}{.}\PY{n}{set\PYZus{}xlabel}\PY{p}{(}\PY{l+s+sa}{r}\PY{l+s+s1}{\PYZsq{}}\PY{l+s+s1}{\PYZdl{}S\PYZus{}}\PY{l+s+s1}{\PYZob{}}\PY{l+s+s1}{\PYZbs{}}\PY{l+s+s1}{mathrm}\PY{l+s+si}{\PYZob{}eff\PYZcb{}}\PY{l+s+s1}{\PYZcb{} }\PY{l+s+s1}{\PYZbs{}}\PY{l+s+s1}{equiv \PYZhy{}}\PY{l+s+s1}{\PYZbs{}}\PY{l+s+s1}{log\PYZti{}r(z)\PYZdl{}}\PY{l+s+s1}{\PYZsq{}}\PY{p}{)}
\PY{n}{ax}\PY{o}{.}\PY{n}{set\PYZus{}ylabel}\PY{p}{(}\PY{l+s+sa}{r}\PY{l+s+s1}{\PYZsq{}}\PY{l+s+s1}{\PYZdl{}S(z)\PYZdl{}}\PY{l+s+s1}{\PYZsq{}}\PY{p}{)}
\PY{n}{ax}\PY{o}{.}\PY{n}{set\PYZus{}aspect}\PY{p}{(}\PY{l+s+s1}{\PYZsq{}}\PY{l+s+s1}{equal}\PY{l+s+s1}{\PYZsq{}}\PY{p}{)}
\PY{n}{plt}\PY{o}{.}\PY{n}{legend}\PY{p}{(}\PY{n}{prop}\PY{o}{=}\PY{p}{\PYZob{}}\PY{l+s+s1}{\PYZsq{}}\PY{l+s+s1}{size}\PY{l+s+s1}{\PYZsq{}}\PY{p}{:} \PY{l+m+mi}{6}\PY{p}{\PYZcb{}}\PY{p}{)}
\PY{n}{plt}\PY{o}{.}\PY{n}{show}\PY{p}{(}\PY{p}{)}
    \end{Verbatim}
  \end{tcolorbox}

{\color{gray}
    \begin{Verbatim}[commandchars=\\\{\},fontsize=\small]
>>> slope 1 linear regression S = -logr + 1455.4647
    \end{Verbatim}
}

    \begin{center}
    \adjustimage{max size={0.9\linewidth}{0.9\paperheight}}{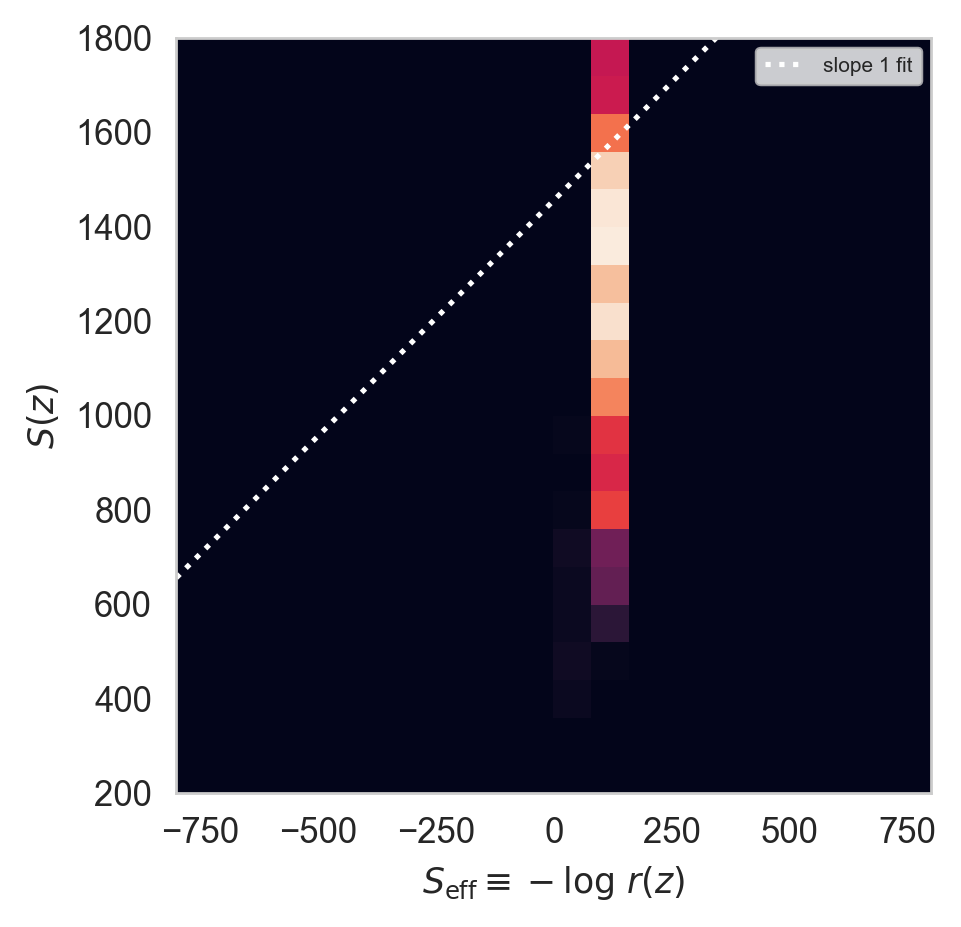}
    \end{center}
    { \hspace*{\fill} \\}
    
  As expected, we are far from that goal because we have not flowed yet!
We will revisit this plot after including and training the
change-of-variables \(f\).

  \hypertarget{affine-coupling-layers}{%
\subsection{\texorpdfstring{\textbf{Affine coupling
layers}}{Affine coupling layers}}\label{affine-coupling-layers}}

As mentioned earlier, an affine transformation is a particularly simple,
yet effective, transform to use within a coupling layer acting on real
degrees of freedom. The transformation of the subset of variables
\(\phi_1\), conditioned on the frozen subset \(\phi_2\), is defined as
\begin{equation}
    g(\phi_1, \phi_2) = \left(e^{s(\phi_2)} \phi_1 + t(\phi_2),  \phi_2\right),
\end{equation} with inverse given by:
\begin{equation} g^{-1}(\phi_1', \phi_2') =  \left((\phi_1' - t(\phi_2')) e^{-s(\phi_2')}, \phi_2'\right)  \end{equation}
where \(s(\phi_2)\) and \(t(\phi_2)\) produce vectors of the same
dimension as \(\phi_1\) and operations above are element-wise on these
vectors. We define the functions \(s\) and \(t\) using a feed-forward
neural network. The coupling layer leaves \(\phi_2\) unchanged. Note
that this is just a simple extension of the scaling transformation
introduced above, with a constant offset \(t(\phi_2)\) added to the
transformation.

  The Jacobian factor for such an affine transformation is easy to compute
(both analytically and numerically). In fact, because
\(\partial [t(\phi_2)] / \partial \phi_1 = 0\), the Jacobian is the same
as for the scaling transformation worked out above. In practice, we work
with log probabilities, so we note that the forward and reverse
transformation return \begin{align}
&&\text{forward:} \qquad &\log J(\phi) &&= \sum_k [s(\phi_2)]_k \\ 
&&\text{reverse:} \qquad &\log J_\text{reverse}(\phi') &&= \sum_k -[s(\phi_2')]_k .
\end{align}

  The subsets \(\phi_1, \phi_2\) are defined by a mask
\(m(\vec{n}) \in \{0, 1\}\). In our conventions \(m(\vec{n}) = 1\)
implies an input to the neural net defining \(s\) and \(t\), and
therefore that the variable on site \(\vec{n}\) is an element of the
frozen subset \(\phi_2\). We choose checkerboard masking, as this
intuitively allows sites to influence the transformation of their direct
neighbors and build local correlations.

  \begin{tcolorbox}[breakable, size=fbox, boxrule=1pt, pad at break*=1mm,colback=cellbackground, colframe=cellborder]
    \begin{Verbatim}[commandchars=\\\{\},fontsize=\small]
\PY{k}{def} \PY{n+nf}{make\PYZus{}checker\PYZus{}mask}\PY{p}{(}\PY{n}{shape}\PY{p}{,} \PY{n}{parity}\PY{p}{)}\PY{p}{:}
    \PY{n}{checker} \PY{o}{=} \PY{n}{torch}\PY{o}{.}\PY{n}{ones}\PY{p}{(}\PY{n}{shape}\PY{p}{,} \PY{n}{dtype}\PY{o}{=}\PY{n}{torch}\PY{o}{.}\PY{n}{uint8}\PY{p}{)} \PY{o}{\PYZhy{}} \PY{n}{parity}
    \PY{n}{checker}\PY{p}{[}\PY{p}{:}\PY{p}{:}\PY{l+m+mi}{2}\PY{p}{,} \PY{p}{:}\PY{p}{:}\PY{l+m+mi}{2}\PY{p}{]} \PY{o}{=} \PY{n}{parity}
    \PY{n}{checker}\PY{p}{[}\PY{l+m+mi}{1}\PY{p}{:}\PY{p}{:}\PY{l+m+mi}{2}\PY{p}{,} \PY{l+m+mi}{1}\PY{p}{:}\PY{p}{:}\PY{l+m+mi}{2}\PY{p}{]} \PY{o}{=} \PY{n}{parity}
    \PY{k}{return} \PY{n}{checker}\PY{o}{.}\PY{n}{to}\PY{p}{(}\PY{n}{torch\PYZus{}device}\PY{p}{)}

\PY{n+nb}{print}\PY{p}{(}\PY{l+s+s2}{\PYZdq{}}\PY{l+s+s2}{For example this is the mask for an 8x8 configuration:}\PY{l+s+se}{\PYZbs{}n}\PY{l+s+s2}{\PYZdq{}}\PY{p}{,}
      \PY{n}{make\PYZus{}checker\PYZus{}mask}\PY{p}{(}\PY{n}{lattice\PYZus{}shape}\PY{p}{,} \PY{l+m+mi}{0}\PY{p}{)}\PY{p}{)}
    \end{Verbatim}
  \end{tcolorbox}

{\color{gray}
    \begin{Verbatim}[commandchars=\\\{\},fontsize=\small]
>>> For example this is the mask for an 8x8 configuration:
...  tensor([[0, 1, 0, 1, 0, 1, 0, 1],
...         [1, 0, 1, 0, 1, 0, 1, 0],
...         [0, 1, 0, 1, 0, 1, 0, 1],
...         [1, 0, 1, 0, 1, 0, 1, 0],
...         [0, 1, 0, 1, 0, 1, 0, 1],
...         [1, 0, 1, 0, 1, 0, 1, 0],
...         [0, 1, 0, 1, 0, 1, 0, 1],
...         [1, 0, 1, 0, 1, 0, 1, 0]], dtype=torch.uint8)
    \end{Verbatim}
}

  We implement the described coupling layer in the code cell below.

For simplicity in our implementation below, we allow \(s(\phi_2)\) and
\(t(\phi_2)\) to produce outputs for \(\phi_2\) as well as \(\phi_1\)
and then mask them out. We use the same NN to parametrize \(s\) and
\(t\), so they have shared parameters; this does not add formal
complications and just makes the model simpler.

\textbf{Technical note:} Pytorch's implementation of 2D CNNs requires
inputs shaped like
\texttt{(batch\_size,\ n\_input\_channels,\ L\_x,\ L\_y)}, hence the use
of \texttt{unsqueeze} below which adds a fake
\texttt{n\_input\_channels} dimension of length 1. The CNNs return an
output shaped like
\texttt{(batch\_size,\ n\_output\_channels,\ L\_x,\ L\_y)}; we use
\texttt{n\_output\_channels\ ==\ 2}, where the two channels are
\(s(x_2)\) and \(t(x_2)\).

  \begin{tcolorbox}[breakable, size=fbox, boxrule=1pt, pad at break*=1mm,colback=cellbackground, colframe=cellborder]
    \begin{Verbatim}[commandchars=\\\{\},fontsize=\small]
\PY{k}{class} \PY{n+nc}{AffineCoupling}\PY{p}{(}\PY{n}{torch}\PY{o}{.}\PY{n}{nn}\PY{o}{.}\PY{n}{Module}\PY{p}{)}\PY{p}{:}
    \PY{k}{def} \PY{n+nf+fm}{\PYZus{}\PYZus{}init\PYZus{}\PYZus{}}\PY{p}{(}\PY{n+nb+bp}{self}\PY{p}{,} \PY{n}{net}\PY{p}{,} \PY{o}{*}\PY{p}{,} \PY{n}{mask\PYZus{}shape}\PY{p}{,} \PY{n}{mask\PYZus{}parity}\PY{p}{)}\PY{p}{:}
        \PY{n+nb}{super}\PY{p}{(}\PY{p}{)}\PY{o}{.}\PY{n+nf+fm}{\PYZus{}\PYZus{}init\PYZus{}\PYZus{}}\PY{p}{(}\PY{p}{)}
        \PY{n+nb+bp}{self}\PY{o}{.}\PY{n}{mask} \PY{o}{=} \PY{n}{make\PYZus{}checker\PYZus{}mask}\PY{p}{(}\PY{n}{mask\PYZus{}shape}\PY{p}{,} \PY{n}{mask\PYZus{}parity}\PY{p}{)}
        \PY{n+nb+bp}{self}\PY{o}{.}\PY{n}{net} \PY{o}{=} \PY{n}{net}

    \PY{k}{def} \PY{n+nf}{forward}\PY{p}{(}\PY{n+nb+bp}{self}\PY{p}{,} \PY{n}{x}\PY{p}{)}\PY{p}{:}
        \PY{n}{x\PYZus{}frozen} \PY{o}{=} \PY{n+nb+bp}{self}\PY{o}{.}\PY{n}{mask} \PY{o}{*} \PY{n}{x}      
        \PY{n}{x\PYZus{}active} \PY{o}{=} \PY{p}{(}\PY{l+m+mi}{1} \PY{o}{\PYZhy{}} \PY{n+nb+bp}{self}\PY{o}{.}\PY{n}{mask}\PY{p}{)} \PY{o}{*} \PY{n}{x}
        \PY{n}{net\PYZus{}out} \PY{o}{=} \PY{n+nb+bp}{self}\PY{o}{.}\PY{n}{net}\PY{p}{(}\PY{n}{x\PYZus{}frozen}\PY{o}{.}\PY{n}{unsqueeze}\PY{p}{(}\PY{l+m+mi}{1}\PY{p}{)}\PY{p}{)}
        \PY{n}{s}\PY{p}{,} \PY{n}{t} \PY{o}{=} \PY{n}{net\PYZus{}out}\PY{p}{[}\PY{p}{:}\PY{p}{,}\PY{l+m+mi}{0}\PY{p}{]}\PY{p}{,} \PY{n}{net\PYZus{}out}\PY{p}{[}\PY{p}{:}\PY{p}{,}\PY{l+m+mi}{1}\PY{p}{]}
        \PY{n}{fx} \PY{o}{=} \PY{p}{(}\PY{l+m+mi}{1} \PY{o}{\PYZhy{}} \PY{n+nb+bp}{self}\PY{o}{.}\PY{n}{mask}\PY{p}{)} \PY{o}{*} \PY{n}{t} \PY{o}{+} \PY{n}{x\PYZus{}active} \PY{o}{*} \PY{n}{torch}\PY{o}{.}\PY{n}{exp}\PY{p}{(}\PY{n}{s}\PY{p}{)} \PY{o}{+} \PY{n}{x\PYZus{}frozen}
        \PY{n}{axes} \PY{o}{=} \PY{n+nb}{range}\PY{p}{(}\PY{l+m+mi}{1}\PY{p}{,}\PY{n+nb}{len}\PY{p}{(}\PY{n}{s}\PY{o}{.}\PY{n}{size}\PY{p}{(}\PY{p}{)}\PY{p}{)}\PY{p}{)}
        \PY{n}{logJ} \PY{o}{=} \PY{n}{torch}\PY{o}{.}\PY{n}{sum}\PY{p}{(}\PY{p}{(}\PY{l+m+mi}{1} \PY{o}{\PYZhy{}} \PY{n+nb+bp}{self}\PY{o}{.}\PY{n}{mask}\PY{p}{)} \PY{o}{*} \PY{n}{s}\PY{p}{,} \PY{n}{dim}\PY{o}{=}\PY{n+nb}{tuple}\PY{p}{(}\PY{n}{axes}\PY{p}{)}\PY{p}{)}
        \PY{k}{return} \PY{n}{fx}\PY{p}{,} \PY{n}{logJ}

    \PY{k}{def} \PY{n+nf}{reverse}\PY{p}{(}\PY{n+nb+bp}{self}\PY{p}{,} \PY{n}{fx}\PY{p}{)}\PY{p}{:}
        \PY{n}{fx\PYZus{}frozen} \PY{o}{=} \PY{n+nb+bp}{self}\PY{o}{.}\PY{n}{mask} \PY{o}{*} \PY{n}{fx}
        \PY{n}{fx\PYZus{}active} \PY{o}{=} \PY{p}{(}\PY{l+m+mi}{1} \PY{o}{\PYZhy{}} \PY{n+nb+bp}{self}\PY{o}{.}\PY{n}{mask}\PY{p}{)} \PY{o}{*} \PY{n}{fx}  
        \PY{n}{net\PYZus{}out} \PY{o}{=} \PY{n+nb+bp}{self}\PY{o}{.}\PY{n}{net}\PY{p}{(}\PY{n}{fx\PYZus{}frozen}\PY{o}{.}\PY{n}{unsqueeze}\PY{p}{(}\PY{l+m+mi}{1}\PY{p}{)}\PY{p}{)}
        \PY{n}{s}\PY{p}{,} \PY{n}{t} \PY{o}{=} \PY{n}{net\PYZus{}out}\PY{p}{[}\PY{p}{:}\PY{p}{,}\PY{l+m+mi}{0}\PY{p}{]}\PY{p}{,} \PY{n}{net\PYZus{}out}\PY{p}{[}\PY{p}{:}\PY{p}{,}\PY{l+m+mi}{1}\PY{p}{]}
        \PY{n}{x} \PY{o}{=} \PY{p}{(}\PY{n}{fx\PYZus{}active} \PY{o}{\PYZhy{}} \PY{p}{(}\PY{l+m+mi}{1} \PY{o}{\PYZhy{}} \PY{n+nb+bp}{self}\PY{o}{.}\PY{n}{mask}\PY{p}{)} \PY{o}{*} \PY{n}{t}\PY{p}{)} \PY{o}{*} \PY{n}{torch}\PY{o}{.}\PY{n}{exp}\PY{p}{(}\PY{o}{\PYZhy{}}\PY{n}{s}\PY{p}{)} \PY{o}{+} \PY{n}{fx\PYZus{}frozen}
        \PY{n}{axes} \PY{o}{=} \PY{n+nb}{range}\PY{p}{(}\PY{l+m+mi}{1}\PY{p}{,}\PY{n+nb}{len}\PY{p}{(}\PY{n}{s}\PY{o}{.}\PY{n}{size}\PY{p}{(}\PY{p}{)}\PY{p}{)}\PY{p}{)}
        \PY{n}{logJ} \PY{o}{=} \PY{n}{torch}\PY{o}{.}\PY{n}{sum}\PY{p}{(}\PY{p}{(}\PY{l+m+mi}{1} \PY{o}{\PYZhy{}} \PY{n+nb+bp}{self}\PY{o}{.}\PY{n}{mask}\PY{p}{)}\PY{o}{*}\PY{p}{(}\PY{o}{\PYZhy{}}\PY{n}{s}\PY{p}{)}\PY{p}{,} \PY{n}{dim}\PY{o}{=}\PY{n+nb}{tuple}\PY{p}{(}\PY{n}{axes}\PY{p}{)}\PY{p}{)}
        \PY{k}{return} \PY{n}{x}\PY{p}{,} \PY{n}{logJ}
    \end{Verbatim}
  \end{tcolorbox}

  \hypertarget{convolutional-neural-nets-cnns}{%
\subsection{\texorpdfstring{\textbf{Convolutional neural nets
(CNNs)}}{Convolutional neural nets (CNNs)}}\label{convolutional-neural-nets-cnns}}

Any continuous function can be used to define the coupling layer
parameters \(s\) and \(t\). We'll use CNNs because they're cheap and
explicitly encode partial spacetime translation symmetry: the
parity-preserving translations are exact symmetries of the output
distribution, due to the checkerboard subsets. (\textbf{Note:} in
\cite{Albergo:2019eim}, fully-connected networks were used as a proof of
principle; we find CNNs are generally the better choice.)

Unlike typical uses of CNNs, our flow formalism requires the input and
output (spatial) shapes to be identical, so there are no pooling
operations and we use stride 1. To implement periodic BCs, we employ
circular padding.

\textbf{Technical note:} In below code we assume PyTorch \textgreater=
1.5.0 where padding semantics changed. If you are willing to downgrade,
the requirement than kernels have odd side length can be dropped, and
padding semantics should be changed to
\texttt{padding\_size\ =\ kernel\_size-1}.

  \begin{tcolorbox}[breakable, size=fbox, boxrule=1pt, pad at break*=1mm,colback=cellbackground, colframe=cellborder]
    \begin{Verbatim}[commandchars=\\\{\},fontsize=\small]
\PY{k}{def} \PY{n+nf}{make\PYZus{}conv\PYZus{}net}\PY{p}{(}\PY{o}{*}\PY{p}{,} \PY{n}{hidden\PYZus{}sizes}\PY{p}{,} \PY{n}{kernel\PYZus{}size}\PY{p}{,} \PY{n}{in\PYZus{}channels}\PY{p}{,} \PY{n}{out\PYZus{}channels}\PY{p}{,} \PY{n}{use\PYZus{}final\PYZus{}tanh}\PY{p}{)}\PY{p}{:}
    \PY{n}{sizes} \PY{o}{=} \PY{p}{[}\PY{n}{in\PYZus{}channels}\PY{p}{]} \PY{o}{+} \PY{n}{hidden\PYZus{}sizes} \PY{o}{+} \PY{p}{[}\PY{n}{out\PYZus{}channels}\PY{p}{]}
    \PY{k}{assert} \PY{n}{packaging}\PY{o}{.}\PY{n}{version}\PY{o}{.}\PY{n}{parse}\PY{p}{(}\PY{n}{torch}\PY{o}{.}\PY{n}{\PYZus{}\PYZus{}version\PYZus{}\PYZus{}}\PY{p}{)} \PY{o}{\PYZgt{}}\PY{o}{=} \PY{n}{packaging}\PY{o}{.}\PY{n}{version}\PY{o}{.}\PY{n}{parse}\PY{p}{(}\PY{l+s+s1}{\PYZsq{}}\PY{l+s+s1}{1.5.0}\PY{l+s+s1}{\PYZsq{}}\PY{p}{)}
    \PY{k}{assert} \PY{n}{kernel\PYZus{}size} \PY{o}{\PYZpc{}} \PY{l+m+mi}{2} \PY{o}{==} \PY{l+m+mi}{1}\PY{p}{,} \PY{l+s+s1}{\PYZsq{}}\PY{l+s+s1}{kernel size must be odd for PyTorch \PYZgt{}= 1.5.0}\PY{l+s+s1}{\PYZsq{}}
    \PY{n}{padding\PYZus{}size} \PY{o}{=} \PY{p}{(}\PY{n}{kernel\PYZus{}size} \PY{o}{/}\PY{o}{/} \PY{l+m+mi}{2}\PY{p}{)}
    \PY{n}{net} \PY{o}{=} \PY{p}{[}\PY{p}{]}
    \PY{k}{for} \PY{n}{i} \PY{o+ow}{in} \PY{n+nb}{range}\PY{p}{(}\PY{n+nb}{len}\PY{p}{(}\PY{n}{sizes}\PY{p}{)} \PY{o}{\PYZhy{}} \PY{l+m+mi}{1}\PY{p}{)}\PY{p}{:}
        \PY{n}{net}\PY{o}{.}\PY{n}{append}\PY{p}{(}\PY{n}{torch}\PY{o}{.}\PY{n}{nn}\PY{o}{.}\PY{n}{Conv2d}\PY{p}{(}
            \PY{n}{sizes}\PY{p}{[}\PY{n}{i}\PY{p}{]}\PY{p}{,} \PY{n}{sizes}\PY{p}{[}\PY{n}{i}\PY{o}{+}\PY{l+m+mi}{1}\PY{p}{]}\PY{p}{,} \PY{n}{kernel\PYZus{}size}\PY{p}{,} \PY{n}{padding}\PY{o}{=}\PY{n}{padding\PYZus{}size}\PY{p}{,}
            \PY{n}{stride}\PY{o}{=}\PY{l+m+mi}{1}\PY{p}{,} \PY{n}{padding\PYZus{}mode}\PY{o}{=}\PY{l+s+s1}{\PYZsq{}}\PY{l+s+s1}{circular}\PY{l+s+s1}{\PYZsq{}}\PY{p}{)}\PY{p}{)}
        \PY{k}{if} \PY{n}{i} \PY{o}{!=} \PY{n+nb}{len}\PY{p}{(}\PY{n}{sizes}\PY{p}{)} \PY{o}{\PYZhy{}} \PY{l+m+mi}{2}\PY{p}{:}
            \PY{n}{net}\PY{o}{.}\PY{n}{append}\PY{p}{(}\PY{n}{torch}\PY{o}{.}\PY{n}{nn}\PY{o}{.}\PY{n}{LeakyReLU}\PY{p}{(}\PY{p}{)}\PY{p}{)}
        \PY{k}{else}\PY{p}{:}
            \PY{k}{if} \PY{n}{use\PYZus{}final\PYZus{}tanh}\PY{p}{:}
                \PY{n}{net}\PY{o}{.}\PY{n}{append}\PY{p}{(}\PY{n}{torch}\PY{o}{.}\PY{n}{nn}\PY{o}{.}\PY{n}{Tanh}\PY{p}{(}\PY{p}{)}\PY{p}{)}
    \PY{k}{return} \PY{n}{torch}\PY{o}{.}\PY{n}{nn}\PY{o}{.}\PY{n}{Sequential}\PY{p}{(}\PY{o}{*}\PY{n}{net}\PY{p}{)}
    \end{Verbatim}
  \end{tcolorbox}

  \hypertarget{assemble-the-model}{%
\subsection{\texorpdfstring{\textbf{Assemble the
model}}{Assemble the model}}\label{assemble-the-model}}

We can construct our model for \(p(\phi)\) by composing a sequence of
these affine coupling layers. We'll use 16 layers, with checkerboard
parity alternating between sites. The CNNs used to compute parameters
have kernel size 3x3 which is sufficient to condition on local
information in each transform, allowing the flow to build up local
correlations in the field configuration. Larger kernel sizes are also
possible at increased cost.

Each network has one input channel \(\phi\) and two output channels
\(\left(s(\phi), t(\phi)\right)\).

  \begin{tcolorbox}[breakable, size=fbox, boxrule=1pt, pad at break*=1mm,colback=cellbackground, colframe=cellborder]
    \begin{Verbatim}[commandchars=\\\{\},fontsize=\small]
\PY{k}{def} \PY{n+nf}{make\PYZus{}phi4\PYZus{}affine\PYZus{}layers}\PY{p}{(}\PY{o}{*}\PY{p}{,} \PY{n}{n\PYZus{}layers}\PY{p}{,} \PY{n}{lattice\PYZus{}shape}\PY{p}{,} \PY{n}{hidden\PYZus{}sizes}\PY{p}{,} \PY{n}{kernel\PYZus{}size}\PY{p}{)}\PY{p}{:}
    \PY{n}{layers} \PY{o}{=} \PY{p}{[}\PY{p}{]}
    \PY{k}{for} \PY{n}{i} \PY{o+ow}{in} \PY{n+nb}{range}\PY{p}{(}\PY{n}{n\PYZus{}layers}\PY{p}{)}\PY{p}{:}
        \PY{n}{parity} \PY{o}{=} \PY{n}{i} \PY{o}{\PYZpc{}} \PY{l+m+mi}{2}
        \PY{n}{net} \PY{o}{=} \PY{n}{make\PYZus{}conv\PYZus{}net}\PY{p}{(}
            \PY{n}{in\PYZus{}channels}\PY{o}{=}\PY{l+m+mi}{1}\PY{p}{,} \PY{n}{out\PYZus{}channels}\PY{o}{=}\PY{l+m+mi}{2}\PY{p}{,} \PY{n}{hidden\PYZus{}sizes}\PY{o}{=}\PY{n}{hidden\PYZus{}sizes}\PY{p}{,}
            \PY{n}{kernel\PYZus{}size}\PY{o}{=}\PY{n}{kernel\PYZus{}size}\PY{p}{,} \PY{n}{use\PYZus{}final\PYZus{}tanh}\PY{o}{=}\PY{k+kc}{True}\PY{p}{)}
        \PY{n}{coupling} \PY{o}{=} \PY{n}{AffineCoupling}\PY{p}{(}\PY{n}{net}\PY{p}{,} \PY{n}{mask\PYZus{}shape}\PY{o}{=}\PY{n}{lattice\PYZus{}shape}\PY{p}{,} \PY{n}{mask\PYZus{}parity}\PY{o}{=}\PY{n}{parity}\PY{p}{)}
        \PY{n}{layers}\PY{o}{.}\PY{n}{append}\PY{p}{(}\PY{n}{coupling}\PY{p}{)}
    \PY{k}{return} \PY{n}{torch}\PY{o}{.}\PY{n}{nn}\PY{o}{.}\PY{n}{ModuleList}\PY{p}{(}\PY{n}{layers}\PY{p}{)}
    \end{Verbatim}
  \end{tcolorbox}

  \begin{tcolorbox}[breakable, size=fbox, boxrule=1pt, pad at break*=1mm,colback=cellbackground, colframe=cellborder]
    \begin{Verbatim}[commandchars=\\\{\},fontsize=\small]
\PY{n}{n\PYZus{}layers} \PY{o}{=} \PY{l+m+mi}{16}
\PY{n}{hidden\PYZus{}sizes} \PY{o}{=} \PY{p}{[}\PY{l+m+mi}{8}\PY{p}{,}\PY{l+m+mi}{8}\PY{p}{]}
\PY{n}{kernel\PYZus{}size} \PY{o}{=} \PY{l+m+mi}{3}
\PY{n}{layers} \PY{o}{=} \PY{n}{make\PYZus{}phi4\PYZus{}affine\PYZus{}layers}\PY{p}{(}
    \PY{n}{lattice\PYZus{}shape}\PY{o}{=}\PY{n}{lattice\PYZus{}shape}\PY{p}{,} \PY{n}{n\PYZus{}layers}\PY{o}{=}\PY{n}{n\PYZus{}layers}\PY{p}{,} 
    \PY{n}{hidden\PYZus{}sizes}\PY{o}{=}\PY{n}{hidden\PYZus{}sizes}\PY{p}{,} \PY{n}{kernel\PYZus{}size}\PY{o}{=}\PY{n}{kernel\PYZus{}size}\PY{p}{)}
\PY{n}{model} \PY{o}{=} \PY{p}{\PYZob{}}\PY{l+s+s1}{\PYZsq{}}\PY{l+s+s1}{layers}\PY{l+s+s1}{\PYZsq{}}\PY{p}{:} \PY{n}{layers}\PY{p}{,} \PY{l+s+s1}{\PYZsq{}}\PY{l+s+s1}{prior}\PY{l+s+s1}{\PYZsq{}}\PY{p}{:} \PY{n}{prior}\PY{p}{\PYZcb{}}
    \end{Verbatim}
  \end{tcolorbox}

  \hypertarget{train-the-model}{%
\subsection{\texorpdfstring{\textbf{Train the
model}}{Train the model}}\label{train-the-model}}

With a model in hand, we need to optimize the coupling layers to improve
the model distribution \(q(\phi)\). To do that, we need a way to measure
how close the model and true distributions are {[}\(q(\phi)\) vs
\(p(\phi)\), respectively{]}. We use a quantity known as the
Kullback-Leibler (KL) divergence to do this. The KL divergence is
minimized when \(p = q\).

Those familiar with flows will note that they are usually trained with
the ``forward direction'' of the KL divergence, \begin{equation}
D_{KL}(p || q) \equiv \int d\phi ~ p(\phi) \left[ \log{p}(\phi) - \log{q}(\phi) \right]
\end{equation} which we can estimate with \(N\) samples drawn from the
target distribution (\(\phi_i \sim p\)) as \begin{equation}
\widehat{D}_{KL}(p || q) = \frac{1}{N} \sum_{i=1}^N \left[ \log{p}(\phi_i) - \log{q}(\phi_i) \right] \quad \left( \phi_i \sim p \right)
\end{equation} corresponding to maximum likelihood estimation with
respect to training data from the true distribution.

Because training data drawn from \(p\) can be scarce in simulations of
lattice field theories, we instead make use of the ``reverse'' KL
divergence, \begin{equation}
D_{KL}(q || p) \equiv \int d\phi ~ q(\phi) \left[ \log{q}(\phi) - \log{p}(\phi) \right]
\end{equation} which we can estimate using \(N\) samples drawn from the
model distribution (\(\phi_i \sim q\)) as \begin{equation}
\widehat{D}_{KL}(q || p) = \frac{1}{N} \sum_{i=1}^N \left[ \log{q}(\phi_i) - \log{p}(\phi_i) \right] \quad \left( \phi_i \sim q \right).
\end{equation} Because data need only be sampled from the \emph{model}
distribution, we can optimize \(q(\phi)\) without data from \(p(\phi)\)
{[}which typically is expensive to generate using standard algorithms
like HMC{]}. This ``self-training'' protocol then consists of 1. Drawing
samples and density estimates from the model 2. Estimating the reverse
KL divergence 3. Using standard stochastic gradient descent methods to
iteratively update neural network weights (we'll use the Adam optimizer)

A possible tradeoff of this approach is that the reverse KL is known as
mode-seeking / zero-forcing (see e.g.~\cite{huszar2015not}) which means
it favors assigning mass to a large mode in the probability density, and
places zero mass elsewhere. This could be disadvantageous for multimodal
target densities. This problem will be investigated in future work.

  \begin{tcolorbox}[breakable, size=fbox, boxrule=1pt, pad at break*=1mm,colback=cellbackground, colframe=cellborder]
    \begin{Verbatim}[commandchars=\\\{\},fontsize=\small]
\PY{k}{def} \PY{n+nf}{calc\PYZus{}dkl}\PY{p}{(}\PY{n}{logp}\PY{p}{,} \PY{n}{logq}\PY{p}{)}\PY{p}{:}
    \PY{k}{return} \PY{p}{(}\PY{n}{logq} \PY{o}{\PYZhy{}} \PY{n}{logp}\PY{p}{)}\PY{o}{.}\PY{n}{mean}\PY{p}{(}\PY{p}{)}  \PY{c+c1}{\PYZsh{} reverse KL, assuming samples from q}
    \end{Verbatim}
  \end{tcolorbox}

  Note that the training step defined below logs a few metrics, including
the effective sample size (ESS) defined and explained later.

  \begin{tcolorbox}[breakable, size=fbox, boxrule=1pt, pad at break*=1mm,colback=cellbackground, colframe=cellborder]
    \begin{Verbatim}[commandchars=\\\{\},fontsize=\small]
\PY{k}{def} \PY{n+nf}{train\PYZus{}step}\PY{p}{(}\PY{n}{model}\PY{p}{,} \PY{n}{action}\PY{p}{,} \PY{n}{loss\PYZus{}fn}\PY{p}{,} \PY{n}{optimizer}\PY{p}{,} \PY{n}{metrics}\PY{p}{)}\PY{p}{:}
    \PY{n}{layers}\PY{p}{,} \PY{n}{prior} \PY{o}{=} \PY{n}{model}\PY{p}{[}\PY{l+s+s1}{\PYZsq{}}\PY{l+s+s1}{layers}\PY{l+s+s1}{\PYZsq{}}\PY{p}{]}\PY{p}{,} \PY{n}{model}\PY{p}{[}\PY{l+s+s1}{\PYZsq{}}\PY{l+s+s1}{prior}\PY{l+s+s1}{\PYZsq{}}\PY{p}{]}
    \PY{n}{optimizer}\PY{o}{.}\PY{n}{zero\PYZus{}grad}\PY{p}{(}\PY{p}{)}

    \PY{n}{x}\PY{p}{,} \PY{n}{logq} \PY{o}{=} \PY{n}{apply\PYZus{}flow\PYZus{}to\PYZus{}prior}\PY{p}{(}\PY{n}{prior}\PY{p}{,} \PY{n}{layers}\PY{p}{,} \PY{n}{batch\PYZus{}size}\PY{o}{=}\PY{n}{batch\PYZus{}size}\PY{p}{)}
    \PY{n}{logp} \PY{o}{=} \PY{o}{\PYZhy{}}\PY{n}{action}\PY{p}{(}\PY{n}{x}\PY{p}{)}
    \PY{n}{loss} \PY{o}{=} \PY{n}{calc\PYZus{}dkl}\PY{p}{(}\PY{n}{logp}\PY{p}{,} \PY{n}{logq}\PY{p}{)}
    \PY{n}{loss}\PY{o}{.}\PY{n}{backward}\PY{p}{(}\PY{p}{)}

    \PY{n}{optimizer}\PY{o}{.}\PY{n}{step}\PY{p}{(}\PY{p}{)}

    \PY{n}{metrics}\PY{p}{[}\PY{l+s+s1}{\PYZsq{}}\PY{l+s+s1}{loss}\PY{l+s+s1}{\PYZsq{}}\PY{p}{]}\PY{o}{.}\PY{n}{append}\PY{p}{(}\PY{n}{grab}\PY{p}{(}\PY{n}{loss}\PY{p}{)}\PY{p}{)}
    \PY{n}{metrics}\PY{p}{[}\PY{l+s+s1}{\PYZsq{}}\PY{l+s+s1}{logp}\PY{l+s+s1}{\PYZsq{}}\PY{p}{]}\PY{o}{.}\PY{n}{append}\PY{p}{(}\PY{n}{grab}\PY{p}{(}\PY{n}{logp}\PY{p}{)}\PY{p}{)}
    \PY{n}{metrics}\PY{p}{[}\PY{l+s+s1}{\PYZsq{}}\PY{l+s+s1}{logq}\PY{l+s+s1}{\PYZsq{}}\PY{p}{]}\PY{o}{.}\PY{n}{append}\PY{p}{(}\PY{n}{grab}\PY{p}{(}\PY{n}{logq}\PY{p}{)}\PY{p}{)}
    \PY{n}{metrics}\PY{p}{[}\PY{l+s+s1}{\PYZsq{}}\PY{l+s+s1}{ess}\PY{l+s+s1}{\PYZsq{}}\PY{p}{]}\PY{o}{.}\PY{n}{append}\PY{p}{(}\PY{n}{grab}\PY{p}{(} \PY{n}{compute\PYZus{}ess}\PY{p}{(}\PY{n}{logp}\PY{p}{,} \PY{n}{logq}\PY{p}{)} \PY{p}{)}\PY{p}{)}
    \end{Verbatim}
  \end{tcolorbox}

  \textbf{Caveat:} \(p(\phi)\) is only known up to normalization,
\(p(\phi) \propto e^{-S[\phi]}\). Using this unnormalized value shifts
the KL divergence by an overall constant. This does not affect training,
but without this normalization we cannot know whether we are converging
to a good estimate directly from the unnormalized KL.

  \hypertarget{telemetry}{%
\subsubsection{\texorpdfstring{\textbf{Telemetry}}{Telemetry}}\label{telemetry}}

We'll measure some observables and diagnostics as we go.

For a batch of samples \(\phi_i\), the effective sample size (ESS) is
defined as \begin{equation}
\frac{ \left(\frac{1}{N} \sum_i p[\phi_i]/q[\phi_i] \right)^2 }{ \frac{1}{N} \sum_i \left( p[\phi_i]/q[\phi_i] \right)^2 }
\end{equation} where \(i\) indexes the samples. This definition
normalizes the ESS to live in the range \([0,1]\). The ESS provides a
useful measure of model quality that doesn't require the overall
normalization of \(p(x)\), where larger values indicate a better
effective sampling of the desired distribution and \(\mathrm{ESS} = 1\)
is a perfect independent draw from the desired distribution for each
sample.

Why not use this directly to train? It's much noisier than the KL
divergences, so in practice we find it's less effective as a loss
function.

\textbf{Caution:} The ESS is biased towards larger values when estimated
using small batches of samples. Much like measures of autocorrelation
time in MCMC approaches, a sufficiently large sample size is needed to
determine whether any regions of sample space are missed.

  \begin{tcolorbox}[breakable, size=fbox, boxrule=1pt, pad at break*=1mm,colback=cellbackground, colframe=cellborder]
    \begin{Verbatim}[commandchars=\\\{\},fontsize=\small]
\PY{k}{def} \PY{n+nf}{compute\PYZus{}ess}\PY{p}{(}\PY{n}{logp}\PY{p}{,} \PY{n}{logq}\PY{p}{)}\PY{p}{:}
    \PY{n}{logw} \PY{o}{=} \PY{n}{logp} \PY{o}{\PYZhy{}} \PY{n}{logq}
    \PY{n}{log\PYZus{}ess} \PY{o}{=} \PY{l+m+mi}{2}\PY{o}{*}\PY{n}{torch}\PY{o}{.}\PY{n}{logsumexp}\PY{p}{(}\PY{n}{logw}\PY{p}{,} \PY{n}{dim}\PY{o}{=}\PY{l+m+mi}{0}\PY{p}{)} \PY{o}{\PYZhy{}} \PY{n}{torch}\PY{o}{.}\PY{n}{logsumexp}\PY{p}{(}\PY{l+m+mi}{2}\PY{o}{*}\PY{n}{logw}\PY{p}{,} \PY{n}{dim}\PY{o}{=}\PY{l+m+mi}{0}\PY{p}{)}
    \PY{n}{ess\PYZus{}per\PYZus{}cfg} \PY{o}{=} \PY{n}{torch}\PY{o}{.}\PY{n}{exp}\PY{p}{(}\PY{n}{log\PYZus{}ess}\PY{p}{)} \PY{o}{/} \PY{n+nb}{len}\PY{p}{(}\PY{n}{logw}\PY{p}{)}
    \PY{k}{return} \PY{n}{ess\PYZus{}per\PYZus{}cfg}
    \end{Verbatim}
  \end{tcolorbox}

  \begin{tcolorbox}[breakable, size=fbox, boxrule=1pt, pad at break*=1mm,colback=cellbackground, colframe=cellborder]
    \begin{Verbatim}[commandchars=\\\{\},fontsize=\small]
\PY{k}{def} \PY{n+nf}{print\PYZus{}metrics}\PY{p}{(}\PY{n}{history}\PY{p}{,} \PY{n}{avg\PYZus{}last\PYZus{}N\PYZus{}epochs}\PY{p}{)}\PY{p}{:}
    \PY{n+nb}{print}\PY{p}{(}\PY{l+s+sa}{f}\PY{l+s+s1}{\PYZsq{}}\PY{l+s+s1}{== Era }\PY{l+s+si}{\PYZob{}}\PY{n}{era}\PY{l+s+si}{\PYZcb{}}\PY{l+s+s1}{ | Epoch }\PY{l+s+si}{\PYZob{}}\PY{n}{epoch}\PY{l+s+si}{\PYZcb{}}\PY{l+s+s1}{ metrics ==}\PY{l+s+s1}{\PYZsq{}}\PY{p}{)}
    \PY{k}{for} \PY{n}{key}\PY{p}{,} \PY{n}{val} \PY{o+ow}{in} \PY{n}{history}\PY{o}{.}\PY{n}{items}\PY{p}{(}\PY{p}{)}\PY{p}{:}
        \PY{n}{avgd} \PY{o}{=} \PY{n}{np}\PY{o}{.}\PY{n}{mean}\PY{p}{(}\PY{n}{val}\PY{p}{[}\PY{o}{\PYZhy{}}\PY{n}{avg\PYZus{}last\PYZus{}N\PYZus{}epochs}\PY{p}{:}\PY{p}{]}\PY{p}{)}
        \PY{n+nb}{print}\PY{p}{(}\PY{l+s+sa}{f}\PY{l+s+s1}{\PYZsq{}}\PY{l+s+se}{\PYZbs{}t}\PY{l+s+si}{\PYZob{}}\PY{n}{key}\PY{l+s+si}{\PYZcb{}}\PY{l+s+s1}{ }\PY{l+s+si}{\PYZob{}}\PY{n}{avgd}\PY{l+s+si}{:}\PY{l+s+s1}{g}\PY{l+s+si}{\PYZcb{}}\PY{l+s+s1}{\PYZsq{}}\PY{p}{)}
    \end{Verbatim}
  \end{tcolorbox}

  \hypertarget{do-the-training}{%
\subsubsection{\texorpdfstring{\textbf{Do the
training!}}{Do the training!}}\label{do-the-training}}

We find that this model trains to achieve an average ESS \(\sim 20 \%\)
after 40 eras with 100 epochs each, which takes \(\sim 30\) minutes on a
Colab GPU. We point out that ESS is a good metric for training but it
fluctuates significantly and can have a bias at finite sample size, so
care should be taken in interpreting the results.

  You can either load a pre-trained model or train your own based on the
flag below.

  \begin{tcolorbox}[breakable, size=fbox, boxrule=1pt, pad at break*=1mm,colback=cellbackground, colframe=cellborder]
    \begin{Verbatim}[commandchars=\\\{\},fontsize=\small]
\PY{n}{use\PYZus{}pretrained} \PY{o}{=} \PY{k+kc}{True}
    \end{Verbatim}
  \end{tcolorbox}

  Below we summarize all parameters discussed so far for the sake of
convenience.

  \begin{tcolorbox}[breakable, size=fbox, boxrule=1pt, pad at break*=1mm,colback=cellbackground, colframe=cellborder]
    \begin{Verbatim}[commandchars=\\\{\},fontsize=\small]
\PY{c+c1}{\PYZsh{} Lattice Theory}
\PY{n}{L} \PY{o}{=} \PY{l+m+mi}{8}
\PY{n}{lattice\PYZus{}shape} \PY{o}{=} \PY{p}{(}\PY{n}{L}\PY{p}{,}\PY{n}{L}\PY{p}{)}
\PY{n}{M2} \PY{o}{=} \PY{o}{\PYZhy{}}\PY{l+m+mf}{4.0}
\PY{n}{lam} \PY{o}{=} \PY{l+m+mf}{8.0}
\PY{n}{phi4\PYZus{}action} \PY{o}{=} \PY{n}{ScalarPhi4Action}\PY{p}{(}\PY{n}{M2}\PY{o}{=}\PY{n}{M2}\PY{p}{,} \PY{n}{lam}\PY{o}{=}\PY{n}{lam}\PY{p}{)}

\PY{c+c1}{\PYZsh{} Model}
\PY{n}{prior} \PY{o}{=} \PY{n}{SimpleNormal}\PY{p}{(}\PY{n}{torch}\PY{o}{.}\PY{n}{zeros}\PY{p}{(}\PY{n}{lattice\PYZus{}shape}\PY{p}{)}\PY{p}{,} \PY{n}{torch}\PY{o}{.}\PY{n}{ones}\PY{p}{(}\PY{n}{lattice\PYZus{}shape}\PY{p}{)}\PY{p}{)}

\PY{n}{n\PYZus{}layers} \PY{o}{=} \PY{l+m+mi}{16}
\PY{n}{hidden\PYZus{}sizes} \PY{o}{=} \PY{p}{[}\PY{l+m+mi}{8}\PY{p}{,}\PY{l+m+mi}{8}\PY{p}{]}
\PY{n}{kernel\PYZus{}size} \PY{o}{=} \PY{l+m+mi}{3}
\PY{n}{layers} \PY{o}{=} \PY{n}{make\PYZus{}phi4\PYZus{}affine\PYZus{}layers}\PY{p}{(}\PY{n}{lattice\PYZus{}shape}\PY{o}{=}\PY{n}{lattice\PYZus{}shape}\PY{p}{,} \PY{n}{n\PYZus{}layers}\PY{o}{=}\PY{n}{n\PYZus{}layers}\PY{p}{,} 
    \PY{n}{hidden\PYZus{}sizes}\PY{o}{=}\PY{n}{hidden\PYZus{}sizes}\PY{p}{,} \PY{n}{kernel\PYZus{}size}\PY{o}{=}\PY{n}{kernel\PYZus{}size}\PY{p}{)}
\PY{n}{model} \PY{o}{=} \PY{p}{\PYZob{}}\PY{l+s+s1}{\PYZsq{}}\PY{l+s+s1}{layers}\PY{l+s+s1}{\PYZsq{}}\PY{p}{:} \PY{n}{layers}\PY{p}{,} \PY{l+s+s1}{\PYZsq{}}\PY{l+s+s1}{prior}\PY{l+s+s1}{\PYZsq{}}\PY{p}{:} \PY{n}{prior}\PY{p}{\PYZcb{}}

\PY{c+c1}{\PYZsh{} Training}
\PY{n}{base\PYZus{}lr} \PY{o}{=} \PY{l+m+mf}{.001}
\PY{n}{optimizer} \PY{o}{=} \PY{n}{torch}\PY{o}{.}\PY{n}{optim}\PY{o}{.}\PY{n}{Adam}\PY{p}{(}\PY{n}{model}\PY{p}{[}\PY{l+s+s1}{\PYZsq{}}\PY{l+s+s1}{layers}\PY{l+s+s1}{\PYZsq{}}\PY{p}{]}\PY{o}{.}\PY{n}{parameters}\PY{p}{(}\PY{p}{)}\PY{p}{,} \PY{n}{lr}\PY{o}{=}\PY{n}{base\PYZus{}lr}\PY{p}{)}
    \end{Verbatim}
  \end{tcolorbox}

  As with any good cooking show, we made a trained version of the model
weights ahead of time (loaded if \texttt{use\_pretrained\ ==\ True}).

  \begin{tcolorbox}[breakable, size=fbox, boxrule=1pt, pad at break*=1mm,colback=cellbackground, colframe=cellborder]
    \begin{Verbatim}[commandchars=\\\{\},fontsize=\small]
\PY{k}{if} \PY{n}{use\PYZus{}pretrained}\PY{p}{:}
    \PY{n+nb}{print}\PY{p}{(}\PY{l+s+s1}{\PYZsq{}}\PY{l+s+s1}{Loading pre\PYZhy{}trained model}\PY{l+s+s1}{\PYZsq{}}\PY{p}{)}
    \PY{n}{phi4\PYZus{}trained\PYZus{}weights} \PY{o}{=} \PY{n}{torch}\PY{o}{.}\PY{n}{load}\PY{p}{(}\PY{n}{io}\PY{o}{.}\PY{n}{BytesIO}\PY{p}{(}\PY{n}{base64}\PY{o}{.}\PY{n}{b64decode}\PY{p}{(}\PY{l+s+sa}{b}\PY{l+s+s2}{\PYZdq{}\PYZdq{}\PYZdq{}}
\PY{l+s+s2}{\PYZlt{}snipped base64 blob\PYZgt{}}
\PY{l+s+s2}{    }\PY{l+s+s2}{\PYZdq{}\PYZdq{}\PYZdq{}}\PY{o}{.}\PY{n}{strip}\PY{p}{(}\PY{p}{)}\PY{p}{)}\PY{p}{)}\PY{p}{,} \PY{n}{map\PYZus{}location}\PY{o}{=}\PY{n}{torch}\PY{o}{.}\PY{n}{device}\PY{p}{(}\PY{l+s+s1}{\PYZsq{}}\PY{l+s+s1}{cpu}\PY{l+s+s1}{\PYZsq{}}\PY{p}{)}\PY{p}{)}
    \PY{n}{model}\PY{p}{[}\PY{l+s+s1}{\PYZsq{}}\PY{l+s+s1}{layers}\PY{l+s+s1}{\PYZsq{}}\PY{p}{]}\PY{o}{.}\PY{n}{load\PYZus{}state\PYZus{}dict}\PY{p}{(}\PY{n}{phi4\PYZus{}trained\PYZus{}weights}\PY{p}{)}
    \PY{k}{if} \PY{n}{torch\PYZus{}device} \PY{o}{==} \PY{l+s+s1}{\PYZsq{}}\PY{l+s+s1}{cuda}\PY{l+s+s1}{\PYZsq{}}\PY{p}{:}
        \PY{n}{model}\PY{p}{[}\PY{l+s+s1}{\PYZsq{}}\PY{l+s+s1}{layers}\PY{l+s+s1}{\PYZsq{}}\PY{p}{]}\PY{o}{.}\PY{n}{cuda}\PY{p}{(}\PY{p}{)}
\PY{k}{else}\PY{p}{:}
    \PY{n+nb}{print}\PY{p}{(}\PY{l+s+s1}{\PYZsq{}}\PY{l+s+s1}{Skipping pre\PYZhy{}trained model}\PY{l+s+s1}{\PYZsq{}}\PY{p}{)}
    \end{Verbatim}
  \end{tcolorbox}

{\color{gray}
    \begin{Verbatim}[commandchars=\\\{\},fontsize=\small]
>>> Loading pre-trained model
    \end{Verbatim}
}

  Main training setup and loop.

  \begin{tcolorbox}[breakable, size=fbox, boxrule=1pt, pad at break*=1mm,colback=cellbackground, colframe=cellborder]
    \begin{Verbatim}[commandchars=\\\{\},fontsize=\small]
\PY{n}{N\PYZus{}era} \PY{o}{=} \PY{l+m+mi}{25}
\PY{n}{N\PYZus{}epoch} \PY{o}{=} \PY{l+m+mi}{100}
\PY{n}{batch\PYZus{}size} \PY{o}{=} \PY{l+m+mi}{64}
\PY{n}{print\PYZus{}freq} \PY{o}{=} \PY{n}{N\PYZus{}epoch}
\PY{n}{plot\PYZus{}freq} \PY{o}{=} \PY{l+m+mi}{1}

\PY{n}{history} \PY{o}{=} \PY{p}{\PYZob{}}
    \PY{l+s+s1}{\PYZsq{}}\PY{l+s+s1}{loss}\PY{l+s+s1}{\PYZsq{}} \PY{p}{:} \PY{p}{[}\PY{p}{]}\PY{p}{,}
    \PY{l+s+s1}{\PYZsq{}}\PY{l+s+s1}{logp}\PY{l+s+s1}{\PYZsq{}} \PY{p}{:} \PY{p}{[}\PY{p}{]}\PY{p}{,}
    \PY{l+s+s1}{\PYZsq{}}\PY{l+s+s1}{logq}\PY{l+s+s1}{\PYZsq{}} \PY{p}{:} \PY{p}{[}\PY{p}{]}\PY{p}{,}
    \PY{l+s+s1}{\PYZsq{}}\PY{l+s+s1}{ess}\PY{l+s+s1}{\PYZsq{}} \PY{p}{:} \PY{p}{[}\PY{p}{]}
\PY{p}{\PYZcb{}}
    \end{Verbatim}
  \end{tcolorbox}

  \begin{tcolorbox}[breakable, size=fbox, boxrule=1pt, pad at break*=1mm,colback=cellbackground, colframe=cellborder]
    \begin{Verbatim}[commandchars=\\\{\},fontsize=\small]
\PY{k}{if} \PY{o+ow}{not} \PY{n}{use\PYZus{}pretrained}\PY{p}{:}
    \PY{p}{[}\PY{n}{plt}\PY{o}{.}\PY{n}{close}\PY{p}{(}\PY{n}{plt}\PY{o}{.}\PY{n}{figure}\PY{p}{(}\PY{n}{fignum}\PY{p}{)}\PY{p}{)} \PY{k}{for} \PY{n}{fignum} \PY{o+ow}{in} \PY{n}{plt}\PY{o}{.}\PY{n}{get\PYZus{}fignums}\PY{p}{(}\PY{p}{)}\PY{p}{]} \PY{c+c1}{\PYZsh{} close all existing figures}
    \PY{n}{live\PYZus{}plot} \PY{o}{=} \PY{n}{init\PYZus{}live\PYZus{}plot}\PY{p}{(}\PY{p}{)}

    \PY{k}{for} \PY{n}{era} \PY{o+ow}{in} \PY{n+nb}{range}\PY{p}{(}\PY{n}{N\PYZus{}era}\PY{p}{)}\PY{p}{:}
        \PY{k}{for} \PY{n}{epoch} \PY{o+ow}{in} \PY{n+nb}{range}\PY{p}{(}\PY{n}{N\PYZus{}epoch}\PY{p}{)}\PY{p}{:}
            \PY{n}{train\PYZus{}step}\PY{p}{(}\PY{n}{model}\PY{p}{,} \PY{n}{phi4\PYZus{}action}\PY{p}{,} \PY{n}{calc\PYZus{}dkl}\PY{p}{,} \PY{n}{optimizer}\PY{p}{,} \PY{n}{history}\PY{p}{)}

            \PY{k}{if} \PY{n}{epoch} \PY{o}{\PYZpc{}} \PY{n}{print\PYZus{}freq} \PY{o}{==} \PY{l+m+mi}{0}\PY{p}{:}
                \PY{n}{print\PYZus{}metrics}\PY{p}{(}\PY{n}{history}\PY{p}{,} \PY{n}{avg\PYZus{}last\PYZus{}N\PYZus{}epochs}\PY{o}{=}\PY{n}{print\PYZus{}freq}\PY{p}{)}

            \PY{k}{if} \PY{n}{epoch} \PY{o}{\PYZpc{}} \PY{n}{plot\PYZus{}freq} \PY{o}{==} \PY{l+m+mi}{0}\PY{p}{:}
                \PY{n}{update\PYZus{}plots}\PY{p}{(}\PY{n}{history}\PY{p}{,} \PY{o}{*}\PY{o}{*}\PY{n}{live\PYZus{}plot}\PY{p}{)}
\PY{k}{else}\PY{p}{:}
    \PY{n+nb}{print}\PY{p}{(}\PY{l+s+s1}{\PYZsq{}}\PY{l+s+s1}{Skipping training}\PY{l+s+s1}{\PYZsq{}}\PY{p}{)}
    \end{Verbatim}
  \end{tcolorbox}

{\color{gray}
    \begin{Verbatim}[commandchars=\\\{\},fontsize=\small]
>>> Skipping training
    \end{Verbatim}
}

  Serialize the weights to distribute the model in this state.

  \begin{tcolorbox}[breakable, size=fbox, boxrule=1pt, pad at break*=1mm,colback=cellbackground, colframe=cellborder]
    \begin{Verbatim}[commandchars=\\\{\},fontsize=\small]
\PY{n+nb}{print}\PY{p}{(}\PY{l+s+s1}{\PYZsq{}}\PY{l+s+s1}{Model weights blob:}\PY{l+s+se}{\PYZbs{}n}\PY{l+s+s1}{===}\PY{l+s+s1}{\PYZsq{}}\PY{p}{)}
\PY{n}{serialized\PYZus{}model} \PY{o}{=} \PY{n}{io}\PY{o}{.}\PY{n}{BytesIO}\PY{p}{(}\PY{p}{)}
\PY{n}{torch}\PY{o}{.}\PY{n}{save}\PY{p}{(}\PY{n}{model}\PY{p}{[}\PY{l+s+s1}{\PYZsq{}}\PY{l+s+s1}{layers}\PY{l+s+s1}{\PYZsq{}}\PY{p}{]}\PY{o}{.}\PY{n}{state\PYZus{}dict}\PY{p}{(}\PY{p}{)}\PY{p}{,} \PY{n}{serialized\PYZus{}model}\PY{p}{)}
\PY{n+nb}{print}\PY{p}{(}\PY{n}{base64}\PY{o}{.}\PY{n}{b64encode}\PY{p}{(}\PY{n}{serialized\PYZus{}model}\PY{o}{.}\PY{n}{getbuffer}\PY{p}{(}\PY{p}{)}\PY{p}{)}\PY{o}{.}\PY{n}{decode}\PY{p}{(}\PY{l+s+s1}{\PYZsq{}}\PY{l+s+s1}{utf\PYZhy{}8}\PY{l+s+s1}{\PYZsq{}}\PY{p}{)}\PY{p}{)}
\PY{n+nb}{print}\PY{p}{(}\PY{l+s+s1}{\PYZsq{}}\PY{l+s+s1}{===}\PY{l+s+s1}{\PYZsq{}}\PY{p}{)}
    \end{Verbatim}
  \end{tcolorbox}

  \hypertarget{evaluate-the-model}{%
\subsection{\texorpdfstring{\textbf{Evaluate the
model}}{Evaluate the model}}\label{evaluate-the-model}}

With a trained model, we now directly draw samples from the model and
check their quality below. We find samples that have regions with
smoother, correlated fluctuations, in comparison to the raw noise from
the prior distribution seen above.

\textbf{Caution:} These samples are drawn from a distribution that only
\emph{approximates} the desired one, so we stress that one should not
measure and report observables directly using these model samples, as
this would introduce bias. However, as discussed later, the reported
probability density from the models allows us to either reweight or
resample, producing \textbf{unbiased estimates of observables} when
these steps are taken.

  \begin{tcolorbox}[breakable, size=fbox, boxrule=1pt, pad at break*=1mm,colback=cellbackground, colframe=cellborder]
    \begin{Verbatim}[commandchars=\\\{\},fontsize=\small]
\PY{n}{torch\PYZus{}x}\PY{p}{,} \PY{n}{torch\PYZus{}logq} \PY{o}{=} \PY{n}{apply\PYZus{}flow\PYZus{}to\PYZus{}prior}\PY{p}{(}\PY{n}{prior}\PY{p}{,} \PY{n}{layers}\PY{p}{,} \PY{n}{batch\PYZus{}size}\PY{o}{=}\PY{l+m+mi}{1024}\PY{p}{)}
\PY{n}{x} \PY{o}{=} \PY{n}{grab}\PY{p}{(}\PY{n}{torch\PYZus{}x}\PY{p}{)}

\PY{n}{fig}\PY{p}{,} \PY{n}{ax} \PY{o}{=} \PY{n}{plt}\PY{o}{.}\PY{n}{subplots}\PY{p}{(}\PY{l+m+mi}{4}\PY{p}{,}\PY{l+m+mi}{4}\PY{p}{,} \PY{n}{dpi}\PY{o}{=}\PY{l+m+mi}{125}\PY{p}{,} \PY{n}{figsize}\PY{o}{=}\PY{p}{(}\PY{l+m+mi}{4}\PY{p}{,}\PY{l+m+mi}{4}\PY{p}{)}\PY{p}{)}
\PY{k}{for} \PY{n}{i} \PY{o+ow}{in} \PY{n+nb}{range}\PY{p}{(}\PY{l+m+mi}{4}\PY{p}{)}\PY{p}{:}
    \PY{k}{for} \PY{n}{j} \PY{o+ow}{in} \PY{n+nb}{range}\PY{p}{(}\PY{l+m+mi}{4}\PY{p}{)}\PY{p}{:}
        \PY{n}{ind} \PY{o}{=} \PY{n}{i}\PY{o}{*}\PY{l+m+mi}{4} \PY{o}{+} \PY{n}{j}
        \PY{n}{ax}\PY{p}{[}\PY{n}{i}\PY{p}{,}\PY{n}{j}\PY{p}{]}\PY{o}{.}\PY{n}{imshow}\PY{p}{(}\PY{n}{np}\PY{o}{.}\PY{n}{tanh}\PY{p}{(}\PY{n}{x}\PY{p}{[}\PY{n}{ind}\PY{p}{]}\PY{p}{)}\PY{p}{,} \PY{n}{vmin}\PY{o}{=}\PY{o}{\PYZhy{}}\PY{l+m+mi}{1}\PY{p}{,} \PY{n}{vmax}\PY{o}{=}\PY{l+m+mi}{1}\PY{p}{,} \PY{n}{cmap}\PY{o}{=}\PY{l+s+s1}{\PYZsq{}}\PY{l+s+s1}{viridis}\PY{l+s+s1}{\PYZsq{}}\PY{p}{)}
        \PY{n}{ax}\PY{p}{[}\PY{n}{i}\PY{p}{,}\PY{n}{j}\PY{p}{]}\PY{o}{.}\PY{n}{axes}\PY{o}{.}\PY{n}{xaxis}\PY{o}{.}\PY{n}{set\PYZus{}visible}\PY{p}{(}\PY{k+kc}{False}\PY{p}{)}
        \PY{n}{ax}\PY{p}{[}\PY{n}{i}\PY{p}{,}\PY{n}{j}\PY{p}{]}\PY{o}{.}\PY{n}{axes}\PY{o}{.}\PY{n}{yaxis}\PY{o}{.}\PY{n}{set\PYZus{}visible}\PY{p}{(}\PY{k+kc}{False}\PY{p}{)}
\PY{n}{plt}\PY{o}{.}\PY{n}{show}\PY{p}{(}\PY{p}{)}
    \end{Verbatim}
  \end{tcolorbox}

    \begin{center}
    \adjustimage{max size={0.9\linewidth}{0.9\paperheight}}{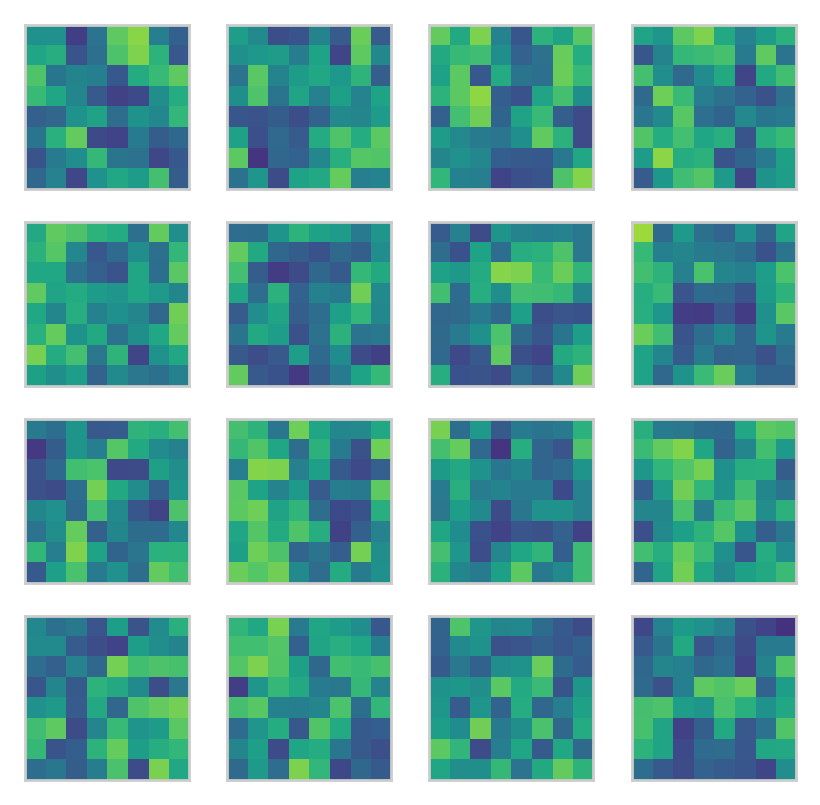}
    \end{center}
    { \hspace*{\fill} \\}
    
  We further see below that the model effective action (\(-\log{q}(x)\))
is very close to the true action, once we account for an overall shift.
This offset corresponds to the unknown multiplicative constant \(1/Z\)
that our training is insensitive to. There is still some variation,
especially in the regions with lower density, indicating the tails of
the distribution are not perfectly modeled.

  \begin{tcolorbox}[breakable, size=fbox, boxrule=1pt, pad at break*=1mm,colback=cellbackground, colframe=cellborder]
    \begin{Verbatim}[commandchars=\\\{\},fontsize=\small]
\PY{n}{S\PYZus{}eff} \PY{o}{=} \PY{o}{\PYZhy{}}\PY{n}{grab}\PY{p}{(}\PY{n}{torch\PYZus{}logq}\PY{p}{)}
\PY{n}{S} \PY{o}{=} \PY{n}{grab}\PY{p}{(}\PY{n}{phi4\PYZus{}action}\PY{p}{(}\PY{n}{torch\PYZus{}x}\PY{p}{)}\PY{p}{)}
\PY{n}{fit\PYZus{}b} \PY{o}{=} \PY{n}{np}\PY{o}{.}\PY{n}{mean}\PY{p}{(}\PY{n}{S}\PY{p}{)} \PY{o}{\PYZhy{}} \PY{n}{np}\PY{o}{.}\PY{n}{mean}\PY{p}{(}\PY{n}{S\PYZus{}eff}\PY{p}{)}
\PY{n+nb}{print}\PY{p}{(}\PY{l+s+sa}{f}\PY{l+s+s1}{\PYZsq{}}\PY{l+s+s1}{slope 1 linear regression S = S\PYZus{}eff + }\PY{l+s+si}{\PYZob{}}\PY{n}{fit\PYZus{}b}\PY{l+s+si}{:}\PY{l+s+s1}{.4f}\PY{l+s+si}{\PYZcb{}}\PY{l+s+s1}{\PYZsq{}}\PY{p}{)}
\PY{n}{fig}\PY{p}{,} \PY{n}{ax} \PY{o}{=} \PY{n}{plt}\PY{o}{.}\PY{n}{subplots}\PY{p}{(}\PY{l+m+mi}{1}\PY{p}{,}\PY{l+m+mi}{1}\PY{p}{,} \PY{n}{dpi}\PY{o}{=}\PY{l+m+mi}{125}\PY{p}{,} \PY{n}{figsize}\PY{o}{=}\PY{p}{(}\PY{l+m+mi}{4}\PY{p}{,}\PY{l+m+mi}{4}\PY{p}{)}\PY{p}{)}
\PY{n}{ax}\PY{o}{.}\PY{n}{hist2d}\PY{p}{(}\PY{n}{S\PYZus{}eff}\PY{p}{,} \PY{n}{S}\PY{p}{,} \PY{n}{bins}\PY{o}{=}\PY{l+m+mi}{20}\PY{p}{,} \PY{n+nb}{range}\PY{o}{=}\PY{p}{[}\PY{p}{[}\PY{l+m+mi}{5}\PY{p}{,} \PY{l+m+mi}{35}\PY{p}{]}\PY{p}{,} \PY{p}{[}\PY{o}{\PYZhy{}}\PY{l+m+mi}{5}\PY{p}{,} \PY{l+m+mi}{25}\PY{p}{]}\PY{p}{]}\PY{p}{)}
\PY{n}{ax}\PY{o}{.}\PY{n}{set\PYZus{}xlabel}\PY{p}{(}\PY{l+s+sa}{r}\PY{l+s+s1}{\PYZsq{}}\PY{l+s+s1}{\PYZdl{}S\PYZus{}}\PY{l+s+s1}{\PYZob{}}\PY{l+s+s1}{\PYZbs{}}\PY{l+s+s1}{mathrm}\PY{l+s+si}{\PYZob{}eff\PYZcb{}}\PY{l+s+s1}{\PYZcb{} = \PYZhy{}}\PY{l+s+s1}{\PYZbs{}}\PY{l+s+s1}{log\PYZti{}q(x)\PYZdl{}}\PY{l+s+s1}{\PYZsq{}}\PY{p}{)}
\PY{n}{ax}\PY{o}{.}\PY{n}{set\PYZus{}ylabel}\PY{p}{(}\PY{l+s+sa}{r}\PY{l+s+s1}{\PYZsq{}}\PY{l+s+s1}{\PYZdl{}S(x)\PYZdl{}}\PY{l+s+s1}{\PYZsq{}}\PY{p}{)}
\PY{n}{ax}\PY{o}{.}\PY{n}{set\PYZus{}aspect}\PY{p}{(}\PY{l+s+s1}{\PYZsq{}}\PY{l+s+s1}{equal}\PY{l+s+s1}{\PYZsq{}}\PY{p}{)}
\PY{n}{xs} \PY{o}{=} \PY{n}{np}\PY{o}{.}\PY{n}{linspace}\PY{p}{(}\PY{l+m+mi}{5}\PY{p}{,} \PY{l+m+mi}{35}\PY{p}{,} \PY{n}{num}\PY{o}{=}\PY{l+m+mi}{4}\PY{p}{,} \PY{n}{endpoint}\PY{o}{=}\PY{k+kc}{True}\PY{p}{)}
\PY{n}{ax}\PY{o}{.}\PY{n}{plot}\PY{p}{(}\PY{n}{xs}\PY{p}{,} \PY{n}{xs} \PY{o}{+} \PY{n}{fit\PYZus{}b}\PY{p}{,} \PY{l+s+s1}{\PYZsq{}}\PY{l+s+s1}{:}\PY{l+s+s1}{\PYZsq{}}\PY{p}{,} \PY{n}{color}\PY{o}{=}\PY{l+s+s1}{\PYZsq{}}\PY{l+s+s1}{w}\PY{l+s+s1}{\PYZsq{}}\PY{p}{,} \PY{n}{label}\PY{o}{=}\PY{l+s+s1}{\PYZsq{}}\PY{l+s+s1}{slope 1 fit}\PY{l+s+s1}{\PYZsq{}}\PY{p}{)}
\PY{n}{plt}\PY{o}{.}\PY{n}{legend}\PY{p}{(}\PY{n}{prop}\PY{o}{=}\PY{p}{\PYZob{}}\PY{l+s+s1}{\PYZsq{}}\PY{l+s+s1}{size}\PY{l+s+s1}{\PYZsq{}}\PY{p}{:} \PY{l+m+mi}{6}\PY{p}{\PYZcb{}}\PY{p}{)}
\PY{n}{plt}\PY{o}{.}\PY{n}{show}\PY{p}{(}\PY{p}{)}
    \end{Verbatim}
  \end{tcolorbox}

{\color{gray}
    \begin{Verbatim}[commandchars=\\\{\},fontsize=\small]
>>> slope 1 linear regression S = S\_eff + -8.6417
    \end{Verbatim}
}

    \begin{center}
    \adjustimage{max size={0.9\linewidth}{0.9\paperheight}}{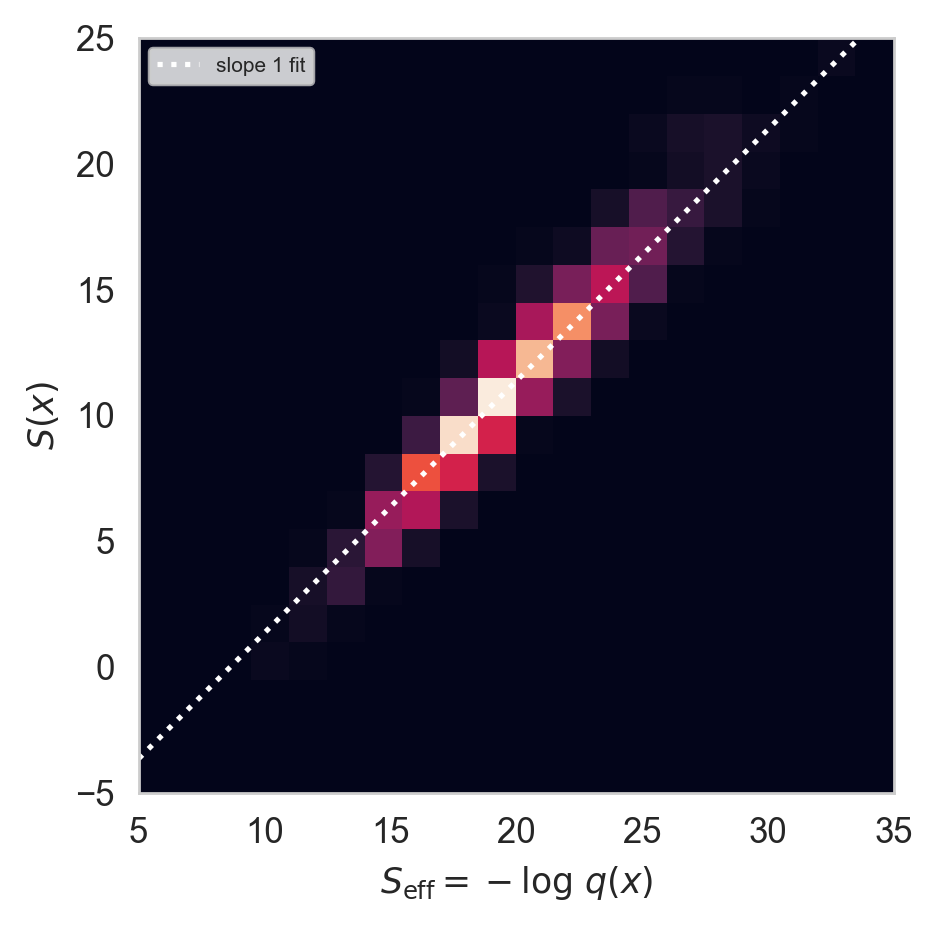}
    \end{center}
    { \hspace*{\fill} \\}
    
  We can see how the model density evolved over training time to become
well-correlated with \(p(x)\) over time (if
\texttt{use\_pretrained\ ==\ False}).

  \begin{tcolorbox}[breakable, size=fbox, boxrule=1pt, pad at break*=1mm,colback=cellbackground, colframe=cellborder]
    \begin{Verbatim}[commandchars=\\\{\},fontsize=\small]
\PY{k}{if} \PY{o+ow}{not} \PY{n}{use\PYZus{}pretrained}\PY{p}{:}
    \PY{n}{fig}\PY{p}{,} \PY{n}{axes} \PY{o}{=} \PY{n}{plt}\PY{o}{.}\PY{n}{subplots}\PY{p}{(}\PY{l+m+mi}{1}\PY{p}{,} \PY{l+m+mi}{10}\PY{p}{,} \PY{n}{dpi}\PY{o}{=}\PY{l+m+mi}{125}\PY{p}{,} \PY{n}{sharey}\PY{o}{=}\PY{k+kc}{True}\PY{p}{,} \PY{n}{figsize}\PY{o}{=}\PY{p}{(}\PY{l+m+mi}{10}\PY{p}{,} \PY{l+m+mi}{1}\PY{p}{)}\PY{p}{)}
    \PY{n}{logq\PYZus{}hist} \PY{o}{=} \PY{n}{np}\PY{o}{.}\PY{n}{array}\PY{p}{(}\PY{n}{history}\PY{p}{[}\PY{l+s+s1}{\PYZsq{}}\PY{l+s+s1}{logq}\PY{l+s+s1}{\PYZsq{}}\PY{p}{]}\PY{p}{)}\PY{o}{.}\PY{n}{reshape}\PY{p}{(}\PY{n}{N\PYZus{}era}\PY{p}{,} \PY{o}{\PYZhy{}}\PY{l+m+mi}{1}\PY{p}{)}\PY{p}{[}\PY{p}{:}\PY{p}{:}\PY{n}{N\PYZus{}era}\PY{o}{/}\PY{o}{/}\PY{l+m+mi}{10}\PY{p}{]}
    \PY{n}{logp\PYZus{}hist} \PY{o}{=} \PY{n}{np}\PY{o}{.}\PY{n}{array}\PY{p}{(}\PY{n}{history}\PY{p}{[}\PY{l+s+s1}{\PYZsq{}}\PY{l+s+s1}{logp}\PY{l+s+s1}{\PYZsq{}}\PY{p}{]}\PY{p}{)}\PY{o}{.}\PY{n}{reshape}\PY{p}{(}\PY{n}{N\PYZus{}era}\PY{p}{,} \PY{o}{\PYZhy{}}\PY{l+m+mi}{1}\PY{p}{)}\PY{p}{[}\PY{p}{:}\PY{p}{:}\PY{n}{N\PYZus{}era}\PY{o}{/}\PY{o}{/}\PY{l+m+mi}{10}\PY{p}{]}
    \PY{k}{for} \PY{n}{i}\PY{p}{,} \PY{p}{(}\PY{n}{ax}\PY{p}{,} \PY{n}{logq}\PY{p}{,} \PY{n}{logp}\PY{p}{)} \PY{o+ow}{in} \PY{n+nb}{enumerate}\PY{p}{(}\PY{n+nb}{zip}\PY{p}{(}\PY{n}{axes}\PY{p}{,} \PY{n}{logq\PYZus{}hist}\PY{p}{,} \PY{n}{logp\PYZus{}hist}\PY{p}{)}\PY{p}{)}\PY{p}{:}
        \PY{n}{ax}\PY{o}{.}\PY{n}{hist2d}\PY{p}{(}\PY{o}{\PYZhy{}}\PY{n}{logq}\PY{p}{,} \PY{o}{\PYZhy{}}\PY{n}{logp}\PY{p}{,} \PY{n}{bins}\PY{o}{=}\PY{l+m+mi}{20}\PY{p}{,} \PY{n+nb}{range}\PY{o}{=}\PY{p}{[}\PY{p}{[}\PY{l+m+mi}{5}\PY{p}{,} \PY{l+m+mi}{35}\PY{p}{]}\PY{p}{,} \PY{p}{[}\PY{o}{\PYZhy{}}\PY{l+m+mi}{5}\PY{p}{,} \PY{l+m+mi}{25}\PY{p}{]}\PY{p}{]}\PY{p}{)}
        \PY{k}{if} \PY{n}{i} \PY{o}{==} \PY{l+m+mi}{0}\PY{p}{:}
            \PY{n}{ax}\PY{o}{.}\PY{n}{set\PYZus{}ylabel}\PY{p}{(}\PY{l+s+sa}{r}\PY{l+s+s1}{\PYZsq{}}\PY{l+s+s1}{\PYZdl{}S(x)\PYZdl{}}\PY{l+s+s1}{\PYZsq{}}\PY{p}{)}
        \PY{n}{ax}\PY{o}{.}\PY{n}{set\PYZus{}xlabel}\PY{p}{(}\PY{l+s+sa}{r}\PY{l+s+s1}{\PYZsq{}}\PY{l+s+s1}{\PYZdl{}S\PYZus{}}\PY{l+s+s1}{\PYZob{}}\PY{l+s+s1}{\PYZbs{}}\PY{l+s+s1}{mathrm}\PY{l+s+si}{\PYZob{}eff\PYZcb{}}\PY{l+s+s1}{\PYZcb{}\PYZdl{}}\PY{l+s+s1}{\PYZsq{}}\PY{p}{)}
        \PY{n}{ax}\PY{o}{.}\PY{n}{set\PYZus{}title}\PY{p}{(}\PY{l+s+sa}{f}\PY{l+s+s1}{\PYZsq{}}\PY{l+s+s1}{Era }\PY{l+s+si}{\PYZob{}}\PY{n}{i} \PY{o}{*} \PY{p}{(}\PY{n}{N\PYZus{}era}\PY{o}{/}\PY{o}{/}\PY{l+m+mi}{10}\PY{p}{)}\PY{l+s+si}{\PYZcb{}}\PY{l+s+s1}{\PYZsq{}}\PY{p}{)}
        \PY{n}{ax}\PY{o}{.}\PY{n}{set\PYZus{}xticks}\PY{p}{(}\PY{p}{[}\PY{p}{]}\PY{p}{)}
        \PY{n}{ax}\PY{o}{.}\PY{n}{set\PYZus{}yticks}\PY{p}{(}\PY{p}{[}\PY{p}{]}\PY{p}{)}
        \PY{n}{ax}\PY{o}{.}\PY{n}{set\PYZus{}aspect}\PY{p}{(}\PY{l+s+s1}{\PYZsq{}}\PY{l+s+s1}{equal}\PY{l+s+s1}{\PYZsq{}}\PY{p}{)}
    \PY{n}{plt}\PY{o}{.}\PY{n}{show}\PY{p}{(}\PY{p}{)}
    \end{Verbatim}
  \end{tcolorbox}

  \hypertarget{independence-metropolis}{%
\subsubsection{\texorpdfstring{\textbf{Independence
Metropolis}}{Independence Metropolis}}\label{independence-metropolis}}

To produce \textbf{unbiased} estimates of observables, either
reweighting or resampling can be performed according to the weights
\(p(\phi_i)/q(\phi_i)\). See Sec. IIA of \cite{Boyda:2020hsi} for a
discussion of the tradeoffs in this choice. There are a number of
possible resampling approaches; we choose to use the model samples as
proposals in a Markov Chain Monte Carlo.

We'll use the Metropolis-Hastings (MH) algorithm to construct the
asymptotically exact Markov chain sampler. Generally, the MH algorithm
consists of proposing an updated configuration \(\phi'\) to the current
configuration \(\phi^{i-1}\) and stochastically accepting or rejecting
the configuration with probability \begin{equation}
p_{\mathrm{accept}}(\phi'|\phi^{i-1}) = \min \left(
    1,\;
    \frac{T(\phi' \rightarrow \phi^{i-1})}{T(\phi^{i-1} \rightarrow \phi')}
    \frac{p(\phi')}{p(\phi^{i-1})}
\right).
\end{equation} Here \(T(x \rightarrow y)\) is the probability of
proposing config \(y\) starting from \(x\). If accepted, we define the
next configuration in the chain to be \(\phi^i = \phi'\); if rejected,
the last configuration is repeated and \(\phi^i = \phi^{i-1}\). Often,
\(p(\phi) \sim e^{-S}\) is computationally tractable but \(T\) is not,
so algorithms are engineered to have symmetric proposal probabilities
such that \(T(x \rightarrow y) = T(y \rightarrow x)\) and the factors in
\(p_{\text{accept}}\) cancel, leading to the familiar Metropolis formula
\(p^{\text{symm}}_{\text{accept}} = \min(1, \exp[-\Delta S])\).

We instead propose updates by drawing samples from our model
independently of the previous configuration, so
\(T(x \rightarrow y) = T(y) = q(y)\), where \(q(y)\) is the model
density computed alongside sample \(y\). The resulting proposal
probability \(T\) is therefore not symmetric but is known. We thus must
accept or reject based on \begin{equation}
p_{\mathrm{accept}}(\phi'|\phi^{i-1}) = \min \left(
    1,\;
    \frac{q(\phi^{i-1})}{p(\phi^{i-1})}
    \frac{p(\phi')}{q(\phi')}
\right).
\end{equation} This procedure is known as the independence Metropolis
sampler. Note that rejections occur proportionally to how poorly the
model density matches the desired density; if \(p(\phi') = q(\phi')\),
all (independent) proposals are accepted, and the chain is a sequence of
totally uncorrelated samples. As the rejection rate increases, the
autocorrelation time does as well.

  Below we build the MH algorithm in two stages. First, we need some way
of generating an ordered list of samples using our model. The code below
defines a generator which does this by drawing batches efficiently in
parallel, then iterating over them one at at time.

  \begin{tcolorbox}[breakable, size=fbox, boxrule=1pt, pad at break*=1mm,colback=cellbackground, colframe=cellborder]
    \begin{Verbatim}[commandchars=\\\{\},fontsize=\small]
\PY{k}{def} \PY{n+nf}{serial\PYZus{}sample\PYZus{}generator}\PY{p}{(}\PY{n}{model}\PY{p}{,} \PY{n}{action}\PY{p}{,} \PY{n}{batch\PYZus{}size}\PY{p}{,} \PY{n}{N\PYZus{}samples}\PY{p}{)}\PY{p}{:}
    \PY{n}{layers}\PY{p}{,} \PY{n}{prior} \PY{o}{=} \PY{n}{model}\PY{p}{[}\PY{l+s+s1}{\PYZsq{}}\PY{l+s+s1}{layers}\PY{l+s+s1}{\PYZsq{}}\PY{p}{]}\PY{p}{,} \PY{n}{model}\PY{p}{[}\PY{l+s+s1}{\PYZsq{}}\PY{l+s+s1}{prior}\PY{l+s+s1}{\PYZsq{}}\PY{p}{]}
    \PY{n}{layers}\PY{o}{.}\PY{n}{eval}\PY{p}{(}\PY{p}{)}
    \PY{n}{x}\PY{p}{,} \PY{n}{logq}\PY{p}{,} \PY{n}{logp} \PY{o}{=} \PY{k+kc}{None}\PY{p}{,} \PY{k+kc}{None}\PY{p}{,} \PY{k+kc}{None}
    \PY{k}{for} \PY{n}{i} \PY{o+ow}{in} \PY{n+nb}{range}\PY{p}{(}\PY{n}{N\PYZus{}samples}\PY{p}{)}\PY{p}{:}
        \PY{n}{batch\PYZus{}i} \PY{o}{=} \PY{n}{i} \PY{o}{\PYZpc{}} \PY{n}{batch\PYZus{}size}
        \PY{k}{if} \PY{n}{batch\PYZus{}i} \PY{o}{==} \PY{l+m+mi}{0}\PY{p}{:}
            \PY{c+c1}{\PYZsh{} we\PYZsq{}re out of samples to propose, generate a new batch}
            \PY{n}{x}\PY{p}{,} \PY{n}{logq} \PY{o}{=} \PY{n}{apply\PYZus{}flow\PYZus{}to\PYZus{}prior}\PY{p}{(}\PY{n}{prior}\PY{p}{,} \PY{n}{layers}\PY{p}{,} \PY{n}{batch\PYZus{}size}\PY{o}{=}\PY{n}{batch\PYZus{}size}\PY{p}{)}
            \PY{n}{logp} \PY{o}{=} \PY{o}{\PYZhy{}}\PY{n}{action}\PY{p}{(}\PY{n}{x}\PY{p}{)}
        \PY{k}{yield} \PY{n}{x}\PY{p}{[}\PY{n}{batch\PYZus{}i}\PY{p}{]}\PY{p}{,} \PY{n}{logq}\PY{p}{[}\PY{n}{batch\PYZus{}i}\PY{p}{]}\PY{p}{,} \PY{n}{logp}\PY{p}{[}\PY{n}{batch\PYZus{}i}\PY{p}{]}
    \end{Verbatim}
  \end{tcolorbox}

  Now we need to iterate over the samples and construct them into a Markov
Chain. The code below implements the Metropolis independence sampler to
do this.

  \begin{tcolorbox}[breakable, size=fbox, boxrule=1pt, pad at break*=1mm,colback=cellbackground, colframe=cellborder]
    \begin{Verbatim}[commandchars=\\\{\},fontsize=\small]
\PY{k}{def} \PY{n+nf}{make\PYZus{}mcmc\PYZus{}ensemble}\PY{p}{(}\PY{n}{model}\PY{p}{,} \PY{n}{action}\PY{p}{,} \PY{n}{batch\PYZus{}size}\PY{p}{,} \PY{n}{N\PYZus{}samples}\PY{p}{)}\PY{p}{:}
    \PY{n}{history} \PY{o}{=} \PY{p}{\PYZob{}}
        \PY{l+s+s1}{\PYZsq{}}\PY{l+s+s1}{x}\PY{l+s+s1}{\PYZsq{}} \PY{p}{:} \PY{p}{[}\PY{p}{]}\PY{p}{,}
        \PY{l+s+s1}{\PYZsq{}}\PY{l+s+s1}{logq}\PY{l+s+s1}{\PYZsq{}} \PY{p}{:} \PY{p}{[}\PY{p}{]}\PY{p}{,}
        \PY{l+s+s1}{\PYZsq{}}\PY{l+s+s1}{logp}\PY{l+s+s1}{\PYZsq{}} \PY{p}{:} \PY{p}{[}\PY{p}{]}\PY{p}{,}
        \PY{l+s+s1}{\PYZsq{}}\PY{l+s+s1}{accepted}\PY{l+s+s1}{\PYZsq{}} \PY{p}{:} \PY{p}{[}\PY{p}{]}
    \PY{p}{\PYZcb{}}

    \PY{c+c1}{\PYZsh{} build Markov chain}
    \PY{n}{sample\PYZus{}gen} \PY{o}{=} \PY{n}{serial\PYZus{}sample\PYZus{}generator}\PY{p}{(}\PY{n}{model}\PY{p}{,} \PY{n}{action}\PY{p}{,} \PY{n}{batch\PYZus{}size}\PY{p}{,} \PY{n}{N\PYZus{}samples}\PY{p}{)}
    \PY{k}{for} \PY{n}{new\PYZus{}x}\PY{p}{,} \PY{n}{new\PYZus{}logq}\PY{p}{,} \PY{n}{new\PYZus{}logp} \PY{o+ow}{in} \PY{n}{sample\PYZus{}gen}\PY{p}{:}
        \PY{k}{if} \PY{n+nb}{len}\PY{p}{(}\PY{n}{history}\PY{p}{[}\PY{l+s+s1}{\PYZsq{}}\PY{l+s+s1}{logp}\PY{l+s+s1}{\PYZsq{}}\PY{p}{]}\PY{p}{)} \PY{o}{==} \PY{l+m+mi}{0}\PY{p}{:}
            \PY{c+c1}{\PYZsh{} always accept first proposal, Markov chain must start somewhere}
            \PY{n}{accepted} \PY{o}{=} \PY{k+kc}{True}
        \PY{k}{else}\PY{p}{:} 
            \PY{c+c1}{\PYZsh{} Metropolis acceptance condition}
            \PY{n}{last\PYZus{}logp} \PY{o}{=} \PY{n}{history}\PY{p}{[}\PY{l+s+s1}{\PYZsq{}}\PY{l+s+s1}{logp}\PY{l+s+s1}{\PYZsq{}}\PY{p}{]}\PY{p}{[}\PY{o}{\PYZhy{}}\PY{l+m+mi}{1}\PY{p}{]}
            \PY{n}{last\PYZus{}logq} \PY{o}{=} \PY{n}{history}\PY{p}{[}\PY{l+s+s1}{\PYZsq{}}\PY{l+s+s1}{logq}\PY{l+s+s1}{\PYZsq{}}\PY{p}{]}\PY{p}{[}\PY{o}{\PYZhy{}}\PY{l+m+mi}{1}\PY{p}{]}
            \PY{n}{p\PYZus{}accept} \PY{o}{=} \PY{n}{torch}\PY{o}{.}\PY{n}{exp}\PY{p}{(}\PY{p}{(}\PY{n}{new\PYZus{}logp} \PY{o}{\PYZhy{}} \PY{n}{new\PYZus{}logq}\PY{p}{)} \PY{o}{\PYZhy{}} \PY{p}{(}\PY{n}{last\PYZus{}logp} \PY{o}{\PYZhy{}} \PY{n}{last\PYZus{}logq}\PY{p}{)}\PY{p}{)}
            \PY{n}{p\PYZus{}accept} \PY{o}{=} \PY{n+nb}{min}\PY{p}{(}\PY{l+m+mi}{1}\PY{p}{,} \PY{n}{p\PYZus{}accept}\PY{p}{)}
            \PY{n}{draw} \PY{o}{=} \PY{n}{torch}\PY{o}{.}\PY{n}{rand}\PY{p}{(}\PY{l+m+mi}{1}\PY{p}{)} \PY{c+c1}{\PYZsh{} \PYZti{} [0,1]}
            \PY{k}{if} \PY{n}{draw} \PY{o}{\PYZlt{}} \PY{n}{p\PYZus{}accept}\PY{p}{:}
                \PY{n}{accepted} \PY{o}{=} \PY{k+kc}{True}
            \PY{k}{else}\PY{p}{:}
                \PY{n}{accepted} \PY{o}{=} \PY{k+kc}{False}
                \PY{n}{new\PYZus{}x} \PY{o}{=} \PY{n}{history}\PY{p}{[}\PY{l+s+s1}{\PYZsq{}}\PY{l+s+s1}{x}\PY{l+s+s1}{\PYZsq{}}\PY{p}{]}\PY{p}{[}\PY{o}{\PYZhy{}}\PY{l+m+mi}{1}\PY{p}{]}
                \PY{n}{new\PYZus{}logp} \PY{o}{=} \PY{n}{last\PYZus{}logp}
                \PY{n}{new\PYZus{}logq} \PY{o}{=} \PY{n}{last\PYZus{}logq}
        \PY{c+c1}{\PYZsh{} Update Markov chain}
        \PY{n}{history}\PY{p}{[}\PY{l+s+s1}{\PYZsq{}}\PY{l+s+s1}{logp}\PY{l+s+s1}{\PYZsq{}}\PY{p}{]}\PY{o}{.}\PY{n}{append}\PY{p}{(}\PY{n}{new\PYZus{}logp}\PY{p}{)}
        \PY{n}{history}\PY{p}{[}\PY{l+s+s1}{\PYZsq{}}\PY{l+s+s1}{logq}\PY{l+s+s1}{\PYZsq{}}\PY{p}{]}\PY{o}{.}\PY{n}{append}\PY{p}{(}\PY{n}{new\PYZus{}logq}\PY{p}{)}
        \PY{n}{history}\PY{p}{[}\PY{l+s+s1}{\PYZsq{}}\PY{l+s+s1}{x}\PY{l+s+s1}{\PYZsq{}}\PY{p}{]}\PY{o}{.}\PY{n}{append}\PY{p}{(}\PY{n}{new\PYZus{}x}\PY{p}{)}
        \PY{n}{history}\PY{p}{[}\PY{l+s+s1}{\PYZsq{}}\PY{l+s+s1}{accepted}\PY{l+s+s1}{\PYZsq{}}\PY{p}{]}\PY{o}{.}\PY{n}{append}\PY{p}{(}\PY{n}{accepted}\PY{p}{)}
    \PY{k}{return} \PY{n}{history}
    \end{Verbatim}
  \end{tcolorbox}

  Finally, the cell below uses the code above to generate an ensemble of
configurations using our trained flow model. You should see a 30-40\%
accept rate.

  \begin{tcolorbox}[breakable, size=fbox, boxrule=1pt, pad at break*=1mm,colback=cellbackground, colframe=cellborder]
    \begin{Verbatim}[commandchars=\\\{\},fontsize=\small]
\PY{n}{ensemble\PYZus{}size} \PY{o}{=} \PY{l+m+mi}{8192}
\PY{n}{phi4\PYZus{}ens} \PY{o}{=} \PY{n}{make\PYZus{}mcmc\PYZus{}ensemble}\PY{p}{(}\PY{n}{model}\PY{p}{,} \PY{n}{phi4\PYZus{}action}\PY{p}{,} \PY{l+m+mi}{64}\PY{p}{,} \PY{n}{ensemble\PYZus{}size}\PY{p}{)}
\PY{n+nb}{print}\PY{p}{(}\PY{l+s+s2}{\PYZdq{}}\PY{l+s+s2}{Accept rate:}\PY{l+s+s2}{\PYZdq{}}\PY{p}{,} \PY{n}{np}\PY{o}{.}\PY{n}{mean}\PY{p}{(}\PY{n}{phi4\PYZus{}ens}\PY{p}{[}\PY{l+s+s1}{\PYZsq{}}\PY{l+s+s1}{accepted}\PY{l+s+s1}{\PYZsq{}}\PY{p}{]}\PY{p}{)}\PY{p}{)}
    \end{Verbatim}
  \end{tcolorbox}

{\color{gray}
    \begin{Verbatim}[commandchars=\\\{\},fontsize=\small]
>>> Accept rate: 0.458984375
    \end{Verbatim}
}

  The generated ensemble is asymptotically unbiased. As an example of an
observable measurements, we measure the two-point susceptibility below
and compare against a value determined from a large HMC ensemble
evaluated at the same choice of parameters.

  \begin{tcolorbox}[breakable, size=fbox, boxrule=1pt, pad at break*=1mm,colback=cellbackground, colframe=cellborder]
    \begin{Verbatim}[commandchars=\\\{\},fontsize=\small]
\PY{n}{n\PYZus{}therm} \PY{o}{=} \PY{l+m+mi}{512}
\PY{n}{cfgs} \PY{o}{=} \PY{n}{np}\PY{o}{.}\PY{n}{stack}\PY{p}{(}\PY{n+nb}{list}\PY{p}{(}\PY{n+nb}{map}\PY{p}{(}\PY{n}{grab}\PY{p}{,} \PY{n}{phi4\PYZus{}ens}\PY{p}{[}\PY{l+s+s1}{\PYZsq{}}\PY{l+s+s1}{x}\PY{l+s+s1}{\PYZsq{}}\PY{p}{]}\PY{p}{)}\PY{p}{)}\PY{p}{,} \PY{n}{axis}\PY{o}{=}\PY{l+m+mi}{0}\PY{p}{)}\PY{p}{[}\PY{n}{n\PYZus{}therm}\PY{p}{:}\PY{p}{]}
\PY{n}{C} \PY{o}{=} \PY{l+m+mi}{0}
\PY{k}{for} \PY{n}{x} \PY{o+ow}{in} \PY{n+nb}{range}\PY{p}{(}\PY{n}{L}\PY{p}{)}\PY{p}{:}
    \PY{k}{for} \PY{n}{y} \PY{o+ow}{in} \PY{n+nb}{range}\PY{p}{(}\PY{n}{L}\PY{p}{)}\PY{p}{:}
        \PY{n}{C} \PY{o}{=} \PY{n}{C} \PY{o}{+} \PY{n}{cfgs}\PY{o}{*}\PY{n}{np}\PY{o}{.}\PY{n}{roll}\PY{p}{(}\PY{n}{cfgs}\PY{p}{,} \PY{p}{(}\PY{o}{\PYZhy{}}\PY{n}{x}\PY{p}{,} \PY{o}{\PYZhy{}}\PY{n}{y}\PY{p}{)}\PY{p}{,} \PY{n}{axis}\PY{o}{=}\PY{p}{(}\PY{l+m+mi}{1}\PY{p}{,}\PY{l+m+mi}{2}\PY{p}{)}\PY{p}{)}
\PY{n}{X} \PY{o}{=} \PY{n}{np}\PY{o}{.}\PY{n}{mean}\PY{p}{(}\PY{n}{C}\PY{p}{,} \PY{n}{axis}\PY{o}{=}\PY{p}{(}\PY{l+m+mi}{1}\PY{p}{,}\PY{l+m+mi}{2}\PY{p}{)}\PY{p}{)}

\PY{k}{def} \PY{n+nf}{bootstrap}\PY{p}{(}\PY{n}{x}\PY{p}{,} \PY{o}{*}\PY{p}{,} \PY{n}{Nboot}\PY{p}{,} \PY{n}{binsize}\PY{p}{)}\PY{p}{:}
    \PY{n}{boots} \PY{o}{=} \PY{p}{[}\PY{p}{]}
    \PY{n}{x} \PY{o}{=} \PY{n}{x}\PY{o}{.}\PY{n}{reshape}\PY{p}{(}\PY{o}{\PYZhy{}}\PY{l+m+mi}{1}\PY{p}{,} \PY{n}{binsize}\PY{p}{,} \PY{o}{*}\PY{n}{x}\PY{o}{.}\PY{n}{shape}\PY{p}{[}\PY{l+m+mi}{1}\PY{p}{:}\PY{p}{]}\PY{p}{)}
    \PY{k}{for} \PY{n}{i} \PY{o+ow}{in} \PY{n+nb}{range}\PY{p}{(}\PY{n}{Nboot}\PY{p}{)}\PY{p}{:}
        \PY{n}{boots}\PY{o}{.}\PY{n}{append}\PY{p}{(}\PY{n}{np}\PY{o}{.}\PY{n}{mean}\PY{p}{(}\PY{n}{x}\PY{p}{[}\PY{n}{np}\PY{o}{.}\PY{n}{random}\PY{o}{.}\PY{n}{randint}\PY{p}{(}\PY{n+nb}{len}\PY{p}{(}\PY{n}{x}\PY{p}{)}\PY{p}{,} \PY{n}{size}\PY{o}{=}\PY{n+nb}{len}\PY{p}{(}\PY{n}{x}\PY{p}{)}\PY{p}{)}\PY{p}{]}\PY{p}{,} \PY{n}{axis}\PY{o}{=}\PY{p}{(}\PY{l+m+mi}{0}\PY{p}{,}\PY{l+m+mi}{1}\PY{p}{)}\PY{p}{)}\PY{p}{)}
    \PY{k}{return} \PY{n}{np}\PY{o}{.}\PY{n}{mean}\PY{p}{(}\PY{n}{boots}\PY{p}{)}\PY{p}{,} \PY{n}{np}\PY{o}{.}\PY{n}{std}\PY{p}{(}\PY{n}{boots}\PY{p}{)}
\PY{n}{X\PYZus{}mean}\PY{p}{,} \PY{n}{X\PYZus{}err} \PY{o}{=} \PY{n}{bootstrap}\PY{p}{(}\PY{n}{X}\PY{p}{,} \PY{n}{Nboot}\PY{o}{=}\PY{l+m+mi}{100}\PY{p}{,} \PY{n}{binsize}\PY{o}{=}\PY{l+m+mi}{4}\PY{p}{)}
\PY{n+nb}{print}\PY{p}{(}\PY{l+s+sa}{f}\PY{l+s+s1}{\PYZsq{}}\PY{l+s+s1}{Two\PYZhy{}point susceptibility = }\PY{l+s+si}{\PYZob{}}\PY{n}{X\PYZus{}mean}\PY{l+s+si}{:}\PY{l+s+s1}{.2f}\PY{l+s+si}{\PYZcb{}}\PY{l+s+s1}{ +/\PYZhy{} }\PY{l+s+si}{\PYZob{}}\PY{n}{X\PYZus{}err}\PY{l+s+si}{:}\PY{l+s+s1}{.2f}\PY{l+s+si}{\PYZcb{}}\PY{l+s+s1}{\PYZsq{}}\PY{p}{)}
\PY{n+nb}{print}\PY{p}{(}\PY{l+s+sa}{f}\PY{l+s+s1}{\PYZsq{}}\PY{l+s+s1}{... vs HMC estimate = 0.75 +/\PYZhy{} 0.01}\PY{l+s+s1}{\PYZsq{}}\PY{p}{)}
    \end{Verbatim}
  \end{tcolorbox}

{\color{gray}
    \begin{Verbatim}[commandchars=\\\{\},fontsize=\small]
>>> Two-point susceptibility = 0.79 +/- 0.02
... {\ldots} vs HMC estimate = 0.75 +/- 0.01
    \end{Verbatim}
}

  \textbf{Caveat:} A poorly trained model can result in a small acceptance
rate and large autocorrelations, in which case a more careful error
analysis is needed to avoid underestimation of errors. Above we employ
binning to reliably estimate error for despite any autocorrelations in
the data.

  \hypertarget{application-2-mathrmu1-gauge-theory-in-2d}{%
\section{\texorpdfstring{Application 2: \(\mathrm{U}(1)\) gauge theory
in
2d}{Application 2: \textbackslash mathrm\{U\}(1) gauge theory in 2d}}\label{application-2-mathrmu1-gauge-theory-in-2d}}

As a second example, we train a flow to sample distributions for
\(\mathrm{U}(1)\) gauge theory in two spacetime dimensions. The desired
physical distributions are symmetric under a large gauge symmetry group.
We can construct flows which explicitly respect this symmetry by
enforcing two requirements:

\begin{enumerate}
\def\labelenumi{\arabic{enumi}.}
\tightlist
\item
  The prior distribution is gauge-invariant. We'll use the uniform
  distribution (with respect to the Haar measure) on each gauge link.
  For \(\mathrm{U}(1)\), this is just the uniform distribution in
  \([0, 2\pi]^{N_d V}\).
\item
  Coupling layers are \textbf{gauge equivariant} (commute with gauge
  transformations).
\end{enumerate}

If both conditions are satisfied, this guarantees a \textbf{gauge
invariant} output distribution. See \cite{Kanwar:2020xzo} for details.

\textbf{CAUTION:} many variable names are reused from the previous
section.

  \hypertarget{physical-theory}{%
\subsection{\texorpdfstring{\textbf{Physical
theory}}{Physical theory}}\label{physical-theory}}

The continuum theory consists of a real-valued field \(A_\mu(x)\) as a
function of 2D coordinate \(x\), with Lorentz index \(\mu\). The lattice
regularization of the theory replaces this field per site with a
collection of parallel transporters \begin{equation}
U_\mu(x) \equiv \exp \left[ {i \int_{x' = x}^{x' = x+\hat{\mu}} A_{\mu}(x')} \right]
\end{equation} with each \(U_\mu(x)\) living on the lattice link
connecting \(x\) to \(x+\hat{\mu}\), such that there are \(2V\)
independent links on a \(V\)-site 2D lattice. As unit-modulus complex
numbers we can consider \(U_\mu(x)\) to live in \(\mathrm{U}(1)\) and
\(A_{\mu}(x)\) to live in the algebra \(\mathfrak{u}(1)\).

  \begin{figure}[H]
      \centering
      \includegraphics[width=5cm]{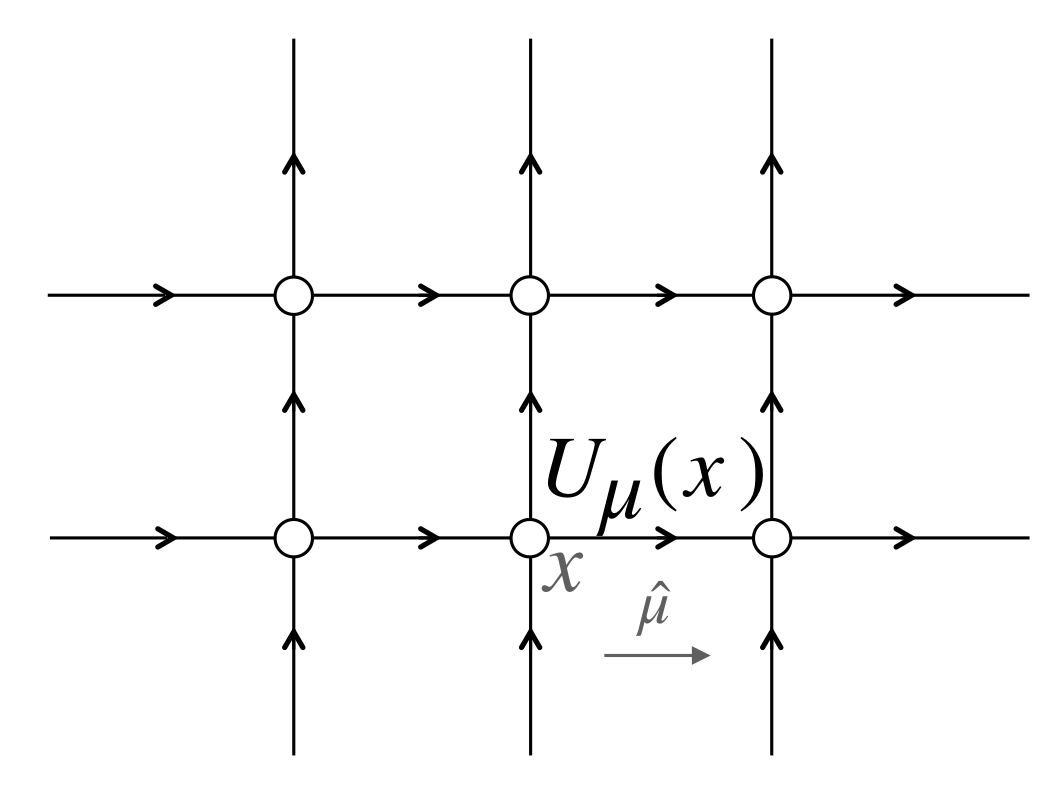}
      \caption{Lattice discretization of a gauge theory.}
    \end{figure}

  We can write in the angular representation
\(U_\mu(\vec{n}) = \exp[i \theta_\mu(\vec{n})]\) where
\(\theta_\mu(\vec{n}) \in \mathbb{R}\). We will work with the
real-valued angles
\(\theta_\mu(\vec{n}) = \arg(U_\mu(\vec{n})) \in [0, 2\pi]\) rather than
the unit-modulus complex number \(U_\mu(\vec{n})\). Using this
parsimonious representation of lattice DOF saves us from having to worry
about maintaining the normalization of the complex \(U\), at the cost of
having to deal with discontinuities at the boundary
\(\theta = 2 \pi \equiv 0\).

  Let's initialize some configurations of an example lattice of size
\(8\times8\) and generate two random configurations:

  \begin{tcolorbox}[breakable, size=fbox, boxrule=1pt, pad at break*=1mm,colback=cellbackground, colframe=cellborder]
    \begin{Verbatim}[commandchars=\\\{\},fontsize=\small]
\PY{n}{L} \PY{o}{=} \PY{l+m+mi}{8}
\PY{n}{lattice\PYZus{}shape} \PY{o}{=} \PY{p}{(}\PY{n}{L}\PY{p}{,}\PY{n}{L}\PY{p}{)}
\PY{n}{link\PYZus{}shape} \PY{o}{=} \PY{p}{(}\PY{l+m+mi}{2}\PY{p}{,}\PY{n}{L}\PY{p}{,}\PY{n}{L}\PY{p}{)}
\PY{c+c1}{\PYZsh{} some arbitrary configurations}
\PY{n}{u1\PYZus{}ex1} \PY{o}{=} \PY{l+m+mi}{2}\PY{o}{*}\PY{n}{np}\PY{o}{.}\PY{n}{pi}\PY{o}{*}\PY{n}{np}\PY{o}{.}\PY{n}{random}\PY{o}{.}\PY{n}{random}\PY{p}{(}\PY{n}{size}\PY{o}{=}\PY{n}{link\PYZus{}shape}\PY{p}{)}\PY{o}{.}\PY{n}{astype}\PY{p}{(}\PY{n}{float\PYZus{}dtype}\PY{p}{)}
\PY{n}{u1\PYZus{}ex2} \PY{o}{=} \PY{l+m+mi}{2}\PY{o}{*}\PY{n}{np}\PY{o}{.}\PY{n}{pi}\PY{o}{*}\PY{n}{np}\PY{o}{.}\PY{n}{random}\PY{o}{.}\PY{n}{random}\PY{p}{(}\PY{n}{size}\PY{o}{=}\PY{n}{link\PYZus{}shape}\PY{p}{)}\PY{o}{.}\PY{n}{astype}\PY{p}{(}\PY{n}{float\PYZus{}dtype}\PY{p}{)}
\PY{n}{cfgs} \PY{o}{=} \PY{n}{torch}\PY{o}{.}\PY{n}{from\PYZus{}numpy}\PY{p}{(}\PY{n}{np}\PY{o}{.}\PY{n}{stack}\PY{p}{(}\PY{p}{(}\PY{n}{u1\PYZus{}ex1}\PY{p}{,} \PY{n}{u1\PYZus{}ex2}\PY{p}{)}\PY{p}{,} \PY{n}{axis}\PY{o}{=}\PY{l+m+mi}{0}\PY{p}{)}\PY{p}{)}\PY{o}{.}\PY{n}{to}\PY{p}{(}\PY{n}{torch\PYZus{}device}\PY{p}{)}
    \end{Verbatim}
  \end{tcolorbox}

  The continuum Euclidean action can be written in terms of the field
strength \(F_{\mu\nu} \equiv \partial_\mu A_\nu - \partial_\nu A_\mu\)
as \begin{equation}
    S^E_{\text{cont}}[A] = \int d^dx \left[ -\frac{1}{2} \sum_{\mu < \nu} F_{\mu\nu}^2 \right]
\end{equation} which can be regularized on the lattice in terms of
parallel transporters \(U_\mu(\vec{n})\) \begin{equation}
\begin{split}
    S^E_{\text{latt}}[U] &= -\beta \sum_{\vec{n}} \left[ \sum_{\mu < \nu} \text{Re} P_{\mu\nu}(\vec{n}) \right] \\
    \text{where} \quad
    P_{\mu\nu}(\vec{n}) &\equiv U_\mu(\vec{n}) ~ U_\nu(\vec{n}+\hat{\mu}) ~ U^\dagger_\mu(\vec{n}+\hat{\nu}) ~ U^\dagger_\nu(\vec{n})
\end{split}
\end{equation} is the ``plaquette'', the simplest possible closed loop
of links on a lattice, a \(1 \times 1\) square. This simple form for
\(S^E_\text{latt}\) is known as the Wilson gauge action.

  \begin{figure}[H]
      \centering
      \includegraphics[width=2cm]{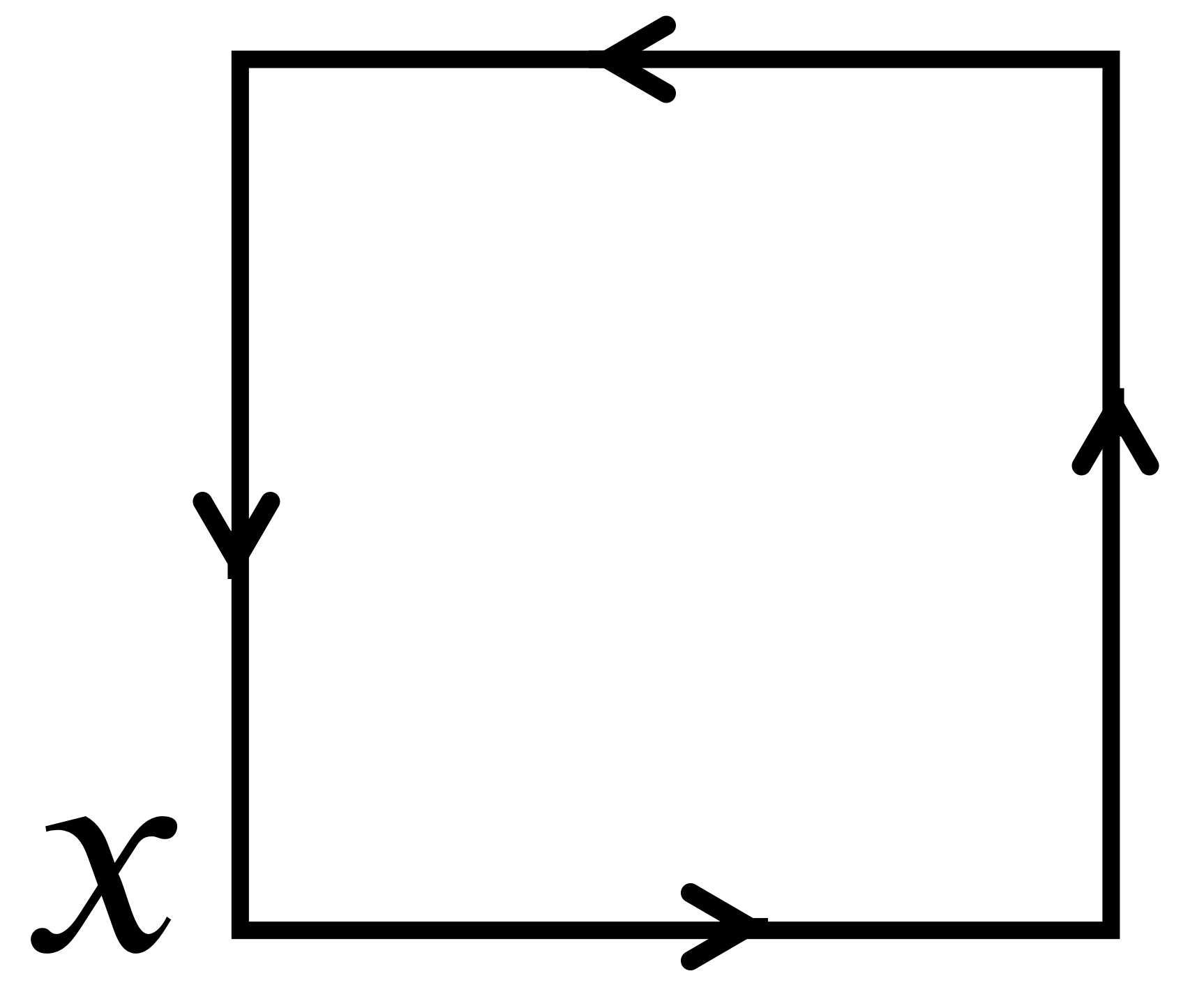}
      \caption{A plaquette.}
    \end{figure}

  The expressions above are valid for non-Abelian gauge theory in an
arbitrary number of spacetime dimensions, but in this notebook we are
interested in Abelian \(\mathrm{U}(1)\) gauge group. This immediately
leads to some simplifications. In 2D, \(P_{\mu\nu}=P_{01}\) is the only
orientation of plaquette and the sum \(\sum_{\mu<\nu}\) is trivial. In
angular representation, as discussed above, the link has form
\(U_\mu(\vec{n})=\exp[i \theta_\mu(\vec{n})]\), hence the plaquette can
be written as \(P_{\mu\nu}(\vec{n}) = \exp[\theta_{\mu\nu}(\vec{n})]\),
where
\(\theta_{\mu\nu}(\vec{n}) = \theta_\mu(\vec{n}) + \theta_\nu(\vec{n}+\hat{\mu}) - \theta_\mu(\vec{n}+\hat{\nu}) - \theta_\nu(\vec{n})\)
and the Wilson gauge action for 2D U(1) gauge theory reduces to
\begin{equation}
    S^E_{\text{latt}}[U] = -\beta \sum_{\vec{n}} \cos \left[
      \theta_{\mu\nu}(\vec{n})
    \right].
\end{equation}

  Since we are working in the angular representation, in the code we are
always dealing with \(\theta_{\mu\nu}\) rather than \(P_{\mu\nu}\). The
function to calculate \(\theta_{\mu\nu}\) in terms of link angles looks
like:

  \begin{tcolorbox}[breakable, size=fbox, boxrule=1pt, pad at break*=1mm,colback=cellbackground, colframe=cellborder]
    \begin{Verbatim}[commandchars=\\\{\},fontsize=\small]
\PY{k}{def} \PY{n+nf}{compute\PYZus{}u1\PYZus{}plaq}\PY{p}{(}\PY{n}{links}\PY{p}{,} \PY{n}{mu}\PY{p}{,} \PY{n}{nu}\PY{p}{)}\PY{p}{:}
    \PY{l+s+sd}{\PYZdq{}\PYZdq{}\PYZdq{}Compute U(1) plaquettes in the (mu,nu) plane given `links` = arg(U)\PYZdq{}\PYZdq{}\PYZdq{}}
    \PY{k}{return} \PY{p}{(}\PY{n}{links}\PY{p}{[}\PY{p}{:}\PY{p}{,}\PY{n}{mu}\PY{p}{]} \PY{o}{+} \PY{n}{torch}\PY{o}{.}\PY{n}{roll}\PY{p}{(}\PY{n}{links}\PY{p}{[}\PY{p}{:}\PY{p}{,}\PY{n}{nu}\PY{p}{]}\PY{p}{,} \PY{o}{\PYZhy{}}\PY{l+m+mi}{1}\PY{p}{,} \PY{n}{mu}\PY{o}{+}\PY{l+m+mi}{1}\PY{p}{)}
            \PY{o}{\PYZhy{}} \PY{n}{torch}\PY{o}{.}\PY{n}{roll}\PY{p}{(}\PY{n}{links}\PY{p}{[}\PY{p}{:}\PY{p}{,}\PY{n}{mu}\PY{p}{]}\PY{p}{,} \PY{o}{\PYZhy{}}\PY{l+m+mi}{1}\PY{p}{,} \PY{n}{nu}\PY{o}{+}\PY{l+m+mi}{1}\PY{p}{)} \PY{o}{\PYZhy{}} \PY{n}{links}\PY{p}{[}\PY{p}{:}\PY{p}{,}\PY{n}{nu}\PY{p}{]}\PY{p}{)}
    \end{Verbatim}
  \end{tcolorbox}

  The gauge action in terms of angular variables is then:

  \begin{tcolorbox}[breakable, size=fbox, boxrule=1pt, pad at break*=1mm,colback=cellbackground, colframe=cellborder]
    \begin{Verbatim}[commandchars=\\\{\},fontsize=\small]
\PY{k}{class} \PY{n+nc}{U1GaugeAction}\PY{p}{:}
    \PY{k}{def} \PY{n+nf+fm}{\PYZus{}\PYZus{}init\PYZus{}\PYZus{}}\PY{p}{(}\PY{n+nb+bp}{self}\PY{p}{,} \PY{n}{beta}\PY{p}{)}\PY{p}{:}
        \PY{n+nb+bp}{self}\PY{o}{.}\PY{n}{beta} \PY{o}{=} \PY{n}{beta}
    \PY{k}{def} \PY{n+nf+fm}{\PYZus{}\PYZus{}call\PYZus{}\PYZus{}}\PY{p}{(}\PY{n+nb+bp}{self}\PY{p}{,} \PY{n}{cfgs}\PY{p}{)}\PY{p}{:}
        \PY{n}{Nd} \PY{o}{=} \PY{n}{cfgs}\PY{o}{.}\PY{n}{shape}\PY{p}{[}\PY{l+m+mi}{1}\PY{p}{]}
        \PY{n}{action\PYZus{}density} \PY{o}{=} \PY{l+m+mi}{0}
        \PY{k}{for} \PY{n}{mu} \PY{o+ow}{in} \PY{n+nb}{range}\PY{p}{(}\PY{n}{Nd}\PY{p}{)}\PY{p}{:}
            \PY{k}{for} \PY{n}{nu} \PY{o+ow}{in} \PY{n+nb}{range}\PY{p}{(}\PY{n}{mu}\PY{o}{+}\PY{l+m+mi}{1}\PY{p}{,}\PY{n}{Nd}\PY{p}{)}\PY{p}{:}
                \PY{n}{action\PYZus{}density} \PY{o}{=} \PY{n}{action\PYZus{}density} \PY{o}{+} \PY{n}{torch}\PY{o}{.}\PY{n}{cos}\PY{p}{(}
                    \PY{n}{compute\PYZus{}u1\PYZus{}plaq}\PY{p}{(}\PY{n}{cfgs}\PY{p}{,} \PY{n}{mu}\PY{p}{,} \PY{n}{nu}\PY{p}{)}\PY{p}{)}
        \PY{k}{return} \PY{o}{\PYZhy{}}\PY{n+nb+bp}{self}\PY{o}{.}\PY{n}{beta} \PY{o}{*} \PY{n}{torch}\PY{o}{.}\PY{n}{sum}\PY{p}{(}\PY{n}{action\PYZus{}density}\PY{p}{,} \PY{n}{dim}\PY{o}{=}\PY{n+nb}{tuple}\PY{p}{(}\PY{n+nb}{range}\PY{p}{(}\PY{l+m+mi}{1}\PY{p}{,}\PY{n}{Nd}\PY{o}{+}\PY{l+m+mi}{1}\PY{p}{)}\PY{p}{)}\PY{p}{)}

\PY{n+nb}{print}\PY{p}{(}\PY{n}{U1GaugeAction}\PY{p}{(}\PY{n}{beta}\PY{o}{=}\PY{l+m+mf}{1.0}\PY{p}{)}\PY{p}{(}\PY{n}{cfgs}\PY{p}{)}\PY{p}{)}
    \end{Verbatim}
  \end{tcolorbox}

{\color{gray}
    \begin{Verbatim}[commandchars=\\\{\},fontsize=\small]
>>> tensor([-2.3814,  3.3746])
    \end{Verbatim}
}

  \begin{tcolorbox}[breakable, size=fbox, boxrule=1pt, pad at break*=1mm,colback=cellbackground, colframe=cellborder]
    \begin{Verbatim}[commandchars=\\\{\},fontsize=\small]
\PY{n}{beta} \PY{o}{=} \PY{l+m+mi}{1}
\PY{n}{u1\PYZus{}action} \PY{o}{=} \PY{n}{U1GaugeAction}\PY{p}{(}\PY{n}{beta}\PY{p}{)}
    \end{Verbatim}
  \end{tcolorbox}

  This action is invariant under \textbf{gauge transformations}
\begin{equation}
    U_\mu(\vec{n}) \rightarrow e^{i \alpha(\vec{n})} U_\mu(\vec{n}) e^{-i \alpha(\vec{n} + \hat{\mu})}
\end{equation} or, in terms of the angular variables, \begin{equation}
    \theta_\mu(\vec{n}) \rightarrow \alpha(\vec{n}) + \theta_\mu(\vec{n}) -\alpha(\vec{n} + \hat{\mu})
\end{equation} for any real-valued lattice field \(\alpha(\vec{n})\)
(i.e.~an independent real number for each lattice site \(\vec{n}\)). The
set of gauge transforms composes a large symmetry group
(\(N_d V\)-dimensional, for U(1)) that we will explicitly encode in the
normalizing flow model. Encoding this symmetry exactly improves data
efficiency of training.

We can check numerically that our Pytorch implementation of the action
is invariant with respect to an arbitrary gauge transformation.

  \begin{tcolorbox}[breakable, size=fbox, boxrule=1pt, pad at break*=1mm,colback=cellbackground, colframe=cellborder]
    \begin{Verbatim}[commandchars=\\\{\},fontsize=\small]
\PY{k}{def} \PY{n+nf}{gauge\PYZus{}transform}\PY{p}{(}\PY{n}{links}\PY{p}{,} \PY{n}{alpha}\PY{p}{)}\PY{p}{:}
    \PY{k}{for} \PY{n}{mu} \PY{o+ow}{in} \PY{n+nb}{range}\PY{p}{(}\PY{n+nb}{len}\PY{p}{(}\PY{n}{links}\PY{o}{.}\PY{n}{shape}\PY{p}{[}\PY{l+m+mi}{2}\PY{p}{:}\PY{p}{]}\PY{p}{)}\PY{p}{)}\PY{p}{:}
        \PY{n}{links}\PY{p}{[}\PY{p}{:}\PY{p}{,}\PY{n}{mu}\PY{p}{]} \PY{o}{=} \PY{n}{alpha} \PY{o}{+} \PY{n}{links}\PY{p}{[}\PY{p}{:}\PY{p}{,}\PY{n}{mu}\PY{p}{]} \PY{o}{\PYZhy{}} \PY{n}{torch}\PY{o}{.}\PY{n}{roll}\PY{p}{(}\PY{n}{alpha}\PY{p}{,} \PY{o}{\PYZhy{}}\PY{l+m+mi}{1}\PY{p}{,} \PY{n}{mu}\PY{o}{+}\PY{l+m+mi}{1}\PY{p}{)}
    \PY{k}{return} \PY{n}{links}
\PY{k}{def} \PY{n+nf}{random\PYZus{}gauge\PYZus{}transform}\PY{p}{(}\PY{n}{x}\PY{p}{)}\PY{p}{:}
    \PY{n}{Nconf}\PY{p}{,} \PY{n}{VolShape} \PY{o}{=} \PY{n}{x}\PY{o}{.}\PY{n}{shape}\PY{p}{[}\PY{l+m+mi}{0}\PY{p}{]}\PY{p}{,} \PY{n}{x}\PY{o}{.}\PY{n}{shape}\PY{p}{[}\PY{l+m+mi}{2}\PY{p}{:}\PY{p}{]}
    \PY{k}{return} \PY{n}{gauge\PYZus{}transform}\PY{p}{(}\PY{n}{x}\PY{p}{,} \PY{l+m+mi}{2}\PY{o}{*}\PY{n}{np}\PY{o}{.}\PY{n}{pi}\PY{o}{*}\PY{n}{torch}\PY{o}{.}\PY{n}{rand}\PY{p}{(}\PY{p}{(}\PY{n}{Nconf}\PY{p}{,}\PY{p}{)} \PY{o}{+} \PY{n}{VolShape}\PY{p}{)}\PY{p}{)}

\PY{c+c1}{\PYZsh{} action is invariant}
\PY{n}{cfgs\PYZus{}transformed} \PY{o}{=} \PY{n}{random\PYZus{}gauge\PYZus{}transform}\PY{p}{(}\PY{n}{cfgs}\PY{p}{)}
\PY{n+nb}{print}\PY{p}{(}\PY{n}{u1\PYZus{}action}\PY{p}{(}\PY{n}{cfgs}\PY{p}{)}\PY{p}{,} \PY{l+s+s1}{\PYZsq{}}\PY{l+s+s1}{vs}\PY{l+s+s1}{\PYZsq{}}\PY{p}{,} \PY{n}{u1\PYZus{}action}\PY{p}{(}\PY{n}{cfgs\PYZus{}transformed}\PY{p}{)}\PY{p}{)}
\PY{k}{assert} \PY{n}{np}\PY{o}{.}\PY{n}{allclose}\PY{p}{(}\PY{n}{grab}\PY{p}{(}\PY{n}{u1\PYZus{}action}\PY{p}{(}\PY{n}{cfgs}\PY{p}{)}\PY{p}{)}\PY{p}{,} \PY{n}{grab}\PY{p}{(}\PY{n}{u1\PYZus{}action}\PY{p}{(}\PY{n}{cfgs\PYZus{}transformed}\PY{p}{)}\PY{p}{)}\PY{p}{)}\PY{p}{,} \PYZbs{}
    \PY{l+s+s1}{\PYZsq{}}\PY{l+s+s1}{gauge transform should be a symmetry of the action}\PY{l+s+s1}{\PYZsq{}}
    \end{Verbatim}
  \end{tcolorbox}

{\color{gray}
    \begin{Verbatim}[commandchars=\\\{\},fontsize=\small]
>>> tensor([-2.3814,  3.3746]) vs tensor([-2.3814,  3.3746])
    \end{Verbatim}
}

  Gauge theory in 2D is a bit peculiar: in the lattice regularization,
each plaquette fluctuates independently except for
exponentially-suppressed correlations due to periodic boundary
conditions (i.e.~in the infinite volume limit the correlation length is
zero). For \(\mathrm{U}(1)\) gauge theory in particular, there is also a
well-defined topological charge on the lattice, \begin{equation}
Q \equiv \frac{1}{2\pi} \sum_{\vec{n}} \arg(P_{01}(\vec{n})), \quad Q \in \mathbb{Z}
\end{equation} where \(\arg(\cdot) \in [-\pi, \pi]\). This topological
charge mixes slowly with usual MCMC techniques, and we find that
directly sampling using flow models vastly improves estimates for
topological quantities.

  \begin{tcolorbox}[breakable, size=fbox, boxrule=1pt, pad at break*=1mm,colback=cellbackground, colframe=cellborder]
    \begin{Verbatim}[commandchars=\\\{\},fontsize=\small]
\PY{k}{def} \PY{n+nf}{topo\PYZus{}charge}\PY{p}{(}\PY{n}{x}\PY{p}{)}\PY{p}{:}
    \PY{n}{P01} \PY{o}{=} \PY{n}{torch\PYZus{}wrap}\PY{p}{(}\PY{n}{compute\PYZus{}u1\PYZus{}plaq}\PY{p}{(}\PY{n}{x}\PY{p}{,} \PY{n}{mu}\PY{o}{=}\PY{l+m+mi}{0}\PY{p}{,} \PY{n}{nu}\PY{o}{=}\PY{l+m+mi}{1}\PY{p}{)}\PY{p}{)}
    \PY{n}{axes} \PY{o}{=} \PY{n+nb}{tuple}\PY{p}{(}\PY{n+nb}{range}\PY{p}{(}\PY{l+m+mi}{1}\PY{p}{,} \PY{n+nb}{len}\PY{p}{(}\PY{n}{P01}\PY{o}{.}\PY{n}{shape}\PY{p}{)}\PY{p}{)}\PY{p}{)}
    \PY{k}{return} \PY{n}{torch}\PY{o}{.}\PY{n}{sum}\PY{p}{(}\PY{n}{P01}\PY{p}{,} \PY{n}{dim}\PY{o}{=}\PY{n}{axes}\PY{p}{)} \PY{o}{/} \PY{p}{(}\PY{l+m+mi}{2}\PY{o}{*}\PY{n}{np}\PY{o}{.}\PY{n}{pi}\PY{p}{)}

\PY{k}{with} \PY{n}{np}\PY{o}{.}\PY{n}{printoptions}\PY{p}{(}\PY{n}{suppress}\PY{o}{=}\PY{k+kc}{True}\PY{p}{)}\PY{p}{:}
    \PY{n+nb}{print}\PY{p}{(}\PY{l+s+sa}{f}\PY{l+s+s1}{\PYZsq{}}\PY{l+s+s1}{cfg topological charges = }\PY{l+s+si}{\PYZob{}}\PY{n}{grab}\PY{p}{(}\PY{n}{topo\PYZus{}charge}\PY{p}{(}\PY{n}{cfgs}\PY{p}{)}\PY{p}{)}\PY{l+s+si}{\PYZcb{}}\PY{l+s+s1}{\PYZsq{}}\PY{p}{)}
\PY{n}{Q} \PY{o}{=} \PY{n}{grab}\PY{p}{(}\PY{n}{topo\PYZus{}charge}\PY{p}{(}\PY{n}{cfgs}\PY{p}{)}\PY{p}{)}
\PY{k}{assert} \PY{n}{np}\PY{o}{.}\PY{n}{allclose}\PY{p}{(}\PY{n}{Q}\PY{p}{,} \PY{n}{np}\PY{o}{.}\PY{n}{around}\PY{p}{(}\PY{n}{Q}\PY{p}{)}\PY{p}{,} \PY{n}{atol}\PY{o}{=}\PY{l+m+mf}{1e\PYZhy{}6}\PY{p}{)}\PY{p}{,} \PY{l+s+s1}{\PYZsq{}}\PY{l+s+s1}{topological charge must be an integer}\PY{l+s+s1}{\PYZsq{}}
    \end{Verbatim}
  \end{tcolorbox}

{\color{gray}
    \begin{Verbatim}[commandchars=\\\{\},fontsize=\small]
>>> cfg topological charges = [3. 0.]
    \end{Verbatim}
}

  Details on the formulation of Lattice Gauge Theory may be found in the
books \cite{Gattringer:2010zz} and \cite{Smit:2002ug}.

  \hypertarget{prior-distribution}{%
\subsection{\texorpdfstring{\textbf{Prior
distribution}}{Prior distribution}}\label{prior-distribution}}

We use a uniform distribution with respect to the Haar measure. For
\(\mathrm{U}(1)\), this just corresponds to a uniform distribution over
\(\arg(U_\mu(\vec{n})) = \theta_\mu(\vec{n}) \in [0, 2\pi]\). This is
easy to sample, as well as gauge invariant. \textbf{Note}: in the
implementation below, we define the Haar measure normalized to total
volume \(2\pi\); this is an irrelevant normalization having no effect on
training or sampling.

  \begin{tcolorbox}[breakable, size=fbox, boxrule=1pt, pad at break*=1mm,colback=cellbackground, colframe=cellborder]
    \begin{Verbatim}[commandchars=\\\{\},fontsize=\small]
\PY{k}{class} \PY{n+nc}{MultivariateUniform}\PY{p}{(}\PY{n}{torch}\PY{o}{.}\PY{n}{nn}\PY{o}{.}\PY{n}{Module}\PY{p}{)}\PY{p}{:}
    \PY{l+s+sd}{\PYZdq{}\PYZdq{}\PYZdq{}Uniformly draw samples from [a,b]\PYZdq{}\PYZdq{}\PYZdq{}}
    \PY{k}{def} \PY{n+nf+fm}{\PYZus{}\PYZus{}init\PYZus{}\PYZus{}}\PY{p}{(}\PY{n+nb+bp}{self}\PY{p}{,} \PY{n}{a}\PY{p}{,} \PY{n}{b}\PY{p}{)}\PY{p}{:}
        \PY{n+nb}{super}\PY{p}{(}\PY{p}{)}\PY{o}{.}\PY{n+nf+fm}{\PYZus{}\PYZus{}init\PYZus{}\PYZus{}}\PY{p}{(}\PY{p}{)}
        \PY{n+nb+bp}{self}\PY{o}{.}\PY{n}{dist} \PY{o}{=} \PY{n}{torch}\PY{o}{.}\PY{n}{distributions}\PY{o}{.}\PY{n}{uniform}\PY{o}{.}\PY{n}{Uniform}\PY{p}{(}\PY{n}{a}\PY{p}{,} \PY{n}{b}\PY{p}{)}
    \PY{k}{def} \PY{n+nf}{log\PYZus{}prob}\PY{p}{(}\PY{n+nb+bp}{self}\PY{p}{,} \PY{n}{x}\PY{p}{)}\PY{p}{:}
        \PY{n}{axes} \PY{o}{=} \PY{n+nb}{range}\PY{p}{(}\PY{l+m+mi}{1}\PY{p}{,} \PY{n+nb}{len}\PY{p}{(}\PY{n}{x}\PY{o}{.}\PY{n}{shape}\PY{p}{)}\PY{p}{)}
        \PY{k}{return} \PY{n}{torch}\PY{o}{.}\PY{n}{sum}\PY{p}{(}\PY{n+nb+bp}{self}\PY{o}{.}\PY{n}{dist}\PY{o}{.}\PY{n}{log\PYZus{}prob}\PY{p}{(}\PY{n}{x}\PY{p}{)}\PY{p}{,} \PY{n}{dim}\PY{o}{=}\PY{n+nb}{tuple}\PY{p}{(}\PY{n}{axes}\PY{p}{)}\PY{p}{)}
    \PY{k}{def} \PY{n+nf}{sample\PYZus{}n}\PY{p}{(}\PY{n+nb+bp}{self}\PY{p}{,} \PY{n}{batch\PYZus{}size}\PY{p}{)}\PY{p}{:}
        \PY{k}{return} \PY{n+nb+bp}{self}\PY{o}{.}\PY{n}{dist}\PY{o}{.}\PY{n}{sample}\PY{p}{(}\PY{p}{(}\PY{n}{batch\PYZus{}size}\PY{p}{,}\PY{p}{)}\PY{p}{)}
    \end{Verbatim}
  \end{tcolorbox}

  \begin{tcolorbox}[breakable, size=fbox, boxrule=1pt, pad at break*=1mm,colback=cellbackground, colframe=cellborder]
    \begin{Verbatim}[commandchars=\\\{\},fontsize=\small]
\PY{n}{prior} \PY{o}{=} \PY{n}{MultivariateUniform}\PY{p}{(}\PY{n}{torch}\PY{o}{.}\PY{n}{zeros}\PY{p}{(}\PY{n}{link\PYZus{}shape}\PY{p}{)}\PY{p}{,} \PY{l+m+mi}{2}\PY{o}{*}\PY{n}{np}\PY{o}{.}\PY{n}{pi}\PY{o}{*}\PY{n}{torch}\PY{o}{.}\PY{n}{ones}\PY{p}{(}\PY{n}{link\PYZus{}shape}\PY{p}{)}\PY{p}{)}
\PY{n}{z} \PY{o}{=} \PY{n}{prior}\PY{o}{.}\PY{n}{sample\PYZus{}n}\PY{p}{(}\PY{l+m+mi}{17}\PY{p}{)}
\PY{n+nb}{print}\PY{p}{(}\PY{l+s+sa}{f}\PY{l+s+s1}{\PYZsq{}}\PY{l+s+s1}{z.shape = }\PY{l+s+si}{\PYZob{}}\PY{n}{z}\PY{o}{.}\PY{n}{shape}\PY{l+s+si}{\PYZcb{}}\PY{l+s+s1}{\PYZsq{}}\PY{p}{)}
\PY{n+nb}{print}\PY{p}{(}\PY{l+s+sa}{f}\PY{l+s+s1}{\PYZsq{}}\PY{l+s+s1}{log r(z) = }\PY{l+s+si}{\PYZob{}}\PY{n}{grab}\PY{p}{(}\PY{n}{prior}\PY{o}{.}\PY{n}{log\PYZus{}prob}\PY{p}{(}\PY{n}{z}\PY{p}{)}\PY{p}{)}\PY{l+s+si}{\PYZcb{}}\PY{l+s+s1}{\PYZsq{}}\PY{p}{)}
    \end{Verbatim}
  \end{tcolorbox}

{\color{gray}
    \begin{Verbatim}[commandchars=\\\{\},fontsize=\small]
>>> z.shape = torch.Size([17, 2, 8, 8])
... log r(z) = [-235.2482645 -235.2482645 -235.2482645 -235.2482645 -235.2482645
...  -235.2482645 -235.2482645 -235.2482645 -235.2482645 -235.2482645
...  -235.2482645 -235.2482645 -235.2482645 -235.2482645 -235.2482645
...  -235.2482645 -235.2482645]
    \end{Verbatim}
}

  \hypertarget{gauge-equivariant-coupling-layers}{%
\subsection{\texorpdfstring{\textbf{Gauge equivariant coupling
layers}}{Gauge equivariant coupling layers}}\label{gauge-equivariant-coupling-layers}}

Recall, our goal is to produce a gauge invariant output distribution,
which can be achieved using gauge equivariant coupling layers combined
with a gauge invariant prior distribution. A coupling layer is
\textbf{gauge equivariant} if applying a gauge transformation commutes
with application of the coupling layer. Consider abstractly factoring
the degrees of freedom in the lattice gauge theory into pure-gauge and
gauge-invariant degrees of freedom. Under this factorization, a gauge
transformation only affects pure-gauge degrees of freedom, and thus a
transformation acting \textbf{only on gauge invariant} quantities will
be a gauge equivariant transformation.

Constructing coupling layers that transform links in a way that only
affects gauge invariant quantities is not obvious, since these
quantities are not necessarily in 1:1 correspondence with the gauge
links, which are the lattice degrees of freedom. In
\cite{Kanwar:2020xzo}, we presented a construction that resolves this
issue by defining how links should be transformed to produce a
transformation of the gauge-invariant spectra of untraced Wilson loops.
The general case is worked out there, but here we consider the special
case of \(\mathrm{U}(1)\) gauge theory where we focus on \(1\times 1\)
Wilson loops (plaquettes).

We define a gauge equivariant coupling layer in terms of an inner
coupling layer which acts on (``active'') plaquettes, which for
\(\mathrm{U}(1)\) gauge theory are scalar, gauge-invariant objects. The
inner coupling layer \(g\) transforms
\(P_{\mu\nu}(\vec{n}) \rightarrow P'_{\mu\nu}(\vec{n})\) (see Figure
below). The update to each plaquette can be uniquely ``pushed onto'' a
corresponding link if we transform few enough plaquettes. For
\(\mathrm{U}(1)\) gauge theory, pushing updates from a plaquette to a
contained link is easy due to the Abelian nature of the group. If the
inner coupling layer maps \(P \rightarrow P'\), then the contained link
\(U\) should be updated as \(U \rightarrow U' = P' P^{-1} U\). This
enacts the desired transformation on the plaquette, \begin{equation}
P = U V \rightarrow P' P^{-1} U V = P' (V^{-1} U^{-1}) U V = P'.
\end{equation} where \(V\) is the remaining product of links
(``staple'') defining the plaquette. A similar expression applies when
the plaquette is defined in terms of \(U^{-1}\) instead. Note that this
``passively'' transforms any other plaquettes containing \(U\).

  \begin{figure}[H]
      \centering
      \includegraphics{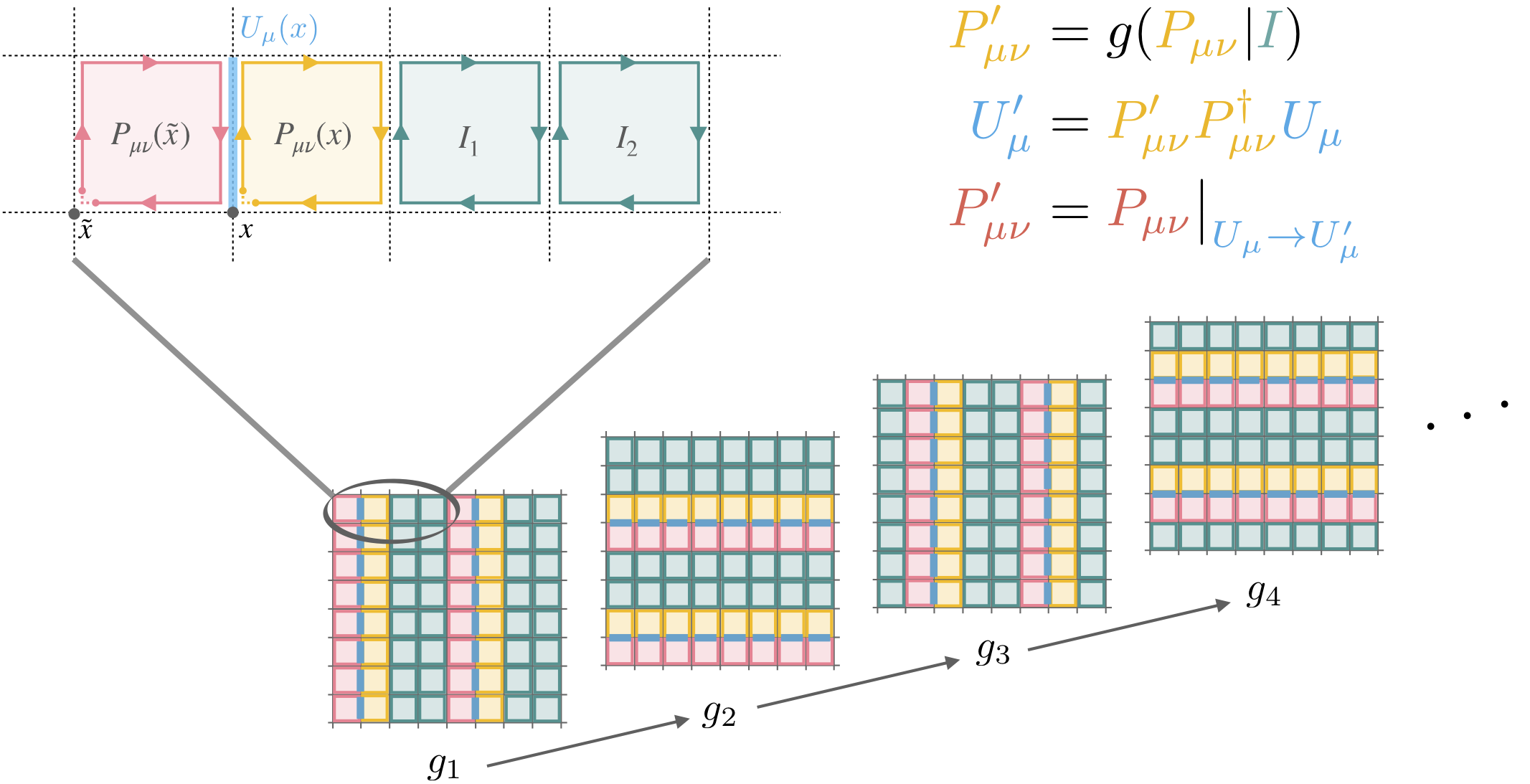}
      \caption{Equivariant action by ``pushing'' plaquette updates onto links in each coupling layer $g_i$.}
    \end{figure}

  This transformation is correctly gauge equivariant if, in addition, the
inner flow updating \(P_{\mu\nu}(\vec{n})\) depends only on frozen gauge
invariant quantities. In comparison to scalar field theory, we must
define \emph{three} disjoint subsets of degrees of freedom to be passed
to the inner coupling layer: \emph{active}, \emph{passive}, and
\emph{frozen}. The frozen subset works the same as before; it is not
updated and can be used as input to the neural nets defining the
parameters. The active subset is actively transformed by the coupling
layer using those parameters, and its transformation will be pushed onto
corresponding links. The passively updated subset contains plaquettes
that include an updated link but are not directly transformed by the
inner layer nor used as inputs to the neural nets defining the
parameters.

A gauge equivariant coupling layer for \(\mathrm{U}(1)\) gauge theory is
defined below, with a specific choice of 1:1 mapping between updated
links and plaquettes: each active plaquette \(P_{01}(\vec{n})\) contains
exactly one link to update, either the link on the left \(U_0(\vec{n})\)
or the link at the bottom \(U_1(\vec{n})^{-1}\), specified implicitly by
masks (we discuss our choice of masking pattern and define the relevant
functions in the next section). Translating to the angular variables
used in code to represent the links, these updates look like

\begin{equation}
\begin{aligned}
\theta_0(\vec{n}) &\rightarrow \theta_0(\vec{n}) + \delta_{01}(\vec{n}) \\
\theta_1(\vec{n}) &\rightarrow \theta_1(\vec{n}) - \delta_{01}(\vec{n}),
\end{aligned}
\end{equation} where
\(\delta_{01}(\vec{n}) = \theta_{01}'(\vec{n}) - \theta_{01}(\vec{n})\).

\textbf{Note:} We define the gauge equivariant coupling layer in terms
of an inner coupling layer \texttt{plaq\_coupling} which we will define
and discuss later.

  \begin{tcolorbox}[breakable, size=fbox, boxrule=1pt, pad at break*=1mm,colback=cellbackground, colframe=cellborder]
    \begin{Verbatim}[commandchars=\\\{\},fontsize=\small]
\PY{k}{class} \PY{n+nc}{GaugeEquivCouplingLayer}\PY{p}{(}\PY{n}{torch}\PY{o}{.}\PY{n}{nn}\PY{o}{.}\PY{n}{Module}\PY{p}{)}\PY{p}{:}
    \PY{l+s+sd}{\PYZdq{}\PYZdq{}\PYZdq{}U(1) gauge equiv coupling layer defined by `plaq\PYZus{}coupling` acting on plaquettes.\PYZdq{}\PYZdq{}\PYZdq{}}
    \PY{k}{def} \PY{n+nf+fm}{\PYZus{}\PYZus{}init\PYZus{}\PYZus{}}\PY{p}{(}\PY{n+nb+bp}{self}\PY{p}{,} \PY{o}{*}\PY{p}{,} \PY{n}{lattice\PYZus{}shape}\PY{p}{,} \PY{n}{mask\PYZus{}mu}\PY{p}{,} \PY{n}{mask\PYZus{}off}\PY{p}{,} \PY{n}{plaq\PYZus{}coupling}\PY{p}{)}\PY{p}{:}
        \PY{n+nb}{super}\PY{p}{(}\PY{p}{)}\PY{o}{.}\PY{n+nf+fm}{\PYZus{}\PYZus{}init\PYZus{}\PYZus{}}\PY{p}{(}\PY{p}{)}
        \PY{n}{link\PYZus{}mask\PYZus{}shape} \PY{o}{=} \PY{p}{(}\PY{n+nb}{len}\PY{p}{(}\PY{n}{lattice\PYZus{}shape}\PY{p}{)}\PY{p}{,}\PY{p}{)} \PY{o}{+} \PY{n}{lattice\PYZus{}shape}
        \PY{n+nb+bp}{self}\PY{o}{.}\PY{n}{active\PYZus{}mask} \PY{o}{=} \PY{n}{make\PYZus{}2d\PYZus{}link\PYZus{}active\PYZus{}stripes}\PY{p}{(}\PY{n}{link\PYZus{}mask\PYZus{}shape}\PY{p}{,} \PY{n}{mask\PYZus{}mu}\PY{p}{,} \PY{n}{mask\PYZus{}off}\PY{p}{)}
        \PY{n+nb+bp}{self}\PY{o}{.}\PY{n}{plaq\PYZus{}coupling} \PY{o}{=} \PY{n}{plaq\PYZus{}coupling}
        
    \PY{k}{def} \PY{n+nf}{forward}\PY{p}{(}\PY{n+nb+bp}{self}\PY{p}{,} \PY{n}{x}\PY{p}{)}\PY{p}{:}
        \PY{n}{plaq} \PY{o}{=} \PY{n}{compute\PYZus{}u1\PYZus{}plaq}\PY{p}{(}\PY{n}{x}\PY{p}{,} \PY{n}{mu}\PY{o}{=}\PY{l+m+mi}{0}\PY{p}{,} \PY{n}{nu}\PY{o}{=}\PY{l+m+mi}{1}\PY{p}{)}
        \PY{n}{new\PYZus{}plaq}\PY{p}{,} \PY{n}{logJ} \PY{o}{=} \PY{n+nb+bp}{self}\PY{o}{.}\PY{n}{plaq\PYZus{}coupling}\PY{p}{(}\PY{n}{plaq}\PY{p}{)}
        \PY{n}{delta\PYZus{}plaq} \PY{o}{=} \PY{n}{new\PYZus{}plaq} \PY{o}{\PYZhy{}} \PY{n}{plaq}
        \PY{n}{delta\PYZus{}links} \PY{o}{=} \PY{n}{torch}\PY{o}{.}\PY{n}{stack}\PY{p}{(}\PY{p}{(}\PY{n}{delta\PYZus{}plaq}\PY{p}{,} \PY{o}{\PYZhy{}}\PY{n}{delta\PYZus{}plaq}\PY{p}{)}\PY{p}{,} \PY{n}{dim}\PY{o}{=}\PY{l+m+mi}{1}\PY{p}{)} \PY{c+c1}{\PYZsh{} signs for U vs Udagger}
        \PY{n}{fx} \PY{o}{=} \PY{n+nb+bp}{self}\PY{o}{.}\PY{n}{active\PYZus{}mask} \PY{o}{*} \PY{n}{torch\PYZus{}mod}\PY{p}{(}\PY{n}{delta\PYZus{}links} \PY{o}{+} \PY{n}{x}\PY{p}{)} \PY{o}{+} \PY{p}{(}\PY{l+m+mi}{1}\PY{o}{\PYZhy{}}\PY{n+nb+bp}{self}\PY{o}{.}\PY{n}{active\PYZus{}mask}\PY{p}{)} \PY{o}{*} \PY{n}{x}
        \PY{k}{return} \PY{n}{fx}\PY{p}{,} \PY{n}{logJ}

    \PY{k}{def} \PY{n+nf}{reverse}\PY{p}{(}\PY{n+nb+bp}{self}\PY{p}{,} \PY{n}{fx}\PY{p}{)}\PY{p}{:}
        \PY{n}{new\PYZus{}plaq} \PY{o}{=} \PY{n}{compute\PYZus{}u1\PYZus{}plaq}\PY{p}{(}\PY{n}{fx}\PY{p}{,} \PY{n}{mu}\PY{o}{=}\PY{l+m+mi}{0}\PY{p}{,} \PY{n}{nu}\PY{o}{=}\PY{l+m+mi}{1}\PY{p}{)}
        \PY{n}{plaq}\PY{p}{,} \PY{n}{logJ} \PY{o}{=} \PY{n+nb+bp}{self}\PY{o}{.}\PY{n}{plaq\PYZus{}coupling}\PY{o}{.}\PY{n}{reverse}\PY{p}{(}\PY{n}{new\PYZus{}plaq}\PY{p}{)}
        \PY{n}{delta\PYZus{}plaq} \PY{o}{=} \PY{n}{plaq} \PY{o}{\PYZhy{}} \PY{n}{new\PYZus{}plaq}
        \PY{n}{delta\PYZus{}links} \PY{o}{=} \PY{n}{torch}\PY{o}{.}\PY{n}{stack}\PY{p}{(}\PY{p}{(}\PY{n}{delta\PYZus{}plaq}\PY{p}{,} \PY{o}{\PYZhy{}}\PY{n}{delta\PYZus{}plaq}\PY{p}{)}\PY{p}{,} \PY{n}{dim}\PY{o}{=}\PY{l+m+mi}{1}\PY{p}{)} \PY{c+c1}{\PYZsh{} signs for U vs Udagger}
        \PY{n}{x} \PY{o}{=} \PY{n+nb+bp}{self}\PY{o}{.}\PY{n}{active\PYZus{}mask} \PY{o}{*} \PY{n}{torch\PYZus{}mod}\PY{p}{(}\PY{n}{delta\PYZus{}links} \PY{o}{+} \PY{n}{fx}\PY{p}{)} \PY{o}{+} \PY{p}{(}\PY{l+m+mi}{1}\PY{o}{\PYZhy{}}\PY{n+nb+bp}{self}\PY{o}{.}\PY{n}{active\PYZus{}mask}\PY{p}{)} \PY{o}{*} \PY{n}{fx}
        \PY{k}{return} \PY{n}{x}\PY{p}{,} \PY{n}{logJ}
    \end{Verbatim}
  \end{tcolorbox}

  Transformation of \(U_0(\vec{n})\) is done according to foluma above but
formula for \(U_1(\vec{n})\) requires additional clarification.
According to the explanation above we would need to update
\(U_1\rightarrow U_1' = P_{10}' P_{10}^{-1} U_1\) but it is generally
accepted in LQCD to use only the positive direction of plaquettes.
Keeping in mind \(P_{01} = P_{10}^{\dagger}\), we can change the
transformation to
\(U_1^\dagger\rightarrow U_1'^\dagger = U_1^\dagger P_{01}^{-1} P_{01}'\).
In the angular representaion it simply has the form
\(\theta_1(\vec{n}) \rightarrow \theta_1(\vec{n}) - \delta \theta_{01}(\vec{n})\),
where \(\delta \theta_{01}(\vec{n}) = \theta_{01}' - \theta_{01}\).

  The masking pattern and choice of gauge invariant quantities for the
inner update could all be generalized. See also \cite{Boyda:2020hsi} for
details on the non-Abelian version of this equivariant construction.

  \hypertarget{gauge-equivariant-masking-patterns}{%
\subsection{\texorpdfstring{\textbf{Gauge-equivariant masking
patterns}}{Gauge-equivariant masking patterns}}\label{gauge-equivariant-masking-patterns}}

There are many choices of masking patterns that allow updates to be
pushed onto links. We used this one because it's simple and strikes a
good balance between updating as many links as possible (number of
active links) and having sufficient info to make well-informed updates
(number of frozen plaquettes). More exploration of optimal masking
pattern structure in higher dimensions will be explored in future work.
We can update all links on the lattice by composing coupling layers with
different mask offsets and directions.

  \begin{figure}[H]
      \centering
      \includegraphics[width=5cm]{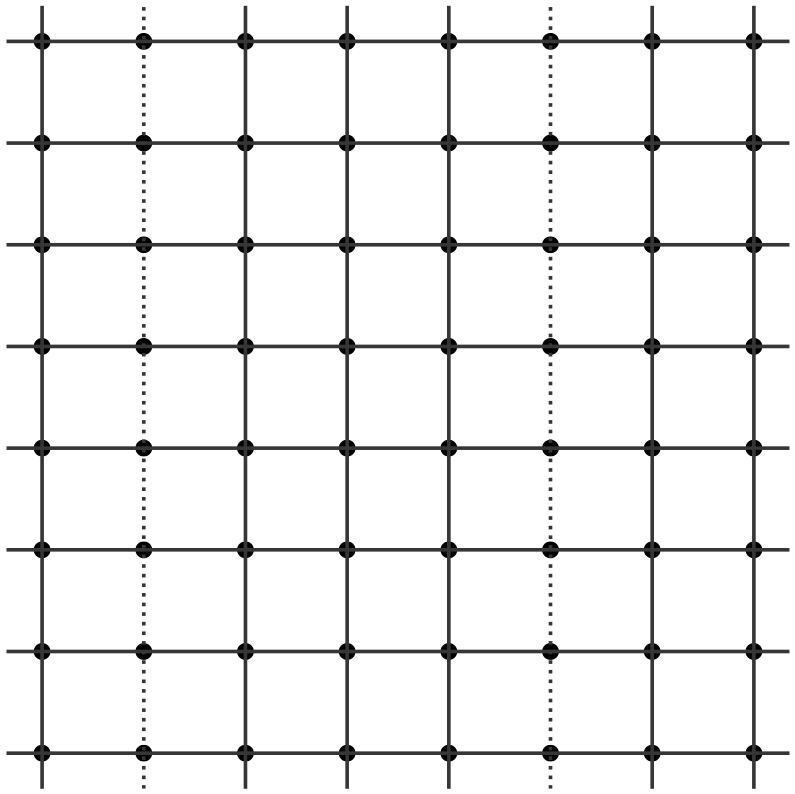}
      \caption{Masking pattern for a single coupling layer, with active links indicated with dotted lines and passive links with solid lines.}
    \end{figure}

  For the links, we need a mask that will pick out the ``active'' links to
be updated. We only update links in one direction at a time, as
indicated in the figure above. The masking pattern for links in the
updated direction looks like stripes along the updated direction, spaced
by 4 lattice units. For links in the other direction, it's all 0s.

  \begin{tcolorbox}[breakable, size=fbox, boxrule=1pt, pad at break*=1mm,colback=cellbackground, colframe=cellborder]
    \begin{Verbatim}[commandchars=\\\{\},fontsize=\small]
\PY{k}{def} \PY{n+nf}{make\PYZus{}2d\PYZus{}link\PYZus{}active\PYZus{}stripes}\PY{p}{(}\PY{n}{shape}\PY{p}{,} \PY{n}{mu}\PY{p}{,} \PY{n}{off}\PY{p}{)}\PY{p}{:}
    \PY{l+s+sd}{\PYZdq{}\PYZdq{}\PYZdq{}}
\PY{l+s+sd}{    Stripes mask looks like in the `mu` channel (mu\PYZhy{}oriented links)::}

\PY{l+s+sd}{      1 0 0 0 1 0 0 0 1 0 0}
\PY{l+s+sd}{      1 0 0 0 1 0 0 0 1 0 0}
\PY{l+s+sd}{      1 0 0 0 1 0 0 0 1 0 0}
\PY{l+s+sd}{      1 0 0 0 1 0 0 0 1 0 0}

\PY{l+s+sd}{    where vertical is the `mu` direction, and the pattern is offset in the nu}
\PY{l+s+sd}{    direction by `off` (mod 4). The other channel is identically 0.}
\PY{l+s+sd}{    \PYZdq{}\PYZdq{}\PYZdq{}}
    \PY{k}{assert} \PY{n+nb}{len}\PY{p}{(}\PY{n}{shape}\PY{p}{)} \PY{o}{==} \PY{l+m+mi}{2}\PY{o}{+}\PY{l+m+mi}{1}\PY{p}{,} \PY{l+s+s1}{\PYZsq{}}\PY{l+s+s1}{need to pass shape suitable for 2D gauge theory}\PY{l+s+s1}{\PYZsq{}}
    \PY{k}{assert} \PY{n}{shape}\PY{p}{[}\PY{l+m+mi}{0}\PY{p}{]} \PY{o}{==} \PY{n+nb}{len}\PY{p}{(}\PY{n}{shape}\PY{p}{[}\PY{l+m+mi}{1}\PY{p}{:}\PY{p}{]}\PY{p}{)}\PY{p}{,} \PY{l+s+s1}{\PYZsq{}}\PY{l+s+s1}{first dim of shape must be Nd}\PY{l+s+s1}{\PYZsq{}}
    \PY{k}{assert} \PY{n}{mu} \PY{o+ow}{in} \PY{p}{(}\PY{l+m+mi}{0}\PY{p}{,}\PY{l+m+mi}{1}\PY{p}{)}\PY{p}{,} \PY{l+s+s1}{\PYZsq{}}\PY{l+s+s1}{mu must be 0 or 1}\PY{l+s+s1}{\PYZsq{}}
    
    \PY{n}{mask} \PY{o}{=} \PY{n}{np}\PY{o}{.}\PY{n}{zeros}\PY{p}{(}\PY{n}{shape}\PY{p}{)}\PY{o}{.}\PY{n}{astype}\PY{p}{(}\PY{n}{np}\PY{o}{.}\PY{n}{uint8}\PY{p}{)}
    \PY{k}{if} \PY{n}{mu} \PY{o}{==} \PY{l+m+mi}{0}\PY{p}{:}
        \PY{n}{mask}\PY{p}{[}\PY{n}{mu}\PY{p}{,}\PY{p}{:}\PY{p}{,}\PY{l+m+mi}{0}\PY{p}{:}\PY{p}{:}\PY{l+m+mi}{4}\PY{p}{]} \PY{o}{=} \PY{l+m+mi}{1}
    \PY{k}{elif} \PY{n}{mu} \PY{o}{==} \PY{l+m+mi}{1}\PY{p}{:}
        \PY{n}{mask}\PY{p}{[}\PY{n}{mu}\PY{p}{,}\PY{l+m+mi}{0}\PY{p}{:}\PY{p}{:}\PY{l+m+mi}{4}\PY{p}{]} \PY{o}{=} \PY{l+m+mi}{1}
    \PY{n}{nu} \PY{o}{=} \PY{l+m+mi}{1}\PY{o}{\PYZhy{}}\PY{n}{mu}
    \PY{n}{mask} \PY{o}{=} \PY{n}{np}\PY{o}{.}\PY{n}{roll}\PY{p}{(}\PY{n}{mask}\PY{p}{,} \PY{n}{off}\PY{p}{,} \PY{n}{axis}\PY{o}{=}\PY{n}{nu}\PY{o}{+}\PY{l+m+mi}{1}\PY{p}{)}
    \PY{k}{return} \PY{n}{torch}\PY{o}{.}\PY{n}{from\PYZus{}numpy}\PY{p}{(}\PY{n}{mask}\PY{o}{.}\PY{n}{astype}\PY{p}{(}\PY{n}{float\PYZus{}dtype}\PY{p}{)}\PY{p}{)}\PY{o}{.}\PY{n}{to}\PY{p}{(}\PY{n}{torch\PYZus{}device}\PY{p}{)}
    \end{Verbatim}
  \end{tcolorbox}

  Before we define the masking patterns for plaquettes, let's define a few
helper functions.

  \begin{tcolorbox}[breakable, size=fbox, boxrule=1pt, pad at break*=1mm,colback=cellbackground, colframe=cellborder]
    \begin{Verbatim}[commandchars=\\\{\},fontsize=\small]
\PY{k}{def} \PY{n+nf}{make\PYZus{}single\PYZus{}stripes}\PY{p}{(}\PY{n}{shape}\PY{p}{,} \PY{n}{mu}\PY{p}{,} \PY{n}{off}\PY{p}{)}\PY{p}{:}
    \PY{l+s+sd}{\PYZdq{}\PYZdq{}\PYZdq{}}
\PY{l+s+sd}{      1 0 0 0 1 0 0 0 1 0 0}
\PY{l+s+sd}{      1 0 0 0 1 0 0 0 1 0 0}
\PY{l+s+sd}{      1 0 0 0 1 0 0 0 1 0 0}
\PY{l+s+sd}{      1 0 0 0 1 0 0 0 1 0 0}

\PY{l+s+sd}{    where vertical is the `mu` direction. Vector of 1 is repeated every 4.}
\PY{l+s+sd}{    The pattern is offset in perpendicular to the mu direction by `off` (mod 4).}
\PY{l+s+sd}{    \PYZdq{}\PYZdq{}\PYZdq{}}
    \PY{k}{assert} \PY{n+nb}{len}\PY{p}{(}\PY{n}{shape}\PY{p}{)} \PY{o}{==} \PY{l+m+mi}{2}\PY{p}{,} \PY{l+s+s1}{\PYZsq{}}\PY{l+s+s1}{need to pass 2D shape}\PY{l+s+s1}{\PYZsq{}}
    \PY{k}{assert} \PY{n}{mu} \PY{o+ow}{in} \PY{p}{(}\PY{l+m+mi}{0}\PY{p}{,}\PY{l+m+mi}{1}\PY{p}{)}\PY{p}{,} \PY{l+s+s1}{\PYZsq{}}\PY{l+s+s1}{mu must be 0 or 1}\PY{l+s+s1}{\PYZsq{}}

    \PY{n}{mask} \PY{o}{=} \PY{n}{np}\PY{o}{.}\PY{n}{zeros}\PY{p}{(}\PY{n}{shape}\PY{p}{)}\PY{o}{.}\PY{n}{astype}\PY{p}{(}\PY{n}{np}\PY{o}{.}\PY{n}{uint8}\PY{p}{)}
    \PY{k}{if} \PY{n}{mu} \PY{o}{==} \PY{l+m+mi}{0}\PY{p}{:}
        \PY{n}{mask}\PY{p}{[}\PY{p}{:}\PY{p}{,}\PY{l+m+mi}{0}\PY{p}{:}\PY{p}{:}\PY{l+m+mi}{4}\PY{p}{]} \PY{o}{=} \PY{l+m+mi}{1}
    \PY{k}{elif} \PY{n}{mu} \PY{o}{==} \PY{l+m+mi}{1}\PY{p}{:}
        \PY{n}{mask}\PY{p}{[}\PY{l+m+mi}{0}\PY{p}{:}\PY{p}{:}\PY{l+m+mi}{4}\PY{p}{]} \PY{o}{=} \PY{l+m+mi}{1}
    \PY{n}{mask} \PY{o}{=} \PY{n}{np}\PY{o}{.}\PY{n}{roll}\PY{p}{(}\PY{n}{mask}\PY{p}{,} \PY{n}{off}\PY{p}{,} \PY{n}{axis}\PY{o}{=}\PY{l+m+mi}{1}\PY{o}{\PYZhy{}}\PY{n}{mu}\PY{p}{)}
    \PY{k}{return} \PY{n}{torch}\PY{o}{.}\PY{n}{from\PYZus{}numpy}\PY{p}{(}\PY{n}{mask}\PY{p}{)}\PY{o}{.}\PY{n}{to}\PY{p}{(}\PY{n}{torch\PYZus{}device}\PY{p}{)}
    \end{Verbatim}
  \end{tcolorbox}

  \begin{tcolorbox}[breakable, size=fbox, boxrule=1pt, pad at break*=1mm,colback=cellbackground, colframe=cellborder]
    \begin{Verbatim}[commandchars=\\\{\},fontsize=\small]
\PY{k}{def} \PY{n+nf}{make\PYZus{}double\PYZus{}stripes}\PY{p}{(}\PY{n}{shape}\PY{p}{,} \PY{n}{mu}\PY{p}{,} \PY{n}{off}\PY{p}{)}\PY{p}{:}
    \PY{l+s+sd}{\PYZdq{}\PYZdq{}\PYZdq{}}
\PY{l+s+sd}{    Double stripes mask looks like::}

\PY{l+s+sd}{      1 1 0 0 1 1 0 0}
\PY{l+s+sd}{      1 1 0 0 1 1 0 0}
\PY{l+s+sd}{      1 1 0 0 1 1 0 0}
\PY{l+s+sd}{      1 1 0 0 1 1 0 0}

\PY{l+s+sd}{    where vertical is the `mu` direction. The pattern is offset in perpendicular}
\PY{l+s+sd}{    to the mu direction by `off` (mod 4).}
\PY{l+s+sd}{    \PYZdq{}\PYZdq{}\PYZdq{}}
    \PY{k}{assert} \PY{n+nb}{len}\PY{p}{(}\PY{n}{shape}\PY{p}{)} \PY{o}{==} \PY{l+m+mi}{2}\PY{p}{,} \PY{l+s+s1}{\PYZsq{}}\PY{l+s+s1}{need to pass 2D shape}\PY{l+s+s1}{\PYZsq{}}
    \PY{k}{assert} \PY{n}{mu} \PY{o+ow}{in} \PY{p}{(}\PY{l+m+mi}{0}\PY{p}{,}\PY{l+m+mi}{1}\PY{p}{)}\PY{p}{,} \PY{l+s+s1}{\PYZsq{}}\PY{l+s+s1}{mu must be 0 or 1}\PY{l+s+s1}{\PYZsq{}}

    \PY{n}{mask} \PY{o}{=} \PY{n}{np}\PY{o}{.}\PY{n}{zeros}\PY{p}{(}\PY{n}{shape}\PY{p}{)}\PY{o}{.}\PY{n}{astype}\PY{p}{(}\PY{n}{np}\PY{o}{.}\PY{n}{uint8}\PY{p}{)}
    \PY{k}{if} \PY{n}{mu} \PY{o}{==} \PY{l+m+mi}{0}\PY{p}{:}
        \PY{n}{mask}\PY{p}{[}\PY{p}{:}\PY{p}{,}\PY{l+m+mi}{0}\PY{p}{:}\PY{p}{:}\PY{l+m+mi}{4}\PY{p}{]} \PY{o}{=} \PY{l+m+mi}{1}
        \PY{n}{mask}\PY{p}{[}\PY{p}{:}\PY{p}{,}\PY{l+m+mi}{1}\PY{p}{:}\PY{p}{:}\PY{l+m+mi}{4}\PY{p}{]} \PY{o}{=} \PY{l+m+mi}{1}
    \PY{k}{elif} \PY{n}{mu} \PY{o}{==} \PY{l+m+mi}{1}\PY{p}{:}
        \PY{n}{mask}\PY{p}{[}\PY{l+m+mi}{0}\PY{p}{:}\PY{p}{:}\PY{l+m+mi}{4}\PY{p}{]} \PY{o}{=} \PY{l+m+mi}{1}
        \PY{n}{mask}\PY{p}{[}\PY{l+m+mi}{1}\PY{p}{:}\PY{p}{:}\PY{l+m+mi}{4}\PY{p}{]} \PY{o}{=} \PY{l+m+mi}{1}
    \PY{n}{mask} \PY{o}{=} \PY{n}{np}\PY{o}{.}\PY{n}{roll}\PY{p}{(}\PY{n}{mask}\PY{p}{,} \PY{n}{off}\PY{p}{,} \PY{n}{axis}\PY{o}{=}\PY{l+m+mi}{1}\PY{o}{\PYZhy{}}\PY{n}{mu}\PY{p}{)}
    \PY{k}{return} \PY{n}{torch}\PY{o}{.}\PY{n}{from\PYZus{}numpy}\PY{p}{(}\PY{n}{mask}\PY{p}{)}\PY{o}{.}\PY{n}{to}\PY{p}{(}\PY{n}{torch\PYZus{}device}\PY{p}{)}
    \end{Verbatim}
  \end{tcolorbox}

  The masking pattern for active, passive, and frozen plaquettes are
stripes in the same direction as the active link mask, with appropriate
relative offsets to ensure these three subsets give a disjoint partition
of all plaquettes. The active plaquettes are the ones ahead of each
active link in the updated direction, while passive plaquettes are the
ones behind. All other plaquettes are frozen.

  \begin{tcolorbox}[breakable, size=fbox, boxrule=1pt, pad at break*=1mm,colback=cellbackground, colframe=cellborder]
    \begin{Verbatim}[commandchars=\\\{\},fontsize=\small]
\PY{k}{def} \PY{n+nf}{make\PYZus{}plaq\PYZus{}masks}\PY{p}{(}\PY{n}{mask\PYZus{}shape}\PY{p}{,} \PY{n}{mask\PYZus{}mu}\PY{p}{,} \PY{n}{mask\PYZus{}off}\PY{p}{)}\PY{p}{:}
    \PY{n}{mask} \PY{o}{=} \PY{p}{\PYZob{}}\PY{p}{\PYZcb{}}
    \PY{n}{mask}\PY{p}{[}\PY{l+s+s1}{\PYZsq{}}\PY{l+s+s1}{frozen}\PY{l+s+s1}{\PYZsq{}}\PY{p}{]} \PY{o}{=} \PY{n}{make\PYZus{}double\PYZus{}stripes}\PY{p}{(}\PY{n}{mask\PYZus{}shape}\PY{p}{,} \PY{n}{mask\PYZus{}mu}\PY{p}{,} \PY{n}{mask\PYZus{}off}\PY{o}{+}\PY{l+m+mi}{1}\PY{p}{)}
    \PY{n}{mask}\PY{p}{[}\PY{l+s+s1}{\PYZsq{}}\PY{l+s+s1}{active}\PY{l+s+s1}{\PYZsq{}}\PY{p}{]} \PY{o}{=} \PY{n}{make\PYZus{}single\PYZus{}stripes}\PY{p}{(}\PY{n}{mask\PYZus{}shape}\PY{p}{,} \PY{n}{mask\PYZus{}mu}\PY{p}{,} \PY{n}{mask\PYZus{}off}\PY{p}{)}
    \PY{n}{mask}\PY{p}{[}\PY{l+s+s1}{\PYZsq{}}\PY{l+s+s1}{passive}\PY{l+s+s1}{\PYZsq{}}\PY{p}{]} \PY{o}{=} \PY{l+m+mi}{1} \PY{o}{\PYZhy{}} \PY{n}{mask}\PY{p}{[}\PY{l+s+s1}{\PYZsq{}}\PY{l+s+s1}{frozen}\PY{l+s+s1}{\PYZsq{}}\PY{p}{]} \PY{o}{\PYZhy{}} \PY{n}{mask}\PY{p}{[}\PY{l+s+s1}{\PYZsq{}}\PY{l+s+s1}{active}\PY{l+s+s1}{\PYZsq{}}\PY{p}{]}
    \PY{k}{return} \PY{n}{mask}

\PY{c+c1}{\PYZsh{} For example}
\PY{n}{\PYZus{}test\PYZus{}plaq\PYZus{}masks} \PY{o}{=} \PY{n}{make\PYZus{}plaq\PYZus{}masks}\PY{p}{(}\PY{p}{(}\PY{l+m+mi}{8}\PY{p}{,}\PY{l+m+mi}{8}\PY{p}{)}\PY{p}{,} \PY{l+m+mi}{0}\PY{p}{,} \PY{n}{mask\PYZus{}off}\PY{o}{=}\PY{l+m+mi}{1}\PY{p}{)}
\PY{n+nb}{print}\PY{p}{(}\PY{l+s+s1}{\PYZsq{}}\PY{l+s+s1}{Frozen (fed into NNs)}\PY{l+s+s1}{\PYZsq{}}\PY{p}{)}
\PY{n+nb}{print}\PY{p}{(}\PY{n}{\PYZus{}test\PYZus{}plaq\PYZus{}masks}\PY{p}{[}\PY{l+s+s1}{\PYZsq{}}\PY{l+s+s1}{frozen}\PY{l+s+s1}{\PYZsq{}}\PY{p}{]}\PY{p}{)}
\PY{n+nb}{print}\PY{p}{(}\PY{l+s+s1}{\PYZsq{}}\PY{l+s+s1}{Active (driving the link update)}\PY{l+s+s1}{\PYZsq{}}\PY{p}{)}
\PY{n+nb}{print}\PY{p}{(}\PY{n}{\PYZus{}test\PYZus{}plaq\PYZus{}masks}\PY{p}{[}\PY{l+s+s1}{\PYZsq{}}\PY{l+s+s1}{active}\PY{l+s+s1}{\PYZsq{}}\PY{p}{]}\PY{p}{)}
\PY{n+nb}{print}\PY{p}{(}\PY{l+s+s1}{\PYZsq{}}\PY{l+s+s1}{Passive (updated as a result of link update)}\PY{l+s+s1}{\PYZsq{}}\PY{p}{)}
\PY{n+nb}{print}\PY{p}{(}\PY{n}{\PYZus{}test\PYZus{}plaq\PYZus{}masks}\PY{p}{[}\PY{l+s+s1}{\PYZsq{}}\PY{l+s+s1}{passive}\PY{l+s+s1}{\PYZsq{}}\PY{p}{]}\PY{p}{)}
    \end{Verbatim}
  \end{tcolorbox}

{\color{gray}
    \begin{Verbatim}[commandchars=\\\{\},fontsize=\small]
>>> Frozen (fed into NNs)
... tensor([[0, 0, 1, 1, 0, 0, 1, 1],
...         [0, 0, 1, 1, 0, 0, 1, 1],
...         [0, 0, 1, 1, 0, 0, 1, 1],
...         [0, 0, 1, 1, 0, 0, 1, 1],
...         [0, 0, 1, 1, 0, 0, 1, 1],
...         [0, 0, 1, 1, 0, 0, 1, 1],
...         [0, 0, 1, 1, 0, 0, 1, 1],
...         [0, 0, 1, 1, 0, 0, 1, 1]], dtype=torch.uint8)
... Active (driving the link update)
... tensor([[0, 1, 0, 0, 0, 1, 0, 0],
...         [0, 1, 0, 0, 0, 1, 0, 0],
...         [0, 1, 0, 0, 0, 1, 0, 0],
...         [0, 1, 0, 0, 0, 1, 0, 0],
...         [0, 1, 0, 0, 0, 1, 0, 0],
...         [0, 1, 0, 0, 0, 1, 0, 0],
...         [0, 1, 0, 0, 0, 1, 0, 0],
...         [0, 1, 0, 0, 0, 1, 0, 0]], dtype=torch.uint8)
... Passive (updated as a result of link update)
... tensor([[1, 0, 0, 0, 1, 0, 0, 0],
...         [1, 0, 0, 0, 1, 0, 0, 0],
...         [1, 0, 0, 0, 1, 0, 0, 0],
...         [1, 0, 0, 0, 1, 0, 0, 0],
...         [1, 0, 0, 0, 1, 0, 0, 0],
...         [1, 0, 0, 0, 1, 0, 0, 0],
...         [1, 0, 0, 0, 1, 0, 0, 0],
...         [1, 0, 0, 0, 1, 0, 0, 0]], dtype=torch.uint8)
    \end{Verbatim}
}

  \hypertarget{flowing-plaquettes-gauge-invariantly}{%
\subsection{\texorpdfstring{\textbf{Flowing plaquettes gauge
invariantly}}{Flowing plaquettes gauge invariantly}}\label{flowing-plaquettes-gauge-invariantly}}

The inner coupling layer simply needs to define an expressive invertible
function on the plaquette angular variables. However, we cannot directly
use scaling or affine transformations because they do not map
\([0,2\pi]\) back into itself.

Here we implement a \textbf{non-compact projection} (NCP) transformation
introduced for angular variables in \cite{Rezende:2020hrd}. The
transform simply changes variables from \(x \in [0,2\pi]\) to
\(\tan(x/2) \in (-\infty, \infty)\) before applying a scaling
transformation, then transforms back to \([0,2\pi]\) afterwards, i.e.:
\begin{equation}
x' = 2 \tan^{-1} \left( e^{s} \tan(x/2) \right)
\end{equation} where all operations are elementwise as usual. The
Jacobian factor for this transformation is \begin{equation}
J(x) = \left[ e^{-s} \cos^2 \left(\frac{x}{2}\right)  + e^s \sin^2 \left(\frac{x}{2}\right) \right]^{-1}.
\end{equation} We define these transformations below.

  \begin{tcolorbox}[breakable, size=fbox, boxrule=1pt, pad at break*=1mm,colback=cellbackground, colframe=cellborder]
    \begin{Verbatim}[commandchars=\\\{\},fontsize=\small]
\PY{k}{def} \PY{n+nf}{tan\PYZus{}transform}\PY{p}{(}\PY{n}{x}\PY{p}{,} \PY{n}{s}\PY{p}{)}\PY{p}{:}
    \PY{k}{return} \PY{n}{torch\PYZus{}mod}\PY{p}{(}\PY{l+m+mi}{2}\PY{o}{*}\PY{n}{torch}\PY{o}{.}\PY{n}{atan}\PY{p}{(}\PY{n}{torch}\PY{o}{.}\PY{n}{exp}\PY{p}{(}\PY{n}{s}\PY{p}{)}\PY{o}{*}\PY{n}{torch}\PY{o}{.}\PY{n}{tan}\PY{p}{(}\PY{n}{x}\PY{o}{/}\PY{l+m+mi}{2}\PY{p}{)}\PY{p}{)}\PY{p}{)}

\PY{k}{def} \PY{n+nf}{tan\PYZus{}transform\PYZus{}logJ}\PY{p}{(}\PY{n}{x}\PY{p}{,} \PY{n}{s}\PY{p}{)}\PY{p}{:}
    \PY{k}{return} \PY{o}{\PYZhy{}}\PY{n}{torch}\PY{o}{.}\PY{n}{log}\PY{p}{(}\PY{n}{torch}\PY{o}{.}\PY{n}{exp}\PY{p}{(}\PY{o}{\PYZhy{}}\PY{n}{s}\PY{p}{)}\PY{o}{*}\PY{n}{torch}\PY{o}{.}\PY{n}{cos}\PY{p}{(}\PY{n}{x}\PY{o}{/}\PY{l+m+mi}{2}\PY{p}{)}\PY{o}{*}\PY{o}{*}\PY{l+m+mi}{2} \PY{o}{+} \PY{n}{torch}\PY{o}{.}\PY{n}{exp}\PY{p}{(}\PY{n}{s}\PY{p}{)}\PY{o}{*}\PY{n}{torch}\PY{o}{.}\PY{n}{sin}\PY{p}{(}\PY{n}{x}\PY{o}{/}\PY{l+m+mi}{2}\PY{p}{)}\PY{o}{*}\PY{o}{*}\PY{l+m+mi}{2}\PY{p}{)}
    \end{Verbatim}
  \end{tcolorbox}

  The average of transformations with different scales \(s_i\) also
defines an invertible transform. This lets use make our models more
expressive, at the cost of needing numerical methods to invert each
coupling layer.

  \begin{tcolorbox}[breakable, size=fbox, boxrule=1pt, pad at break*=1mm,colback=cellbackground, colframe=cellborder]
    \begin{Verbatim}[commandchars=\\\{\},fontsize=\small]
\PY{k}{def} \PY{n+nf}{mixture\PYZus{}tan\PYZus{}transform}\PY{p}{(}\PY{n}{x}\PY{p}{,} \PY{n}{s}\PY{p}{)}\PY{p}{:}
    \PY{k}{assert} \PY{n+nb}{len}\PY{p}{(}\PY{n}{x}\PY{o}{.}\PY{n}{shape}\PY{p}{)} \PY{o}{==} \PY{n+nb}{len}\PY{p}{(}\PY{n}{s}\PY{o}{.}\PY{n}{shape}\PY{p}{)}\PY{p}{,} \PYZbs{}
        \PY{l+s+sa}{f}\PY{l+s+s1}{\PYZsq{}}\PY{l+s+s1}{Dimension mismatch between x and s }\PY{l+s+si}{\PYZob{}}\PY{n}{x}\PY{o}{.}\PY{n}{shape}\PY{l+s+si}{\PYZcb{}}\PY{l+s+s1}{ vs }\PY{l+s+si}{\PYZob{}}\PY{n}{s}\PY{o}{.}\PY{n}{shape}\PY{l+s+si}{\PYZcb{}}\PY{l+s+s1}{\PYZsq{}}
    \PY{k}{return} \PY{n}{torch}\PY{o}{.}\PY{n}{mean}\PY{p}{(}\PY{n}{tan\PYZus{}transform}\PY{p}{(}\PY{n}{x}\PY{p}{,} \PY{n}{s}\PY{p}{)}\PY{p}{,} \PY{n}{dim}\PY{o}{=}\PY{l+m+mi}{1}\PY{p}{,} \PY{n}{keepdim}\PY{o}{=}\PY{k+kc}{True}\PY{p}{)}

\PY{k}{def} \PY{n+nf}{mixture\PYZus{}tan\PYZus{}transform\PYZus{}logJ}\PY{p}{(}\PY{n}{x}\PY{p}{,} \PY{n}{s}\PY{p}{)}\PY{p}{:}
    \PY{k}{assert} \PY{n+nb}{len}\PY{p}{(}\PY{n}{x}\PY{o}{.}\PY{n}{shape}\PY{p}{)} \PY{o}{==} \PY{n+nb}{len}\PY{p}{(}\PY{n}{s}\PY{o}{.}\PY{n}{shape}\PY{p}{)}\PY{p}{,} \PYZbs{}
        \PY{l+s+sa}{f}\PY{l+s+s1}{\PYZsq{}}\PY{l+s+s1}{Dimension mismatch between x and s }\PY{l+s+si}{\PYZob{}}\PY{n}{x}\PY{o}{.}\PY{n}{shape}\PY{l+s+si}{\PYZcb{}}\PY{l+s+s1}{ vs }\PY{l+s+si}{\PYZob{}}\PY{n}{s}\PY{o}{.}\PY{n}{shape}\PY{l+s+si}{\PYZcb{}}\PY{l+s+s1}{\PYZsq{}}
    \PY{k}{return} \PY{n}{torch}\PY{o}{.}\PY{n}{logsumexp}\PY{p}{(}\PY{n}{tan\PYZus{}transform\PYZus{}logJ}\PY{p}{(}\PY{n}{x}\PY{p}{,} \PY{n}{s}\PY{p}{)}\PY{p}{,} \PY{n}{dim}\PY{o}{=}\PY{l+m+mi}{1}\PY{p}{)} \PY{o}{\PYZhy{}} \PY{n}{np}\PY{o}{.}\PY{n}{log}\PY{p}{(}\PY{n}{s}\PY{o}{.}\PY{n}{shape}\PY{p}{[}\PY{l+m+mi}{1}\PY{p}{]}\PY{p}{)}
    \end{Verbatim}
  \end{tcolorbox}

  Unfortunately, the NCP transformation does not have an analytic inverse
transformation but it is easy to calculate it numerically. It is worth
noting that inverse transformation is required only to measure the model
density on new data (not used for training or evaluation), so slow
numerical inversion is not an issue. There are alternative coupling
layers that avoid this issue, if such measurements are needed.

We implement numerical inversion using the bisection algorithm below.

  \begin{tcolorbox}[breakable, size=fbox, boxrule=1pt, pad at break*=1mm,colback=cellbackground, colframe=cellborder]
    \begin{Verbatim}[commandchars=\\\{\},fontsize=\small]
\PY{k}{def} \PY{n+nf}{invert\PYZus{}transform\PYZus{}bisect}\PY{p}{(}\PY{n}{y}\PY{p}{,} \PY{o}{*}\PY{p}{,} \PY{n}{f}\PY{p}{,} \PY{n}{tol}\PY{p}{,} \PY{n}{max\PYZus{}iter}\PY{p}{,} \PY{n}{a}\PY{o}{=}\PY{l+m+mi}{0}\PY{p}{,} \PY{n}{b}\PY{o}{=}\PY{l+m+mi}{2}\PY{o}{*}\PY{n}{np}\PY{o}{.}\PY{n}{pi}\PY{p}{)}\PY{p}{:}
    \PY{n}{min\PYZus{}x} \PY{o}{=} \PY{n}{a}\PY{o}{*}\PY{n}{torch}\PY{o}{.}\PY{n}{ones\PYZus{}like}\PY{p}{(}\PY{n}{y}\PY{p}{)}
    \PY{n}{max\PYZus{}x} \PY{o}{=} \PY{n}{b}\PY{o}{*}\PY{n}{torch}\PY{o}{.}\PY{n}{ones\PYZus{}like}\PY{p}{(}\PY{n}{y}\PY{p}{)}
    \PY{n}{min\PYZus{}val} \PY{o}{=} \PY{n}{f}\PY{p}{(}\PY{n}{min\PYZus{}x}\PY{p}{)}
    \PY{n}{max\PYZus{}val} \PY{o}{=} \PY{n}{f}\PY{p}{(}\PY{n}{max\PYZus{}x}\PY{p}{)}
    \PY{k}{with} \PY{n}{torch}\PY{o}{.}\PY{n}{no\PYZus{}grad}\PY{p}{(}\PY{p}{)}\PY{p}{:}
        \PY{k}{for} \PY{n}{i} \PY{o+ow}{in} \PY{n+nb}{range}\PY{p}{(}\PY{n}{max\PYZus{}iter}\PY{p}{)}\PY{p}{:}
            \PY{n}{mid\PYZus{}x} \PY{o}{=} \PY{p}{(}\PY{n}{min\PYZus{}x} \PY{o}{+} \PY{n}{max\PYZus{}x}\PY{p}{)} \PY{o}{/} \PY{l+m+mi}{2}
            \PY{n}{mid\PYZus{}val} \PY{o}{=} \PY{n}{f}\PY{p}{(}\PY{n}{mid\PYZus{}x}\PY{p}{)}
            \PY{n}{greater\PYZus{}mask} \PY{o}{=} \PY{p}{(}\PY{n}{y} \PY{o}{\PYZgt{}} \PY{n}{mid\PYZus{}val}\PY{p}{)}\PY{o}{.}\PY{n}{int}\PY{p}{(}\PY{p}{)}
            \PY{n}{greater\PYZus{}mask} \PY{o}{=} \PY{n}{greater\PYZus{}mask}\PY{o}{.}\PY{n}{float}\PY{p}{(}\PY{p}{)}
            \PY{n}{err} \PY{o}{=} \PY{n}{torch}\PY{o}{.}\PY{n}{max}\PY{p}{(}\PY{n}{torch}\PY{o}{.}\PY{n}{abs}\PY{p}{(}\PY{n}{y} \PY{o}{\PYZhy{}} \PY{n}{mid\PYZus{}val}\PY{p}{)}\PY{p}{)}
            \PY{k}{if} \PY{n}{err} \PY{o}{\PYZlt{}} \PY{n}{tol}\PY{p}{:} \PY{k}{return} \PY{n}{mid\PYZus{}x}
            \PY{k}{if} \PY{n}{torch}\PY{o}{.}\PY{n}{all}\PY{p}{(}\PY{p}{(}\PY{n}{mid\PYZus{}x} \PY{o}{==} \PY{n}{min\PYZus{}x}\PY{p}{)} \PY{o}{+} \PY{p}{(}\PY{n}{mid\PYZus{}x} \PY{o}{==} \PY{n}{max\PYZus{}x}\PY{p}{)}\PY{p}{)}\PY{p}{:}
                \PY{n+nb}{print}\PY{p}{(}\PY{l+s+s1}{\PYZsq{}}\PY{l+s+s1}{WARNING: Reached floating point precision before tolerance }\PY{l+s+s1}{\PYZsq{}}
                      \PY{l+s+sa}{f}\PY{l+s+s1}{\PYZsq{}}\PY{l+s+s1}{(iter }\PY{l+s+si}{\PYZob{}}\PY{n}{i}\PY{l+s+si}{\PYZcb{}}\PY{l+s+s1}{, err }\PY{l+s+si}{\PYZob{}}\PY{n}{err}\PY{l+s+si}{\PYZcb{}}\PY{l+s+s1}{)}\PY{l+s+s1}{\PYZsq{}}\PY{p}{)}
                \PY{k}{return} \PY{n}{mid\PYZus{}x}
            \PY{n}{min\PYZus{}x} \PY{o}{=} \PY{n}{greater\PYZus{}mask}\PY{o}{*}\PY{n}{mid\PYZus{}x} \PY{o}{+} \PY{p}{(}\PY{l+m+mi}{1}\PY{o}{\PYZhy{}}\PY{n}{greater\PYZus{}mask}\PY{p}{)}\PY{o}{*}\PY{n}{min\PYZus{}x}
            \PY{n}{min\PYZus{}val} \PY{o}{=} \PY{n}{greater\PYZus{}mask}\PY{o}{*}\PY{n}{mid\PYZus{}val} \PY{o}{+} \PY{p}{(}\PY{l+m+mi}{1}\PY{o}{\PYZhy{}}\PY{n}{greater\PYZus{}mask}\PY{p}{)}\PY{o}{*}\PY{n}{min\PYZus{}val}
            \PY{n}{max\PYZus{}x} \PY{o}{=} \PY{p}{(}\PY{l+m+mi}{1}\PY{o}{\PYZhy{}}\PY{n}{greater\PYZus{}mask}\PY{p}{)}\PY{o}{*}\PY{n}{mid\PYZus{}x} \PY{o}{+} \PY{n}{greater\PYZus{}mask}\PY{o}{*}\PY{n}{max\PYZus{}x}
            \PY{n}{max\PYZus{}val} \PY{o}{=} \PY{p}{(}\PY{l+m+mi}{1}\PY{o}{\PYZhy{}}\PY{n}{greater\PYZus{}mask}\PY{p}{)}\PY{o}{*}\PY{n}{mid\PYZus{}val} \PY{o}{+} \PY{n}{greater\PYZus{}mask}\PY{o}{*}\PY{n}{max\PYZus{}val}
        \PY{n+nb}{print}\PY{p}{(}\PY{l+s+sa}{f}\PY{l+s+s1}{\PYZsq{}}\PY{l+s+s1}{WARNING: Did not converge to tol }\PY{l+s+si}{\PYZob{}}\PY{n}{tol}\PY{l+s+si}{\PYZcb{}}\PY{l+s+s1}{ in }\PY{l+s+si}{\PYZob{}}\PY{n}{max\PYZus{}iter}\PY{l+s+si}{\PYZcb{}}\PY{l+s+s1}{ iters! Error was }\PY{l+s+si}{\PYZob{}}\PY{n}{err}\PY{l+s+si}{\PYZcb{}}\PY{l+s+s1}{\PYZsq{}}\PY{p}{)}
        \PY{k}{return} \PY{n}{mid\PYZus{}x}
    \end{Verbatim}
  \end{tcolorbox}

  As before, we'll use neural nets to parametrize the scales \(s_i\). We
will preprocess the input angles (of the frozen plaquettes) as
\(x \rightarrow (\sin(x), \cos(x))\) to ensure the neural nets have
continous outputs with respect to the angular inputs.

  \begin{tcolorbox}[breakable, size=fbox, boxrule=1pt, pad at break*=1mm,colback=cellbackground, colframe=cellborder]
    \begin{Verbatim}[commandchars=\\\{\},fontsize=\small]
\PY{k}{def} \PY{n+nf}{stack\PYZus{}cos\PYZus{}sin}\PY{p}{(}\PY{n}{x}\PY{p}{)}\PY{p}{:}
    \PY{k}{return} \PY{n}{torch}\PY{o}{.}\PY{n}{stack}\PY{p}{(}\PY{p}{(}\PY{n}{torch}\PY{o}{.}\PY{n}{cos}\PY{p}{(}\PY{n}{x}\PY{p}{)}\PY{p}{,} \PY{n}{torch}\PY{o}{.}\PY{n}{sin}\PY{p}{(}\PY{n}{x}\PY{p}{)}\PY{p}{)}\PY{p}{,} \PY{n}{dim}\PY{o}{=}\PY{l+m+mi}{1}\PY{p}{)}
    \end{Verbatim}
  \end{tcolorbox}

  Altogether, the coupling layer uses the average of NCP transforms as
defined above, composed with simple offsets
\(x \rightarrow x + t \pmod{2\pi}\). As discussed above, we partition
the plaquettes into three sets: active, passive, and frozen, with the
exact partitioning scheme defined by our choice of masking pattern. The
inner coupling layer updates the active plaquettes, uses the frozen
plaquettes as inputs for the neural nets defining \(s_i\) and \(t\), and
ignores the passive plaquettes completely.

  \begin{tcolorbox}[breakable, size=fbox, boxrule=1pt, pad at break*=1mm,colback=cellbackground, colframe=cellborder]
    \begin{Verbatim}[commandchars=\\\{\},fontsize=\small]
\PY{k}{class} \PY{n+nc}{NCPPlaqCouplingLayer}\PY{p}{(}\PY{n}{torch}\PY{o}{.}\PY{n}{nn}\PY{o}{.}\PY{n}{Module}\PY{p}{)}\PY{p}{:}
    \PY{k}{def} \PY{n+nf+fm}{\PYZus{}\PYZus{}init\PYZus{}\PYZus{}}\PY{p}{(}\PY{n+nb+bp}{self}\PY{p}{,} \PY{n}{net}\PY{p}{,} \PY{o}{*}\PY{p}{,} \PY{n}{mask\PYZus{}shape}\PY{p}{,} \PY{n}{mask\PYZus{}mu}\PY{p}{,} \PY{n}{mask\PYZus{}off}\PY{p}{,}
                 \PY{n}{inv\PYZus{}prec}\PY{o}{=}\PY{l+m+mf}{1e\PYZhy{}6}\PY{p}{,} \PY{n}{inv\PYZus{}max\PYZus{}iter}\PY{o}{=}\PY{l+m+mi}{1000}\PY{p}{)}\PY{p}{:}
        \PY{n+nb}{super}\PY{p}{(}\PY{p}{)}\PY{o}{.}\PY{n+nf+fm}{\PYZus{}\PYZus{}init\PYZus{}\PYZus{}}\PY{p}{(}\PY{p}{)}
        \PY{k}{assert} \PY{n+nb}{len}\PY{p}{(}\PY{n}{mask\PYZus{}shape}\PY{p}{)} \PY{o}{==} \PY{l+m+mi}{2}\PY{p}{,} \PY{p}{(}
            \PY{l+s+sa}{f}\PY{l+s+s1}{\PYZsq{}}\PY{l+s+s1}{NCPPlaqCouplingLayer is implemented only in 2D, }\PY{l+s+s1}{\PYZsq{}}
            \PY{l+s+sa}{f}\PY{l+s+s1}{\PYZsq{}}\PY{l+s+s1}{mask shape }\PY{l+s+si}{\PYZob{}}\PY{n}{mask\PYZus{}shape}\PY{l+s+si}{\PYZcb{}}\PY{l+s+s1}{ is invalid}\PY{l+s+s1}{\PYZsq{}}\PY{p}{)}
        \PY{n+nb+bp}{self}\PY{o}{.}\PY{n}{mask} \PY{o}{=} \PY{n}{make\PYZus{}plaq\PYZus{}masks}\PY{p}{(}\PY{n}{mask\PYZus{}shape}\PY{p}{,} \PY{n}{mask\PYZus{}mu}\PY{p}{,} \PY{n}{mask\PYZus{}off}\PY{p}{)}
        \PY{n+nb+bp}{self}\PY{o}{.}\PY{n}{net} \PY{o}{=} \PY{n}{net}
        \PY{n+nb+bp}{self}\PY{o}{.}\PY{n}{inv\PYZus{}prec} \PY{o}{=} \PY{n}{inv\PYZus{}prec}
        \PY{n+nb+bp}{self}\PY{o}{.}\PY{n}{inv\PYZus{}max\PYZus{}iter} \PY{o}{=} \PY{n}{inv\PYZus{}max\PYZus{}iter}

    \PY{k}{def} \PY{n+nf}{forward}\PY{p}{(}\PY{n+nb+bp}{self}\PY{p}{,} \PY{n}{x}\PY{p}{)}\PY{p}{:}
        \PY{n}{x2} \PY{o}{=} \PY{n+nb+bp}{self}\PY{o}{.}\PY{n}{mask}\PY{p}{[}\PY{l+s+s1}{\PYZsq{}}\PY{l+s+s1}{frozen}\PY{l+s+s1}{\PYZsq{}}\PY{p}{]} \PY{o}{*} \PY{n}{x}
        \PY{n}{net\PYZus{}out} \PY{o}{=} \PY{n+nb+bp}{self}\PY{o}{.}\PY{n}{net}\PY{p}{(}\PY{n}{stack\PYZus{}cos\PYZus{}sin}\PY{p}{(}\PY{n}{x2}\PY{p}{)}\PY{p}{)}
        \PY{k}{assert} \PY{n}{net\PYZus{}out}\PY{o}{.}\PY{n}{shape}\PY{p}{[}\PY{l+m+mi}{1}\PY{p}{]} \PY{o}{\PYZgt{}}\PY{o}{=} \PY{l+m+mi}{2}\PY{p}{,} \PY{l+s+s1}{\PYZsq{}}\PY{l+s+s1}{CNN must output n\PYZus{}mix (s\PYZus{}i) + 1 (t) channels}\PY{l+s+s1}{\PYZsq{}}
        \PY{n}{s}\PY{p}{,} \PY{n}{t} \PY{o}{=} \PY{n}{net\PYZus{}out}\PY{p}{[}\PY{p}{:}\PY{p}{,}\PY{p}{:}\PY{o}{\PYZhy{}}\PY{l+m+mi}{1}\PY{p}{]}\PY{p}{,} \PY{n}{net\PYZus{}out}\PY{p}{[}\PY{p}{:}\PY{p}{,}\PY{o}{\PYZhy{}}\PY{l+m+mi}{1}\PY{p}{]}

        \PY{n}{x1} \PY{o}{=} \PY{n+nb+bp}{self}\PY{o}{.}\PY{n}{mask}\PY{p}{[}\PY{l+s+s1}{\PYZsq{}}\PY{l+s+s1}{active}\PY{l+s+s1}{\PYZsq{}}\PY{p}{]} \PY{o}{*} \PY{n}{x}
        \PY{n}{x1} \PY{o}{=} \PY{n}{x1}\PY{o}{.}\PY{n}{unsqueeze}\PY{p}{(}\PY{l+m+mi}{1}\PY{p}{)}
        \PY{n}{local\PYZus{}logJ} \PY{o}{=} \PY{n+nb+bp}{self}\PY{o}{.}\PY{n}{mask}\PY{p}{[}\PY{l+s+s1}{\PYZsq{}}\PY{l+s+s1}{active}\PY{l+s+s1}{\PYZsq{}}\PY{p}{]} \PY{o}{*} \PY{n}{mixture\PYZus{}tan\PYZus{}transform\PYZus{}logJ}\PY{p}{(}\PY{n}{x1}\PY{p}{,} \PY{n}{s}\PY{p}{)}
        \PY{n}{axes} \PY{o}{=} \PY{n+nb}{tuple}\PY{p}{(}\PY{n+nb}{range}\PY{p}{(}\PY{l+m+mi}{1}\PY{p}{,} \PY{n+nb}{len}\PY{p}{(}\PY{n}{local\PYZus{}logJ}\PY{o}{.}\PY{n}{shape}\PY{p}{)}\PY{p}{)}\PY{p}{)}
        \PY{n}{logJ} \PY{o}{=} \PY{n}{torch}\PY{o}{.}\PY{n}{sum}\PY{p}{(}\PY{n}{local\PYZus{}logJ}\PY{p}{,} \PY{n}{dim}\PY{o}{=}\PY{n}{axes}\PY{p}{)}
        \PY{n}{fx1} \PY{o}{=} \PY{n+nb+bp}{self}\PY{o}{.}\PY{n}{mask}\PY{p}{[}\PY{l+s+s1}{\PYZsq{}}\PY{l+s+s1}{active}\PY{l+s+s1}{\PYZsq{}}\PY{p}{]} \PY{o}{*} \PY{n}{mixture\PYZus{}tan\PYZus{}transform}\PY{p}{(}\PY{n}{x1}\PY{p}{,} \PY{n}{s}\PY{p}{)}\PY{o}{.}\PY{n}{squeeze}\PY{p}{(}\PY{l+m+mi}{1}\PY{p}{)}

        \PY{n}{fx} \PY{o}{=} \PY{p}{(}
            \PY{n+nb+bp}{self}\PY{o}{.}\PY{n}{mask}\PY{p}{[}\PY{l+s+s1}{\PYZsq{}}\PY{l+s+s1}{active}\PY{l+s+s1}{\PYZsq{}}\PY{p}{]} \PY{o}{*} \PY{n}{torch\PYZus{}mod}\PY{p}{(}\PY{n}{fx1} \PY{o}{+} \PY{n}{t}\PY{p}{)} \PY{o}{+}
            \PY{n+nb+bp}{self}\PY{o}{.}\PY{n}{mask}\PY{p}{[}\PY{l+s+s1}{\PYZsq{}}\PY{l+s+s1}{passive}\PY{l+s+s1}{\PYZsq{}}\PY{p}{]} \PY{o}{*} \PY{n}{x} \PY{o}{+}
            \PY{n+nb+bp}{self}\PY{o}{.}\PY{n}{mask}\PY{p}{[}\PY{l+s+s1}{\PYZsq{}}\PY{l+s+s1}{frozen}\PY{l+s+s1}{\PYZsq{}}\PY{p}{]} \PY{o}{*} \PY{n}{x}\PY{p}{)}
        \PY{k}{return} \PY{n}{fx}\PY{p}{,} \PY{n}{logJ}
        
    \PY{k}{def} \PY{n+nf}{reverse}\PY{p}{(}\PY{n+nb+bp}{self}\PY{p}{,} \PY{n}{fx}\PY{p}{)}\PY{p}{:}
        \PY{n}{fx2} \PY{o}{=} \PY{n+nb+bp}{self}\PY{o}{.}\PY{n}{mask}\PY{p}{[}\PY{l+s+s1}{\PYZsq{}}\PY{l+s+s1}{frozen}\PY{l+s+s1}{\PYZsq{}}\PY{p}{]} \PY{o}{*} \PY{n}{fx}
        \PY{n}{net\PYZus{}out} \PY{o}{=} \PY{n+nb+bp}{self}\PY{o}{.}\PY{n}{net}\PY{p}{(}\PY{n}{stack\PYZus{}cos\PYZus{}sin}\PY{p}{(}\PY{n}{fx2}\PY{p}{)}\PY{p}{)}
        \PY{k}{assert} \PY{n}{net\PYZus{}out}\PY{o}{.}\PY{n}{shape}\PY{p}{[}\PY{l+m+mi}{1}\PY{p}{]} \PY{o}{\PYZgt{}}\PY{o}{=} \PY{l+m+mi}{2}\PY{p}{,} \PY{l+s+s1}{\PYZsq{}}\PY{l+s+s1}{CNN must output n\PYZus{}mix (s\PYZus{}i) + 1 (t) channels}\PY{l+s+s1}{\PYZsq{}}
        \PY{n}{s}\PY{p}{,} \PY{n}{t} \PY{o}{=} \PY{n}{net\PYZus{}out}\PY{p}{[}\PY{p}{:}\PY{p}{,}\PY{p}{:}\PY{o}{\PYZhy{}}\PY{l+m+mi}{1}\PY{p}{]}\PY{p}{,} \PY{n}{net\PYZus{}out}\PY{p}{[}\PY{p}{:}\PY{p}{,}\PY{o}{\PYZhy{}}\PY{l+m+mi}{1}\PY{p}{]}

        \PY{n}{x1} \PY{o}{=} \PY{n}{torch\PYZus{}mod}\PY{p}{(}\PY{n+nb+bp}{self}\PY{o}{.}\PY{n}{mask}\PY{p}{[}\PY{l+s+s1}{\PYZsq{}}\PY{l+s+s1}{active}\PY{l+s+s1}{\PYZsq{}}\PY{p}{]} \PY{o}{*} \PY{p}{(}\PY{n}{fx} \PY{o}{\PYZhy{}} \PY{n}{t}\PY{p}{)}\PY{o}{.}\PY{n}{unsqueeze}\PY{p}{(}\PY{l+m+mi}{1}\PY{p}{)}\PY{p}{)}
        \PY{n}{transform} \PY{o}{=} \PY{k}{lambda} \PY{n}{x}\PY{p}{:} \PY{n+nb+bp}{self}\PY{o}{.}\PY{n}{mask}\PY{p}{[}\PY{l+s+s1}{\PYZsq{}}\PY{l+s+s1}{active}\PY{l+s+s1}{\PYZsq{}}\PY{p}{]} \PY{o}{*} \PY{n}{mixture\PYZus{}tan\PYZus{}transform}\PY{p}{(}\PY{n}{x}\PY{p}{,} \PY{n}{s}\PY{p}{)}
        \PY{n}{x1} \PY{o}{=} \PY{n}{invert\PYZus{}transform\PYZus{}bisect}\PY{p}{(}
            \PY{n}{x1}\PY{p}{,} \PY{n}{f}\PY{o}{=}\PY{n}{transform}\PY{p}{,} \PY{n}{tol}\PY{o}{=}\PY{n+nb+bp}{self}\PY{o}{.}\PY{n}{inv\PYZus{}prec}\PY{p}{,} \PY{n}{max\PYZus{}iter}\PY{o}{=}\PY{n+nb+bp}{self}\PY{o}{.}\PY{n}{inv\PYZus{}max\PYZus{}iter}\PY{p}{)}
        \PY{n}{local\PYZus{}logJ} \PY{o}{=} \PY{n+nb+bp}{self}\PY{o}{.}\PY{n}{mask}\PY{p}{[}\PY{l+s+s1}{\PYZsq{}}\PY{l+s+s1}{active}\PY{l+s+s1}{\PYZsq{}}\PY{p}{]} \PY{o}{*} \PY{n}{mixture\PYZus{}tan\PYZus{}transform\PYZus{}logJ}\PY{p}{(}\PY{n}{x1}\PY{p}{,} \PY{n}{s}\PY{p}{)}
        \PY{n}{axes} \PY{o}{=} \PY{n+nb}{tuple}\PY{p}{(}\PY{n+nb}{range}\PY{p}{(}\PY{l+m+mi}{1}\PY{p}{,} \PY{n+nb}{len}\PY{p}{(}\PY{n}{local\PYZus{}logJ}\PY{o}{.}\PY{n}{shape}\PY{p}{)}\PY{p}{)}\PY{p}{)}
        \PY{n}{logJ} \PY{o}{=} \PY{o}{\PYZhy{}}\PY{n}{torch}\PY{o}{.}\PY{n}{sum}\PY{p}{(}\PY{n}{local\PYZus{}logJ}\PY{p}{,} \PY{n}{dim}\PY{o}{=}\PY{n}{axes}\PY{p}{)}
        \PY{n}{x1} \PY{o}{=} \PY{n}{x1}\PY{o}{.}\PY{n}{squeeze}\PY{p}{(}\PY{l+m+mi}{1}\PY{p}{)}

        \PY{n}{x} \PY{o}{=} \PY{p}{(}
            \PY{n+nb+bp}{self}\PY{o}{.}\PY{n}{mask}\PY{p}{[}\PY{l+s+s1}{\PYZsq{}}\PY{l+s+s1}{active}\PY{l+s+s1}{\PYZsq{}}\PY{p}{]} \PY{o}{*} \PY{n}{x1} \PY{o}{+}
            \PY{n+nb+bp}{self}\PY{o}{.}\PY{n}{mask}\PY{p}{[}\PY{l+s+s1}{\PYZsq{}}\PY{l+s+s1}{passive}\PY{l+s+s1}{\PYZsq{}}\PY{p}{]} \PY{o}{*} \PY{n}{fx} \PY{o}{+}
            \PY{n+nb+bp}{self}\PY{o}{.}\PY{n}{mask}\PY{p}{[}\PY{l+s+s1}{\PYZsq{}}\PY{l+s+s1}{frozen}\PY{l+s+s1}{\PYZsq{}}\PY{p}{]} \PY{o}{*} \PY{n}{fx2}\PY{p}{)}
        \PY{k}{return} \PY{n}{x}\PY{p}{,} \PY{n}{logJ}
    \end{Verbatim}
  \end{tcolorbox}

  \hypertarget{assemble-the-model}{%
\subsection{\texorpdfstring{\textbf{Assemble the
model}}{Assemble the model}}\label{assemble-the-model}}

Finally, we'll use CNNs for this application as well. To summarize, each
coupling layer:

\begin{itemize}
\tightlist
\item
  Computes plaquettes from the gauge links, and partitions them into
  active, passive, and frozen subsets
\item
  Provides the frozen plaquette angles \(x\) as inputs
  \((\cos x, \sin x)\) to a CNN
\item
  Uses the resulting scales \(s_i\) and offset \(t\) to update the
  active plaquettes with mixed NCP
\item
  Updates the active links to induce the transformation of the active
  plaquettes (updating the passive plaquettes as a side effect)
\end{itemize}

The flow as a whole is made by stacking coupling layers, repeatedly
scanning the masking pattern across all four distinct offsets and both
directions (see the figure above). Eight layers are required to update
each link once.

  \begin{tcolorbox}[breakable, size=fbox, boxrule=1pt, pad at break*=1mm,colback=cellbackground, colframe=cellborder]
    \begin{Verbatim}[commandchars=\\\{\},fontsize=\small]
\PY{k}{def} \PY{n+nf}{make\PYZus{}u1\PYZus{}equiv\PYZus{}layers}\PY{p}{(}\PY{o}{*}\PY{p}{,} \PY{n}{n\PYZus{}layers}\PY{p}{,} \PY{n}{n\PYZus{}mixture\PYZus{}comps}\PY{p}{,} \PY{n}{lattice\PYZus{}shape}\PY{p}{,} \PY{n}{hidden\PYZus{}sizes}\PY{p}{,} \PY{n}{kernel\PYZus{}size}\PY{p}{)}\PY{p}{:}
    \PY{n}{layers} \PY{o}{=} \PY{p}{[}\PY{p}{]}
    \PY{k}{for} \PY{n}{i} \PY{o+ow}{in} \PY{n+nb}{range}\PY{p}{(}\PY{n}{n\PYZus{}layers}\PY{p}{)}\PY{p}{:}
        \PY{c+c1}{\PYZsh{} periodically loop through all arrangements of maskings}
        \PY{n}{mu} \PY{o}{=} \PY{n}{i} \PY{o}{\PYZpc{}} \PY{l+m+mi}{2}
        \PY{n}{off} \PY{o}{=} \PY{p}{(}\PY{n}{i}\PY{o}{/}\PY{o}{/}\PY{l+m+mi}{2}\PY{p}{)} \PY{o}{\PYZpc{}} \PY{l+m+mi}{4}
        \PY{n}{in\PYZus{}channels} \PY{o}{=} \PY{l+m+mi}{2} \PY{c+c1}{\PYZsh{} x \PYZhy{} \PYZgt{} (cos(x), sin(x))}
        \PY{n}{out\PYZus{}channels} \PY{o}{=} \PY{n}{n\PYZus{}mixture\PYZus{}comps} \PY{o}{+} \PY{l+m+mi}{1} \PY{c+c1}{\PYZsh{} for mixture s and t, respectively}
        \PY{n}{net} \PY{o}{=} \PY{n}{make\PYZus{}conv\PYZus{}net}\PY{p}{(}\PY{n}{in\PYZus{}channels}\PY{o}{=}\PY{n}{in\PYZus{}channels}\PY{p}{,} \PY{n}{out\PYZus{}channels}\PY{o}{=}\PY{n}{out\PYZus{}channels}\PY{p}{,}
            \PY{n}{hidden\PYZus{}sizes}\PY{o}{=}\PY{n}{hidden\PYZus{}sizes}\PY{p}{,} \PY{n}{kernel\PYZus{}size}\PY{o}{=}\PY{n}{kernel\PYZus{}size}\PY{p}{,}
            \PY{n}{use\PYZus{}final\PYZus{}tanh}\PY{o}{=}\PY{k+kc}{False}\PY{p}{)}
        \PY{n}{plaq\PYZus{}coupling} \PY{o}{=} \PY{n}{NCPPlaqCouplingLayer}\PY{p}{(}
            \PY{n}{net}\PY{p}{,} \PY{n}{mask\PYZus{}shape}\PY{o}{=}\PY{n}{lattice\PYZus{}shape}\PY{p}{,} \PY{n}{mask\PYZus{}mu}\PY{o}{=}\PY{n}{mu}\PY{p}{,} \PY{n}{mask\PYZus{}off}\PY{o}{=}\PY{n}{off}\PY{p}{)}
        \PY{n}{link\PYZus{}coupling} \PY{o}{=} \PY{n}{GaugeEquivCouplingLayer}\PY{p}{(}
            \PY{n}{lattice\PYZus{}shape}\PY{o}{=}\PY{n}{lattice\PYZus{}shape}\PY{p}{,} \PY{n}{mask\PYZus{}mu}\PY{o}{=}\PY{n}{mu}\PY{p}{,} \PY{n}{mask\PYZus{}off}\PY{o}{=}\PY{n}{off}\PY{p}{,} 
            \PY{n}{plaq\PYZus{}coupling}\PY{o}{=}\PY{n}{plaq\PYZus{}coupling}\PY{p}{)}
        \PY{n}{layers}\PY{o}{.}\PY{n}{append}\PY{p}{(}\PY{n}{link\PYZus{}coupling}\PY{p}{)}
    \PY{k}{return} \PY{n}{torch}\PY{o}{.}\PY{n}{nn}\PY{o}{.}\PY{n}{ModuleList}\PY{p}{(}\PY{n}{layers}\PY{p}{)}
    \end{Verbatim}
  \end{tcolorbox}

  \hypertarget{train-the-model}{%
\subsection{\texorpdfstring{\textbf{Train the
model}}{Train the model}}\label{train-the-model}}

We use the same self-training scheme with the reverse KL divergence as
for \(\phi^4\) theory. You should find that this model trains much
faster than the model for \(\phi^4\) theory above, achieving
\(\sim 20\%\) ESS after 10 eras of training, which takes 8 minutes on a
Google Colab GPU.

As with before, if you don't want to train the model, we have provided a
pre-trained example; just set the flag below to use it.

  \begin{tcolorbox}[breakable, size=fbox, boxrule=1pt, pad at break*=1mm,colback=cellbackground, colframe=cellborder]
    \begin{Verbatim}[commandchars=\\\{\},fontsize=\small]
\PY{n}{use\PYZus{}pretrained} \PY{o}{=} \PY{k+kc}{True}
    \end{Verbatim}
  \end{tcolorbox}

  For convenience, this cell reproduces all of the setup code from above:

  \begin{tcolorbox}[breakable, size=fbox, boxrule=1pt, pad at break*=1mm,colback=cellbackground, colframe=cellborder]
    \begin{Verbatim}[commandchars=\\\{\},fontsize=\small]
\PY{c+c1}{\PYZsh{} Theory}
\PY{n}{L} \PY{o}{=} \PY{l+m+mi}{8}
\PY{n}{lattice\PYZus{}shape} \PY{o}{=} \PY{p}{(}\PY{n}{L}\PY{p}{,}\PY{n}{L}\PY{p}{)}
\PY{n}{link\PYZus{}shape} \PY{o}{=} \PY{p}{(}\PY{l+m+mi}{2}\PY{p}{,}\PY{n}{L}\PY{p}{,}\PY{n}{L}\PY{p}{)}
\PY{n}{beta} \PY{o}{=} \PY{l+m+mf}{2.0}
\PY{n}{u1\PYZus{}action} \PY{o}{=} \PY{n}{U1GaugeAction}\PY{p}{(}\PY{n}{beta}\PY{p}{)}

\PY{c+c1}{\PYZsh{} Model}
\PY{n}{prior} \PY{o}{=} \PY{n}{MultivariateUniform}\PY{p}{(}\PY{n}{torch}\PY{o}{.}\PY{n}{zeros}\PY{p}{(}\PY{n}{link\PYZus{}shape}\PY{p}{)}\PY{p}{,} \PY{l+m+mi}{2}\PY{o}{*}\PY{n}{np}\PY{o}{.}\PY{n}{pi}\PY{o}{*}\PY{n}{torch}\PY{o}{.}\PY{n}{ones}\PY{p}{(}\PY{n}{link\PYZus{}shape}\PY{p}{)}\PY{p}{)}

\PY{n}{n\PYZus{}layers} \PY{o}{=} \PY{l+m+mi}{16}
\PY{n}{n\PYZus{}s\PYZus{}nets} \PY{o}{=} \PY{l+m+mi}{2}
\PY{n}{hidden\PYZus{}sizes} \PY{o}{=} \PY{p}{[}\PY{l+m+mi}{8}\PY{p}{,}\PY{l+m+mi}{8}\PY{p}{]}
\PY{n}{kernel\PYZus{}size} \PY{o}{=} \PY{l+m+mi}{3}
\PY{n}{layers} \PY{o}{=} \PY{n}{make\PYZus{}u1\PYZus{}equiv\PYZus{}layers}\PY{p}{(}\PY{n}{lattice\PYZus{}shape}\PY{o}{=}\PY{n}{lattice\PYZus{}shape}\PY{p}{,} \PY{n}{n\PYZus{}layers}\PY{o}{=}\PY{n}{n\PYZus{}layers}\PY{p}{,} \PY{n}{n\PYZus{}mixture\PYZus{}comps}\PY{o}{=}\PY{n}{n\PYZus{}s\PYZus{}nets}\PY{p}{,}
                             \PY{n}{hidden\PYZus{}sizes}\PY{o}{=}\PY{n}{hidden\PYZus{}sizes}\PY{p}{,} \PY{n}{kernel\PYZus{}size}\PY{o}{=}\PY{n}{kernel\PYZus{}size}\PY{p}{)}
\PY{n}{set\PYZus{}weights}\PY{p}{(}\PY{n}{layers}\PY{p}{)}
\PY{n}{model} \PY{o}{=} \PY{p}{\PYZob{}}\PY{l+s+s1}{\PYZsq{}}\PY{l+s+s1}{layers}\PY{l+s+s1}{\PYZsq{}}\PY{p}{:} \PY{n}{layers}\PY{p}{,} \PY{l+s+s1}{\PYZsq{}}\PY{l+s+s1}{prior}\PY{l+s+s1}{\PYZsq{}}\PY{p}{:} \PY{n}{prior}\PY{p}{\PYZcb{}}

\PY{c+c1}{\PYZsh{} Training}
\PY{n}{base\PYZus{}lr} \PY{o}{=} \PY{l+m+mf}{.001}
\PY{n}{optimizer} \PY{o}{=} \PY{n}{torch}\PY{o}{.}\PY{n}{optim}\PY{o}{.}\PY{n}{Adam}\PY{p}{(}\PY{n}{model}\PY{p}{[}\PY{l+s+s1}{\PYZsq{}}\PY{l+s+s1}{layers}\PY{l+s+s1}{\PYZsq{}}\PY{p}{]}\PY{o}{.}\PY{n}{parameters}\PY{p}{(}\PY{p}{)}\PY{p}{,} \PY{n}{lr}\PY{o}{=}\PY{n}{base\PYZus{}lr}\PY{p}{)}
    \end{Verbatim}
  \end{tcolorbox}

  \begin{tcolorbox}[breakable, size=fbox, boxrule=1pt, pad at break*=1mm,colback=cellbackground, colframe=cellborder]
    \begin{Verbatim}[commandchars=\\\{\},fontsize=\small]
\PY{k}{if} \PY{n}{use\PYZus{}pretrained}\PY{p}{:}
    \PY{n+nb}{print}\PY{p}{(}\PY{l+s+s1}{\PYZsq{}}\PY{l+s+s1}{Loading pre\PYZhy{}trained model}\PY{l+s+s1}{\PYZsq{}}\PY{p}{)}
    \PY{n}{u1\PYZus{}trained\PYZus{}weights} \PY{o}{=} \PY{n}{torch}\PY{o}{.}\PY{n}{load}\PY{p}{(}\PY{n}{io}\PY{o}{.}\PY{n}{BytesIO}\PY{p}{(}\PY{n}{base64}\PY{o}{.}\PY{n}{b64decode}\PY{p}{(}\PY{l+s+sa}{b}\PY{l+s+s2}{\PYZdq{}\PYZdq{}\PYZdq{}}
\PY{l+s+s2}{\PYZlt{}snipped base64 blob\PYZgt{}}
\PY{l+s+s2}{    }\PY{l+s+s2}{\PYZdq{}\PYZdq{}\PYZdq{}}\PY{o}{.}\PY{n}{strip}\PY{p}{(}\PY{p}{)}\PY{p}{)}\PY{p}{)}\PY{p}{,} \PY{n}{map\PYZus{}location}\PY{o}{=}\PY{n}{torch}\PY{o}{.}\PY{n}{device}\PY{p}{(}\PY{l+s+s1}{\PYZsq{}}\PY{l+s+s1}{cpu}\PY{l+s+s1}{\PYZsq{}}\PY{p}{)}\PY{p}{)}
    \PY{n}{model}\PY{p}{[}\PY{l+s+s1}{\PYZsq{}}\PY{l+s+s1}{layers}\PY{l+s+s1}{\PYZsq{}}\PY{p}{]}\PY{o}{.}\PY{n}{load\PYZus{}state\PYZus{}dict}\PY{p}{(}\PY{n}{u1\PYZus{}trained\PYZus{}weights}\PY{p}{)}
    \PY{k}{if} \PY{n}{torch\PYZus{}device} \PY{o}{==} \PY{l+s+s1}{\PYZsq{}}\PY{l+s+s1}{cuda}\PY{l+s+s1}{\PYZsq{}}\PY{p}{:}
        \PY{n}{model}\PY{p}{[}\PY{l+s+s1}{\PYZsq{}}\PY{l+s+s1}{layers}\PY{l+s+s1}{\PYZsq{}}\PY{p}{]}\PY{o}{.}\PY{n}{cuda}\PY{p}{(}\PY{p}{)}
\PY{k}{else}\PY{p}{:}
    \PY{n+nb}{print}\PY{p}{(}\PY{l+s+s1}{\PYZsq{}}\PY{l+s+s1}{Skipping pre\PYZhy{}trained model}\PY{l+s+s1}{\PYZsq{}}\PY{p}{)}
    \end{Verbatim}
  \end{tcolorbox}

{\color{gray}
    \begin{Verbatim}[commandchars=\\\{\},fontsize=\small]
>>> Loading pre-trained model
    \end{Verbatim}
}

  \begin{tcolorbox}[breakable, size=fbox, boxrule=1pt, pad at break*=1mm,colback=cellbackground, colframe=cellborder]
    \begin{Verbatim}[commandchars=\\\{\},fontsize=\small]
\PY{n}{N\PYZus{}era} \PY{o}{=} \PY{l+m+mi}{10}
\PY{n}{N\PYZus{}epoch} \PY{o}{=} \PY{l+m+mi}{100}
\PY{n}{batch\PYZus{}size} \PY{o}{=} \PY{l+m+mi}{64}
\PY{n}{print\PYZus{}freq} \PY{o}{=} \PY{n}{N\PYZus{}epoch} \PY{c+c1}{\PYZsh{} epochs}
\PY{n}{plot\PYZus{}freq} \PY{o}{=} \PY{l+m+mi}{1} \PY{c+c1}{\PYZsh{} epochs}

\PY{n}{history} \PY{o}{=} \PY{p}{\PYZob{}}
    \PY{l+s+s1}{\PYZsq{}}\PY{l+s+s1}{loss}\PY{l+s+s1}{\PYZsq{}} \PY{p}{:} \PY{p}{[}\PY{p}{]}\PY{p}{,}
    \PY{l+s+s1}{\PYZsq{}}\PY{l+s+s1}{logp}\PY{l+s+s1}{\PYZsq{}} \PY{p}{:} \PY{p}{[}\PY{p}{]}\PY{p}{,}
    \PY{l+s+s1}{\PYZsq{}}\PY{l+s+s1}{logq}\PY{l+s+s1}{\PYZsq{}} \PY{p}{:} \PY{p}{[}\PY{p}{]}\PY{p}{,}
    \PY{l+s+s1}{\PYZsq{}}\PY{l+s+s1}{ess}\PY{l+s+s1}{\PYZsq{}} \PY{p}{:} \PY{p}{[}\PY{p}{]}
\PY{p}{\PYZcb{}}
    \end{Verbatim}
  \end{tcolorbox}

  \begin{tcolorbox}[breakable, size=fbox, boxrule=1pt, pad at break*=1mm,colback=cellbackground, colframe=cellborder]
    \begin{Verbatim}[commandchars=\\\{\},fontsize=\small]
\PY{k}{if} \PY{o+ow}{not} \PY{n}{use\PYZus{}pretrained}\PY{p}{:}
    \PY{p}{[}\PY{n}{plt}\PY{o}{.}\PY{n}{close}\PY{p}{(}\PY{n}{plt}\PY{o}{.}\PY{n}{figure}\PY{p}{(}\PY{n}{fignum}\PY{p}{)}\PY{p}{)} \PY{k}{for} \PY{n}{fignum} \PY{o+ow}{in} \PY{n}{plt}\PY{o}{.}\PY{n}{get\PYZus{}fignums}\PY{p}{(}\PY{p}{)}\PY{p}{]} \PY{c+c1}{\PYZsh{} close all existing figures}
    \PY{n}{live\PYZus{}plot} \PY{o}{=} \PY{n}{init\PYZus{}live\PYZus{}plot}\PY{p}{(}\PY{p}{)}

    \PY{k}{for} \PY{n}{era} \PY{o+ow}{in} \PY{n+nb}{range}\PY{p}{(}\PY{n}{N\PYZus{}era}\PY{p}{)}\PY{p}{:}
        \PY{k}{for} \PY{n}{epoch} \PY{o+ow}{in} \PY{n+nb}{range}\PY{p}{(}\PY{n}{N\PYZus{}epoch}\PY{p}{)}\PY{p}{:}
            \PY{n}{train\PYZus{}step}\PY{p}{(}\PY{n}{model}\PY{p}{,} \PY{n}{u1\PYZus{}action}\PY{p}{,} \PY{n}{calc\PYZus{}dkl}\PY{p}{,} \PY{n}{optimizer}\PY{p}{,} \PY{n}{history}\PY{p}{)}

            \PY{k}{if} \PY{n}{epoch} \PY{o}{\PYZpc{}} \PY{n}{print\PYZus{}freq} \PY{o}{==} \PY{l+m+mi}{0}\PY{p}{:}
                \PY{n}{print\PYZus{}metrics}\PY{p}{(}\PY{n}{history}\PY{p}{,} \PY{n}{avg\PYZus{}last\PYZus{}N\PYZus{}epochs}\PY{o}{=}\PY{n}{print\PYZus{}freq}\PY{p}{)}
                  
            \PY{k}{if} \PY{n}{epoch} \PY{o}{\PYZpc{}} \PY{n}{plot\PYZus{}freq} \PY{o}{==} \PY{l+m+mi}{0}\PY{p}{:}
                \PY{n}{update\PYZus{}plots}\PY{p}{(}\PY{n}{history}\PY{p}{,} \PY{o}{*}\PY{o}{*}\PY{n}{live\PYZus{}plot}\PY{p}{)}
\PY{k}{else}\PY{p}{:}
    \PY{n+nb}{print}\PY{p}{(}\PY{l+s+s1}{\PYZsq{}}\PY{l+s+s1}{Skipping training}\PY{l+s+s1}{\PYZsq{}}\PY{p}{)}
    \end{Verbatim}
  \end{tcolorbox}

{\color{gray}
    \begin{Verbatim}[commandchars=\\\{\},fontsize=\small]
>>> Skipping training
    \end{Verbatim}
}

  \begin{tcolorbox}[breakable, size=fbox, boxrule=1pt, pad at break*=1mm,colback=cellbackground, colframe=cellborder]
    \begin{Verbatim}[commandchars=\\\{\},fontsize=\small]
\PY{n+nb}{print}\PY{p}{(}\PY{l+s+s1}{\PYZsq{}}\PY{l+s+s1}{Model weights blob:}\PY{l+s+se}{\PYZbs{}n}\PY{l+s+s1}{===}\PY{l+s+s1}{\PYZsq{}}\PY{p}{)}
\PY{n}{serialized\PYZus{}model} \PY{o}{=} \PY{n}{io}\PY{o}{.}\PY{n}{BytesIO}\PY{p}{(}\PY{p}{)}
\PY{n}{torch}\PY{o}{.}\PY{n}{save}\PY{p}{(}\PY{n}{model}\PY{p}{[}\PY{l+s+s1}{\PYZsq{}}\PY{l+s+s1}{layers}\PY{l+s+s1}{\PYZsq{}}\PY{p}{]}\PY{o}{.}\PY{n}{state\PYZus{}dict}\PY{p}{(}\PY{p}{)}\PY{p}{,} \PY{n}{serialized\PYZus{}model}\PY{p}{)}
\PY{n+nb}{print}\PY{p}{(}\PY{n}{base64}\PY{o}{.}\PY{n}{b64encode}\PY{p}{(}\PY{n}{serialized\PYZus{}model}\PY{o}{.}\PY{n}{getbuffer}\PY{p}{(}\PY{p}{)}\PY{p}{)}\PY{o}{.}\PY{n}{decode}\PY{p}{(}\PY{l+s+s1}{\PYZsq{}}\PY{l+s+s1}{utf\PYZhy{}8}\PY{l+s+s1}{\PYZsq{}}\PY{p}{)}\PY{p}{)}
\PY{n+nb}{print}\PY{p}{(}\PY{l+s+s1}{\PYZsq{}}\PY{l+s+s1}{===}\PY{l+s+s1}{\PYZsq{}}\PY{p}{)}
    \end{Verbatim}
  \end{tcolorbox}

  \hypertarget{evaluate-the-model}{%
\subsection{\texorpdfstring{\textbf{Evaluate the
model}}{Evaluate the model}}\label{evaluate-the-model}}

We can apply the same checks to our trained U(1) model as we did to the
\(\phi^4\) model above.

  It's harder to visualize gauge fields because the variables are angular
and there are multiple degrees of freedom per site. However, we can
double-check that a random draw from the trained model produces samples
with \(\log q\) correlated with \(\log p\).

  \begin{tcolorbox}[breakable, size=fbox, boxrule=1pt, pad at break*=1mm,colback=cellbackground, colframe=cellborder]
    \begin{Verbatim}[commandchars=\\\{\},fontsize=\small]
\PY{n}{layers}\PY{p}{,} \PY{n}{prior} \PY{o}{=} \PY{n}{model}\PY{p}{[}\PY{l+s+s1}{\PYZsq{}}\PY{l+s+s1}{layers}\PY{l+s+s1}{\PYZsq{}}\PY{p}{]}\PY{p}{,} \PY{n}{model}\PY{p}{[}\PY{l+s+s1}{\PYZsq{}}\PY{l+s+s1}{prior}\PY{l+s+s1}{\PYZsq{}}\PY{p}{]}
\PY{n}{torch\PYZus{}x}\PY{p}{,} \PY{n}{torch\PYZus{}logq} \PY{o}{=} \PY{n}{apply\PYZus{}flow\PYZus{}to\PYZus{}prior}\PY{p}{(}\PY{n}{prior}\PY{p}{,} \PY{n}{layers}\PY{p}{,} \PY{n}{batch\PYZus{}size}\PY{o}{=}\PY{l+m+mi}{1024}\PY{p}{)}

\PY{n}{S\PYZus{}eff} \PY{o}{=} \PY{o}{\PYZhy{}}\PY{n}{grab}\PY{p}{(}\PY{n}{torch\PYZus{}logq}\PY{p}{)}
\PY{n}{S} \PY{o}{=} \PY{n}{grab}\PY{p}{(}\PY{n}{u1\PYZus{}action}\PY{p}{(}\PY{n}{torch\PYZus{}x}\PY{p}{)}\PY{p}{)}
\PY{n}{fit\PYZus{}b} \PY{o}{=} \PY{n}{np}\PY{o}{.}\PY{n}{mean}\PY{p}{(}\PY{n}{S}\PY{p}{)} \PY{o}{\PYZhy{}} \PY{n}{np}\PY{o}{.}\PY{n}{mean}\PY{p}{(}\PY{n}{S\PYZus{}eff}\PY{p}{)}
\PY{n+nb}{print}\PY{p}{(}\PY{l+s+sa}{f}\PY{l+s+s1}{\PYZsq{}}\PY{l+s+s1}{slope 1 linear regression S = S\PYZus{}eff + }\PY{l+s+si}{\PYZob{}}\PY{n}{fit\PYZus{}b}\PY{l+s+si}{:}\PY{l+s+s1}{.4f}\PY{l+s+si}{\PYZcb{}}\PY{l+s+s1}{\PYZsq{}}\PY{p}{)}
\PY{n}{fig}\PY{p}{,} \PY{n}{ax} \PY{o}{=} \PY{n}{plt}\PY{o}{.}\PY{n}{subplots}\PY{p}{(}\PY{l+m+mi}{1}\PY{p}{,}\PY{l+m+mi}{1}\PY{p}{,} \PY{n}{dpi}\PY{o}{=}\PY{l+m+mi}{125}\PY{p}{,} \PY{n}{figsize}\PY{o}{=}\PY{p}{(}\PY{l+m+mi}{4}\PY{p}{,}\PY{l+m+mi}{4}\PY{p}{)}\PY{p}{)}
\PY{n}{ax}\PY{o}{.}\PY{n}{hist2d}\PY{p}{(}\PY{n}{S\PYZus{}eff}\PY{p}{,} \PY{n}{S}\PY{p}{,} \PY{n}{bins}\PY{o}{=}\PY{l+m+mi}{20}\PY{p}{,} \PY{n+nb}{range}\PY{o}{=}\PY{p}{[}\PY{p}{[}\PY{l+m+mi}{175}\PY{p}{,} \PY{l+m+mi}{225}\PY{p}{]}\PY{p}{,} \PY{p}{[}\PY{o}{\PYZhy{}}\PY{l+m+mi}{110}\PY{p}{,} \PY{o}{\PYZhy{}}\PY{l+m+mi}{60}\PY{p}{]}\PY{p}{]}\PY{p}{)}
\PY{n}{ax}\PY{o}{.}\PY{n}{set\PYZus{}xlabel}\PY{p}{(}\PY{l+s+sa}{r}\PY{l+s+s1}{\PYZsq{}}\PY{l+s+s1}{\PYZdl{}S\PYZus{}}\PY{l+s+s1}{\PYZob{}}\PY{l+s+s1}{\PYZbs{}}\PY{l+s+s1}{mathrm}\PY{l+s+si}{\PYZob{}eff\PYZcb{}}\PY{l+s+s1}{\PYZcb{} = \PYZhy{}}\PY{l+s+s1}{\PYZbs{}}\PY{l+s+s1}{log\PYZti{}q(x)\PYZdl{}}\PY{l+s+s1}{\PYZsq{}}\PY{p}{)}
\PY{n}{ax}\PY{o}{.}\PY{n}{set\PYZus{}ylabel}\PY{p}{(}\PY{l+s+sa}{r}\PY{l+s+s1}{\PYZsq{}}\PY{l+s+s1}{\PYZdl{}S(x)\PYZdl{}}\PY{l+s+s1}{\PYZsq{}}\PY{p}{)}
\PY{n}{ax}\PY{o}{.}\PY{n}{set\PYZus{}aspect}\PY{p}{(}\PY{l+s+s1}{\PYZsq{}}\PY{l+s+s1}{equal}\PY{l+s+s1}{\PYZsq{}}\PY{p}{)}
\PY{n}{xs} \PY{o}{=} \PY{n}{np}\PY{o}{.}\PY{n}{linspace}\PY{p}{(}\PY{l+m+mi}{175}\PY{p}{,} \PY{l+m+mi}{225}\PY{p}{,} \PY{n}{num}\PY{o}{=}\PY{l+m+mi}{4}\PY{p}{,} \PY{n}{endpoint}\PY{o}{=}\PY{k+kc}{True}\PY{p}{)}
\PY{n}{ax}\PY{o}{.}\PY{n}{plot}\PY{p}{(}\PY{n}{xs}\PY{p}{,} \PY{n}{xs} \PY{o}{+} \PY{n}{fit\PYZus{}b}\PY{p}{,} \PY{l+s+s1}{\PYZsq{}}\PY{l+s+s1}{:}\PY{l+s+s1}{\PYZsq{}}\PY{p}{,} \PY{n}{color}\PY{o}{=}\PY{l+s+s1}{\PYZsq{}}\PY{l+s+s1}{w}\PY{l+s+s1}{\PYZsq{}}\PY{p}{,} \PY{n}{label}\PY{o}{=}\PY{l+s+s1}{\PYZsq{}}\PY{l+s+s1}{slope 1 fit}\PY{l+s+s1}{\PYZsq{}}\PY{p}{)}
\PY{n}{plt}\PY{o}{.}\PY{n}{legend}\PY{p}{(}\PY{n}{prop}\PY{o}{=}\PY{p}{\PYZob{}}\PY{l+s+s1}{\PYZsq{}}\PY{l+s+s1}{size}\PY{l+s+s1}{\PYZsq{}}\PY{p}{:} \PY{l+m+mi}{6}\PY{p}{\PYZcb{}}\PY{p}{)}
\PY{n}{plt}\PY{o}{.}\PY{n}{show}\PY{p}{(}\PY{p}{)}
    \end{Verbatim}
  \end{tcolorbox}

{\color{gray}
    \begin{Verbatim}[commandchars=\\\{\},fontsize=\small]
>>> slope 1 linear regression S = S\_eff + -286.5096
    \end{Verbatim}
}

    \begin{center}
    \adjustimage{max size={0.9\linewidth}{0.9\paperheight}}{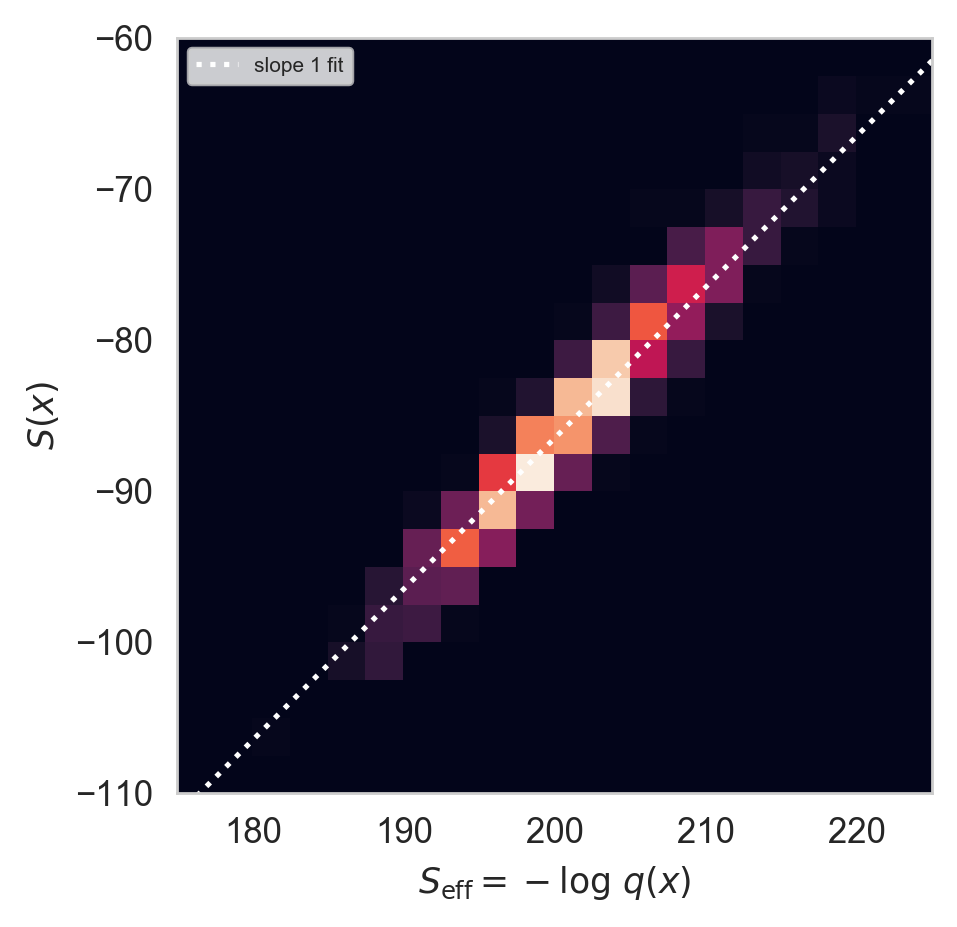}
    \end{center}
    { \hspace*{\fill} \\}
    
  We can see how the model density evolved over training time to become
well-correlated with \(p(x)\) over time (if
\texttt{use\_pretrained\ ==\ False}).

  \begin{tcolorbox}[breakable, size=fbox, boxrule=1pt, pad at break*=1mm,colback=cellbackground, colframe=cellborder]
    \begin{Verbatim}[commandchars=\\\{\},fontsize=\small]
\PY{k}{if} \PY{o+ow}{not} \PY{n}{use\PYZus{}pretrained}\PY{p}{:}
    \PY{n}{fig}\PY{p}{,} \PY{n}{axes} \PY{o}{=} \PY{n}{plt}\PY{o}{.}\PY{n}{subplots}\PY{p}{(}\PY{l+m+mi}{1}\PY{p}{,} \PY{l+m+mi}{10}\PY{p}{,} \PY{n}{dpi}\PY{o}{=}\PY{l+m+mi}{125}\PY{p}{,} \PY{n}{sharey}\PY{o}{=}\PY{k+kc}{True}\PY{p}{,} \PY{n}{figsize}\PY{o}{=}\PY{p}{(}\PY{l+m+mi}{10}\PY{p}{,} \PY{l+m+mi}{1}\PY{p}{)}\PY{p}{)}
    \PY{n}{logq\PYZus{}hist} \PY{o}{=} \PY{n}{np}\PY{o}{.}\PY{n}{array}\PY{p}{(}\PY{n}{history}\PY{p}{[}\PY{l+s+s1}{\PYZsq{}}\PY{l+s+s1}{logq}\PY{l+s+s1}{\PYZsq{}}\PY{p}{]}\PY{p}{)}\PY{o}{.}\PY{n}{reshape}\PY{p}{(}\PY{n}{N\PYZus{}era}\PY{p}{,} \PY{o}{\PYZhy{}}\PY{l+m+mi}{1}\PY{p}{)}\PY{p}{[}\PY{p}{:}\PY{p}{:}\PY{n}{N\PYZus{}era}\PY{o}{/}\PY{o}{/}\PY{l+m+mi}{10}\PY{p}{]}
    \PY{n}{logp\PYZus{}hist} \PY{o}{=} \PY{n}{np}\PY{o}{.}\PY{n}{array}\PY{p}{(}\PY{n}{history}\PY{p}{[}\PY{l+s+s1}{\PYZsq{}}\PY{l+s+s1}{logp}\PY{l+s+s1}{\PYZsq{}}\PY{p}{]}\PY{p}{)}\PY{o}{.}\PY{n}{reshape}\PY{p}{(}\PY{n}{N\PYZus{}era}\PY{p}{,} \PY{o}{\PYZhy{}}\PY{l+m+mi}{1}\PY{p}{)}\PY{p}{[}\PY{p}{:}\PY{p}{:}\PY{n}{N\PYZus{}era}\PY{o}{/}\PY{o}{/}\PY{l+m+mi}{10}\PY{p}{]}
    \PY{k}{for} \PY{n}{i}\PY{p}{,} \PY{p}{(}\PY{n}{ax}\PY{p}{,} \PY{n}{logq}\PY{p}{,} \PY{n}{logp}\PY{p}{)} \PY{o+ow}{in} \PY{n+nb}{enumerate}\PY{p}{(}\PY{n+nb}{zip}\PY{p}{(}\PY{n}{axes}\PY{p}{,} \PY{n}{logq\PYZus{}hist}\PY{p}{,} \PY{n}{logp\PYZus{}hist}\PY{p}{)}\PY{p}{)}\PY{p}{:}
        \PY{n}{ax}\PY{o}{.}\PY{n}{hist2d}\PY{p}{(}\PY{o}{\PYZhy{}}\PY{n}{logq}\PY{p}{,} \PY{o}{\PYZhy{}}\PY{n}{logp}\PY{p}{,} \PY{n}{bins}\PY{o}{=}\PY{l+m+mi}{20}\PY{p}{,} \PY{n+nb}{range}\PY{o}{=}\PY{p}{[}\PY{p}{[}\PY{l+m+mi}{175}\PY{p}{,} \PY{l+m+mi}{225}\PY{p}{]}\PY{p}{,} \PY{p}{[}\PY{o}{\PYZhy{}}\PY{l+m+mi}{110}\PY{p}{,} \PY{o}{\PYZhy{}}\PY{l+m+mi}{60}\PY{p}{]}\PY{p}{]}\PY{p}{)}
        \PY{k}{if} \PY{n}{i} \PY{o}{==} \PY{l+m+mi}{0}\PY{p}{:}
            \PY{n}{ax}\PY{o}{.}\PY{n}{set\PYZus{}ylabel}\PY{p}{(}\PY{l+s+sa}{r}\PY{l+s+s1}{\PYZsq{}}\PY{l+s+s1}{\PYZdl{}S(x)\PYZdl{}}\PY{l+s+s1}{\PYZsq{}}\PY{p}{)}
        \PY{n}{ax}\PY{o}{.}\PY{n}{set\PYZus{}xlabel}\PY{p}{(}\PY{l+s+sa}{r}\PY{l+s+s1}{\PYZsq{}}\PY{l+s+s1}{\PYZdl{}S\PYZus{}}\PY{l+s+s1}{\PYZob{}}\PY{l+s+s1}{\PYZbs{}}\PY{l+s+s1}{mathrm}\PY{l+s+si}{\PYZob{}eff\PYZcb{}}\PY{l+s+s1}{\PYZcb{}\PYZdl{}}\PY{l+s+s1}{\PYZsq{}}\PY{p}{)}
        \PY{n}{ax}\PY{o}{.}\PY{n}{set\PYZus{}title}\PY{p}{(}\PY{l+s+sa}{f}\PY{l+s+s1}{\PYZsq{}}\PY{l+s+s1}{Era }\PY{l+s+si}{\PYZob{}}\PY{n}{i} \PY{o}{*} \PY{p}{(}\PY{n}{N\PYZus{}era}\PY{o}{/}\PY{o}{/}\PY{l+m+mi}{10}\PY{p}{)}\PY{l+s+si}{\PYZcb{}}\PY{l+s+s1}{\PYZsq{}}\PY{p}{)}
        \PY{n}{ax}\PY{o}{.}\PY{n}{set\PYZus{}xticks}\PY{p}{(}\PY{p}{[}\PY{p}{]}\PY{p}{)}
        \PY{n}{ax}\PY{o}{.}\PY{n}{set\PYZus{}yticks}\PY{p}{(}\PY{p}{[}\PY{p}{]}\PY{p}{)}
        \PY{n}{ax}\PY{o}{.}\PY{n}{set\PYZus{}aspect}\PY{p}{(}\PY{l+s+s1}{\PYZsq{}}\PY{l+s+s1}{equal}\PY{l+s+s1}{\PYZsq{}}\PY{p}{)}
    \PY{n}{plt}\PY{o}{.}\PY{n}{show}\PY{p}{(}\PY{p}{)}
\PY{k}{else}\PY{p}{:}
    \PY{n+nb}{print}\PY{p}{(}\PY{l+s+s1}{\PYZsq{}}\PY{l+s+s1}{Skipping plot because use\PYZus{}pretrained == True}\PY{l+s+s1}{\PYZsq{}}\PY{p}{)}
    \end{Verbatim}
  \end{tcolorbox}

{\color{gray}
    \begin{Verbatim}[commandchars=\\\{\},fontsize=\small]
>>> Skipping plot because use\_pretrained == True
    \end{Verbatim}
}

  We can reuse our Metropolis independence sampler from above to sample
the theory using our model and check that we get a good acceptance rate.
You should see an accept rate around 40-50\% for the model trained
above.

  \begin{tcolorbox}[breakable, size=fbox, boxrule=1pt, pad at break*=1mm,colback=cellbackground, colframe=cellborder]
    \begin{Verbatim}[commandchars=\\\{\},fontsize=\small]
\PY{n}{ensemble\PYZus{}size} \PY{o}{=} \PY{l+m+mi}{8192}
\PY{n}{u1\PYZus{}ens} \PY{o}{=} \PY{n}{make\PYZus{}mcmc\PYZus{}ensemble}\PY{p}{(}\PY{n}{model}\PY{p}{,} \PY{n}{u1\PYZus{}action}\PY{p}{,} \PY{l+m+mi}{64}\PY{p}{,} \PY{n}{ensemble\PYZus{}size}\PY{p}{)}
\PY{n+nb}{print}\PY{p}{(}\PY{l+s+s2}{\PYZdq{}}\PY{l+s+s2}{Accept rate:}\PY{l+s+s2}{\PYZdq{}}\PY{p}{,} \PY{n}{np}\PY{o}{.}\PY{n}{mean}\PY{p}{(}\PY{n}{u1\PYZus{}ens}\PY{p}{[}\PY{l+s+s1}{\PYZsq{}}\PY{l+s+s1}{accepted}\PY{l+s+s1}{\PYZsq{}}\PY{p}{]}\PY{p}{)}\PY{p}{)}
    \end{Verbatim}
  \end{tcolorbox}

{\color{gray}
    \begin{Verbatim}[commandchars=\\\{\},fontsize=\small]
>>> Accept rate: 0.244140625
    \end{Verbatim}
}

  Algorithms like HMC have a difficult time sampling from different
topological sectors, exhibiting ``topological freezing'' where the
topological charge \(Q\) moves very slowly in Markov chain time. We can
measure this quantity on the ensemble of U(1) configurations we just
generated and see that it mixes quickly using our direct sampling
approach. See \cite{Kanwar:2020xzo} for a detailed comparison against
two standard approaches.

  \begin{tcolorbox}[breakable, size=fbox, boxrule=1pt, pad at break*=1mm,colback=cellbackground, colframe=cellborder]
    \begin{Verbatim}[commandchars=\\\{\},fontsize=\small]
\PY{n}{Q} \PY{o}{=} \PY{n}{grab}\PY{p}{(}\PY{n}{topo\PYZus{}charge}\PY{p}{(}\PY{n}{torch}\PY{o}{.}\PY{n}{stack}\PY{p}{(}\PY{n}{u1\PYZus{}ens}\PY{p}{[}\PY{l+s+s1}{\PYZsq{}}\PY{l+s+s1}{x}\PY{l+s+s1}{\PYZsq{}}\PY{p}{]}\PY{p}{,} \PY{n}{axis}\PY{o}{=}\PY{l+m+mi}{0}\PY{p}{)}\PY{p}{)}\PY{p}{)}
\PY{n}{plt}\PY{o}{.}\PY{n}{figure}\PY{p}{(}\PY{n}{figsize}\PY{o}{=}\PY{p}{(}\PY{l+m+mi}{5}\PY{p}{,}\PY{l+m+mf}{3.5}\PY{p}{)}\PY{p}{,} \PY{n}{dpi}\PY{o}{=}\PY{l+m+mi}{125}\PY{p}{)}
\PY{n}{plt}\PY{o}{.}\PY{n}{plot}\PY{p}{(}\PY{n}{Q}\PY{p}{)}
\PY{n}{plt}\PY{o}{.}\PY{n}{xlabel}\PY{p}{(}\PY{l+s+sa}{r}\PY{l+s+s1}{\PYZsq{}}\PY{l+s+s1}{\PYZdl{}t\PYZus{}}\PY{l+s+si}{\PYZob{}MC\PYZcb{}}\PY{l+s+s1}{\PYZdl{}}\PY{l+s+s1}{\PYZsq{}}\PY{p}{)}
\PY{n}{plt}\PY{o}{.}\PY{n}{ylabel}\PY{p}{(}\PY{l+s+sa}{r}\PY{l+s+s1}{\PYZsq{}}\PY{l+s+s1}{topological charge \PYZdl{}Q\PYZdl{}}\PY{l+s+s1}{\PYZsq{}}\PY{p}{)}
\PY{n}{plt}\PY{o}{.}\PY{n}{show}\PY{p}{(}\PY{p}{)}
    \end{Verbatim}
  \end{tcolorbox}

    \begin{center}
    \adjustimage{max size={0.9\linewidth}{0.9\paperheight}}{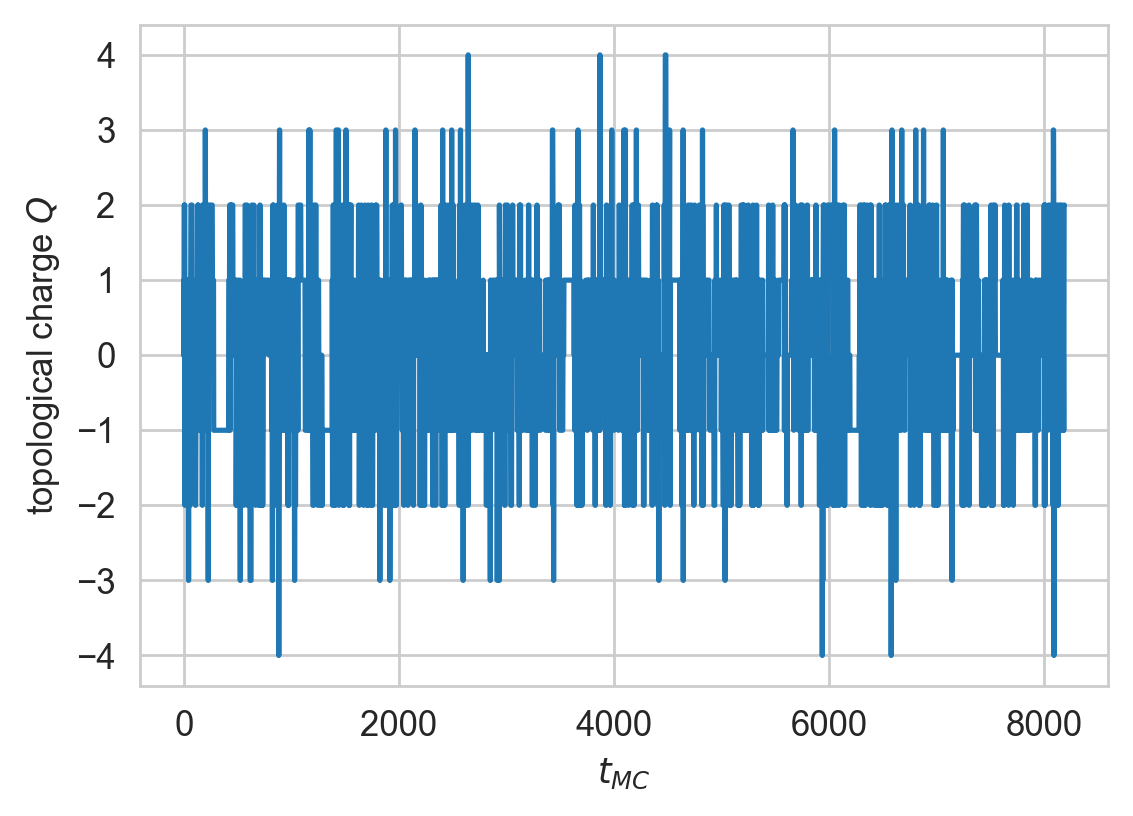}
    \end{center}
    { \hspace*{\fill} \\}
    
  As with scalar theory, the generated ensemble is asymptotically
unbiased. As an example of an observable measurement, we compute the
topological susceptibility below and compare against a value determined
from a large HMC ensemble evaluated at the same choice of parameters.

  \begin{tcolorbox}[breakable, size=fbox, boxrule=1pt, pad at break*=1mm,colback=cellbackground, colframe=cellborder]
    \begin{Verbatim}[commandchars=\\\{\},fontsize=\small]
\PY{n}{X\PYZus{}mean}\PY{p}{,} \PY{n}{X\PYZus{}err} \PY{o}{=} \PY{n}{bootstrap}\PY{p}{(}\PY{n}{Q}\PY{o}{*}\PY{o}{*}\PY{l+m+mi}{2}\PY{p}{,} \PY{n}{Nboot}\PY{o}{=}\PY{l+m+mi}{100}\PY{p}{,} \PY{n}{binsize}\PY{o}{=}\PY{l+m+mi}{16}\PY{p}{)}
\PY{n+nb}{print}\PY{p}{(}\PY{l+s+sa}{f}\PY{l+s+s1}{\PYZsq{}}\PY{l+s+s1}{Topological susceptibility = }\PY{l+s+si}{\PYZob{}}\PY{n}{X\PYZus{}mean}\PY{l+s+si}{:}\PY{l+s+s1}{.2f}\PY{l+s+si}{\PYZcb{}}\PY{l+s+s1}{ +/\PYZhy{} }\PY{l+s+si}{\PYZob{}}\PY{n}{X\PYZus{}err}\PY{l+s+si}{:}\PY{l+s+s1}{.2f}\PY{l+s+si}{\PYZcb{}}\PY{l+s+s1}{\PYZsq{}}\PY{p}{)}
\PY{n+nb}{print}\PY{p}{(}\PY{l+s+sa}{f}\PY{l+s+s1}{\PYZsq{}}\PY{l+s+s1}{... vs HMC estimate = 1.23 +/\PYZhy{} 0.02}\PY{l+s+s1}{\PYZsq{}}\PY{p}{)}
    \end{Verbatim}
  \end{tcolorbox}

{\color{gray}
    \begin{Verbatim}[commandchars=\\\{\},fontsize=\small]
>>> Topological susceptibility = 1.22 +/- 0.05
... {\ldots} vs HMC estimate = 1.23 +/- 0.02
    \end{Verbatim}
}

  \hypertarget{additional-references}{%
\section{Additional References}\label{additional-references}}

Below we provide a selection of references to related works.
\textbf{NOTE}: Please see the PDF version for linked bibliography
entries.

\begin{itemize}
\tightlist
\item
  \textbf{Normalizing flows:} Agnelli, et al.~(2010)
  \cite{Agnelli2010ClusteringAC}; Tabak and Vanden-Eijnden (2010)
  \cite{tabak2010}; Dinh, et al.~(2014) \cite{Dinh:2014}; Dinh, et
  al.~(2016) \cite{dinh2016density}; Papamakarios, et al.~(2019)
  \cite{papamakarios2019normalizing}
\item
  \textbf{Symmetries and equivariance:} Cohen and Welling (2016)
  \cite{cohen2016group}; Cohen, et al.~(2019) \cite{cohen2019gauge};
  Rezende, et al.~(2019) \cite{rezende2019equivariant}; Köhler, et
  al.~(2020) \cite{kohler2020equivariant}; Luo, et al.~(2020)
  \cite{luo2020gauge}; Favoni, et al.~(2020) \cite{favoni2020lattice}
\item
  \textbf{Flows on manifolds:} Gemici, et al.~(2016)
  \cite{gemici2016normalizing}; Falorsi, et al.~(2019)
  \cite{falorsi19lie}; Finzi, et al.~(2020) \cite{Finzi:2020}; Mathieu
  and Nickel (2020) \cite{mathieu2020riemannian}; Falorsi and Forré
  (2020) \cite{falorsi2020neural}
\item
  \textbf{Applications of flows:} Müller, et al.~(2018)
  \cite{mller2018neural}; Noé, et al.~(2019) \cite{noe2019boltzmann};
  Wu, et al.~(2020) \cite{wu2020stochastic}; Dibak, et al.~(2020)
  \cite{dibak2020temperaturesteerable}; Nicoli, et al.~(2021)
  \cite{PhysRevLett.126.032001}
\end{itemize}

  \hypertarget{acknowledgments}{%
\section{Acknowledgments}\label{acknowledgments}}

GK, DB, DCH, and PES are supported in part by the U.S. Department of
Energy, Office of Science, Office of Nuclear Physics, under grant
Contract Number DESC0011090. PES is additionally supported by the
National Science Foundation under EAGER grant 2035015, by the U.S. DOE
Early Career Award DE-SC0021006, by a NEC research award, and by the
Carl G and Shirley Sontheimer Research Fund. KC is supported by the
National Science Foundation under the awards ACI1450310, OAC1836650, and
OAC-1841471 and by the Moore-Sloan data science environment at NYU. MSA
thanks the Flatiron Institute and is supported by the Carl Feinberg
Fellowship in Theoretical Physics and the James Arthur Fellowship. This
work is associated with an ALCF Aurora Early Science Program project and
was supported by the Argonne Leadership Computing Facility, which is a
U.S. Department of Energy Office of Science User Facility operated under
contract DE-AC02-06CH11357. This work is supported by the U.S. National
Science Foundation under Cooperative Agreement PHY2019786 (The NSF AI
Institute for Artificial Intelligence and Fundamental Interactions,
http://iaifi.org/).


\bibliographystyle{apsrev4-1}
\bibliography{citations}

\end{document}